\documentclass[%
reprint,
superscriptaddress,
amsmath,amssymb,
aps,
prd,
]{revtex4-2}

\usepackage{natbib}                                         
\usepackage[T1]{fontenc}                                    
\usepackage[colorlinks=true,citecolor=royalBlue,linkcolor=bordeauxRed,urlcolor=oliveGreen]{hyperref}                                       
\usepackage{url}                                            
\usepackage[lcgreekalpha]{stix}
\usepackage{bm}                                             
\usepackage{cleveref}                                       
\usepackage{xcolor}                                         
\usepackage{physics}
\usepackage{dsfont}
\usepackage{multirow}
\usepackage{pifont}
\usepackage{adjustbox}
\usepackage{blindtext}

\graphicspath{{./pic/}}

\definecolor{bordeauxRed}{RGB}{178, 34, 52} 
\definecolor{royalBlue}{RGB}{65,105,225} 
\definecolor{oliveGreen}{RGB}{128,128,0} 

\hypersetup{
    colorlinks,
    linkcolor=bordeauxRed,
    citecolor=royalBlue,
    urlcolor=oliveGreen,
    breaklinks=true 
}

\begin{document}

\title{Superluminal modes in a quantum field simulator for cosmology from analog Transplanckian physics}

\author{Christian F.\@ Schmidt}
    \email{christian.schmidt@uni-jena.de}
    \affiliation{Theoretisch-Physikalisches Institut, Friedrich-Schiller-Universit\"{a}t Jena,\\
    Max-Wien-Platz 1, 07743 Jena, Germany}

\author{Stefan Floerchinger}
    \email{stefan.floerchinger@uni-jena.de}
    \affiliation{Theoretisch-Physikalisches Institut, Friedrich-Schiller-Universit\"{a}t Jena,\\
    Max-Wien-Platz 1, 07743 Jena, Germany}

\begin{abstract}
The quantum-field-theoretic description for the U(1)-Goldstone boson of a scalar Bose-Einstein condensate with time-dependent contact interactions is developed beyond the acoustic approximation in accordance with Bogoliubov theory.
The resulting effective action is mapped to a relativistic quantum field theory on a dispersive (or rainbow) cosmological spacetime which has a superluminal Corley-Jacobson dispersion relation.
Time-dependent changes of the s-wave scattering length to quantum-simulate cosmological particle production are accompanied by a time-dependent healing length that can be interpreted as an analog Planck length in the comoving frame. 
Non-adiabatic transitions acquire a dispersive character, which is thoroughly discussed.
The framework is applied to exponentially expanding or power-law contracting $(2+1)$-dimensional spacetimes which are known to produce scale-invariant cosmological power spectra. 
The sensitivity of these scenarios to the time-dependence of the Bogoliubov dispersion is investigated: 
We find a violation of scale-invariance via analytically trackable Transplanckian damping effects if the cut-off scale is not well separated from the horizon-crossing scale.
In case of the exponential expansion, these damping effects remarkably settle and converge to another scale-invariant plateau in the far ultraviolet regime where non-adiabatic transitions are suppressed by the high dispersion.
The developed framework enables quantitative access to more drastic analog cosmological scenarios with improved predictability in the ultraviolet regime that ultimately may lead to the observation of a scale-invariant cosmological power spectrum in the laboratory.
\end{abstract}
\maketitle

\section{Introduction}
Effective field theories (EFTs) for (linearized) perturbations in condensed matter systems exhibit structural analogies to quantum field theories on curved spacetimes (QFTCS) that can be utilized to perform analog quantum simulations of hallmark phenomena such as Hawking radiation \cite{Hawking1974,Hawking1975}, cosmological particle production \cite{Parker1969}, black-hole superradiance \cite{Zeldovich1971,Unruh1974} or the dynamical Casimir-effect \cite{Moore1970,FullingDavies1976} (consider \cite{Jacquet2020,BarceloLiberatiVisser2024,Schuetzhold2025,Grass2025} for recent reviews on this field of research).

It is a non-trivial question what happens when the aforementioned phenomena exceed the range of validity of the EFT from which they are derived.
For instance, the formalism of QFTCS that underlies both Hawking radiation and inflationary particle production \cite{Starobinsky1979,MukhanovChibisov1981,MukhanovFeldmanBrandenberger1992} (or alternative bouncing scenarios \cite{Wands1999,FinelliBrandenberger2002,Brandenberger2012}) is expected to break down on scales close to or below the Planck length, the so-called \emph{Transplanckian} scales, where quantum-gravitational effects need to be taken into account.
However, the outcome of both processes (asymptotic thermal radiation in the black-hole system, or a scale-invariant spectrum in case of primordial cosmology) relies on modes that become strongly gravitationally blue-shifted into the Transplanckian regime when traced back to their origin (the event horizon in the black-hole case, or the initial causally connected patch in the cosmological case) which led to concerns regarding their viability \cite{Jacobson1991,Brandenberger2000}.
Nevertheless, recent measurements of the Cosmological Microwave Background \cite{Planck2018} still confirm an almost scale-invariant spectrum of cosmological perturbations and therefore the cosmological EFT.

A phenomenological approach to analyze these so-called \emph{Transplanckian problems} (reviewed among others in \cite{MartinBrandenberger2013}) consists of imposing a modified dispersion relation (MDR) which deviates from linearity above some cut-off momentum to model Lorentz-invariance-violations (LIV) which characteristically arise in quantum gravity models such as Ho\v{r}ava-Lifshitz gravity \cite{Horava2009,Visser2009,SotiriouVisserWeinfurtner2009,Volovik2009HLG,Sotiriou2011} or Einstein-\AE{}ether gravity \cite{JacobsonMattingly2001,Cropp2014} (the tight constrains on such LIV are discussed in \cite{Liberati2013}).
Inspired by the subluminal dispersion relation of his condensed-matter-analog \cite{Unruh1981}, Unruh carried out a pioneering numerical study \cite{Unruh1995} that was investigated analytically and generalized in \cite{Brout1995}. 
These developments led Corley \& Jacobson to analyze the issue via the lowest-order (quartic) expansion of a subluminal dispersion relation \cite{CorleyJacobson1996} and later-on the corresponding superluminal form \cite{Corley1998} (consider \cite{Jacobson1999} for a review on these early developments).

For these models, the Transplanckian problem of cosmology was analyzed by Martin {\&} Brandenberger \cite{MartinBrandenberger2000,MartinBrandenberger2001,MartinBrandenberger2001b,MartinBrandenberger2002,BrandenbergerJorasMartin2002} and Niemeyer {\&} Parentani \cite{Niemeyer2001,NiemeyerParentani2001} 
who found a preservation of scale-invariance if the Hubble horizon is well separated from the cut-off scale \cite{NiemeyerParentani2001,Parentani2002,MacherParentani2008} (or equivalently adiabatic evolution in the ultraviolet \cite{MartinBrandenberger2000,NiemeyerParentani2001,Starobinsky2001,MartinBrandenberger2013}).
Similar concepts also showed a robustness of the Hawking effect for radiation from black holes if the surface gravity is well separated from the cut-off scale \cite{UnruhSchuetzhold2005,MacherParentani2009,CoutantParentani2014}.
However, there are also physically plausible constructions that violate these criteria \cite{BrandenbergerMartin2002,BrandenbergerJorasMartin2002,UnruhSchuetzhold2005} such that the MDR-approach does not seem to entirely resolve the original issues which arguably only a full theory of quantum gravity with phenomenological prowess could achieve.

In recent years, the condensed matter systems that inspired the MDR-approach \cite{Parentani2002} found success as simulators for dynamics in curved spacetime \cite{Belgiorno2010,Weinfurtner2011,Jaskula2012,Hung2013,Steinhauer2016,Torres2017,Eckel2018,Wittemer2019,Chen2021,Braidotti2022,Steinhauer2022,Viermann2022,Sparn2024,Gondret2024,Svancara2024},
such that the lessons and techniques obtained from dispersive field theory are still applied to this day \cite{Weinfurtner2007,Weinfurtner2009,MacherParentani2009a,MacherParentani2009b,BarceloGarayJannes2009,FinazziParentani2012,FinazziParentani2012b,CoutantParentaniFinazzi2012,CoutantWeinfurtner2017,ChaFischer2017,RibeiroFischer2023,DelPorro2023,DelPorro2025,ChandranFischer2025} 
where considerable interest also stems from their potential as table-top testing grounds for quantum gravity phenomenology \cite{JacobsonLiberatiMattingly2005b,Liberati2006,Weinfurtner2006,Visser2009,SotiriouVisserWeinfurtner2009,Volovik2009HLG,AmelinoCamelia2013}.
As a result of the MDR in the analog quantum simulators, the emergent spacetime itself is modified \cite{Weinfurtner2007,Weinfurtner2009} which leads to a so-called rainbow spacetime \cite{MagueijoSmolin2004,Ling2007,Lobo2017} whose geometry and symmetries were discussed in further detail \cite{Girelli2007b}.
Recently, a study on analog Transplanckian cosmology using MDRs occuring in dipolar Bose-Einstein condensates has been conducted \cite{ChandranFischer2025} with an emphasis on the minimal length principle \cite{KMM1995,Garay1995,Brout1999,Kempf2001,KempfNiemeyer2001,Hassan2003}.

Motivated by these considerations we generalize the theoretical description of the quantum field simulator for cosmology introduced in \cite{Viermann2022,Tolosa2022} from the acoustic regime \cite{Schmidt2024} to the dispersive regime by accounting for the full Bogoliubov dispersion relation (which is analog to a first-order expansion of a generic superluminal MDR).
Our theoretical findings were previously compared to experimental results by the Oberthaler group \cite{Sparn2024}.

\emph{The remainder of this paper is organized as follows:}
In \cref{sec:BeyondAcoustic} we derive an effective field theory for the U(1) Goldstone boson and its conjugate momentum in full Bogoliubov theory under the assumption of a time-dependent, homogeneous background. 
In \cref{sec:RainbowSpacetime} the effective field theory is interpreted as a quantum field theory on a dispersive, cosmological spacetime and properties of the quantum mode evolution are discussed in terms of a dispersive, parametric oscillator.
We furthermore contextualize the time-dependent Bogoliubov dispersion relation with common modifications of dispersion relations in the literature including a quantitative perspective on (non)-adiabaticity.
In \cref{sec:ScalInv} we theoretically analyze the implications of the Bogoliubov dispersion relation on the emergence of scale-invariance via an analog exponential expansion or power-law contraction.
In the final \cref{sec:analogParticleProduction} we introduce laboratory boundary conditions and examine out-of-equilibrium field correlations in context of \emph{in-situ} density measurements.
The impact of these boundary conditions on scale-invariant power spectra are discussed, and previous experimental scenarios \cite{Sparn2024} are analyzed via the concepts developed in this work.
We formulate some conclusions in \cref{sec:Conclusion}.

\emph{Notation}
We use the following abbreviation for real-space and momentum-space integrals
\begin{equation}
        \int_{t,\bm{x}} \equiv \int \mathrm{d}t \, \mathrm{d^2} \bm{x}, \quad \int_{\bm{k}} \equiv \int \frac{\mathrm{d^2}\bm{k}}{(2\pi)^2},  \quad   \int_k \equiv \int_0^\infty \frac{\mathrm{d}k \, k}{2\pi}. 
\end{equation}

\section{Beyond the acoustic approximation}
\label{sec:BeyondAcoustic}
In this section, we extend the field-theory methodology of \cite{Tolosa2022} beyond the acoustic approximation.
For a comparison to full Bogoliubov theory in the operator formalism, which is oftentimes used in the literature, consider \cref{app:BogTheoryComp}.

\subsection{Condensate background}
\label{subsec:CondensateBackground}
An effectively two-dimensional Bose-Einstein condensate with repulsive contact interactions can be described in the dilute regime by the effective action \cite{Floerchinger2008}
\begin{equation}
    \begin{aligned}
    &\Gamma[\Phi] = \int \text{d}t \, \text{d}^2 \mathbf{x} \Big\{ \hbar \Phi^*(t,\mathbf x) \left[ \mathrm{i} \frac{\partial}{\partial t} - V_\mathrm{ext}(t,\mathbf x) \right]  \Phi(t,\mathbf x) \\
    &- \frac{\hbar^2}{2 m} \boldsymbol{\nabla}\Phi^*(t, \mathbf x) \boldsymbol{\nabla} \Phi(t, \mathbf x) - \frac{\lambda(t)}{2} \left[ \Phi^*(t,\mathbf x) \Phi(t,\mathbf x) \right]^2 \Big\}.
    \end{aligned} 
\label{eq:ActionBEC}
\end{equation}
It is convenient to split the field $\Phi$ into a background part $\phi_0$ and a fluctuating part parametrized by two real fields $\phi_1$ and $\phi_2$, such that
\begin{equation}
    \Phi(t, \bm{r}) = \phi_0 (t, \bm{r}) + \frac{1}{\sqrt{2}} \left[ \phi_1 (t, \bm{r}) + \mathrm{i} \phi_2 (t, \bm{r}) \right].
    \label{eq:BackgroundSplit}
\end{equation}
The background field is specified as a solution to the Gross-Pitaevskii equation \cite{Lifshitz1980},
\begin{equation}
        i \hbar \pdv{t} \phi_0 = \left( - \frac{\hbar^2}{2m} \bm{\nabla}^2 + V + \lambda \abs{\phi_0}^2 \right) \phi_0,
    \label{eq:GrossPitaevskii2D}
\end{equation}
and corresponds to a condensate mean field, which is discussed in further detail in reference \cite{Tolosa2022}.
In terms of the Madelung representation \cite{Madelung1927},
\begin{equation}
    \phi_0 (t, \bm{r}) = \sqrt{n_0 (t, \bm{r})} \exp {\mathrm{i} \theta_0 (t, \bm{r})},
    \label{eq:MadelungRepresentation}
\end{equation}
the Gross-Pitaevskii equation \eqref{eq:GrossPitaevskii2D} is equivalent to a pair of hydrodynamic equations involving the background particle number density $n_0 (t, \bm{r}) = \abs{\phi_0 (t, \bm{r})}^2$ and the background phase $\theta_0 (t, \bm{r})$. 
More concretely, one finds the continuity equation
\begin{equation}
    0 = \partial_t n_0 + \bm{\nabla} (n_0 \bm{v}),
    \label{eq:ContinuityEquation}
\end{equation}
and the Euler equation
\begin{equation}
    0 = \hbar \partial_t \theta_0 + V + \lambda n_0 + \frac{m}{2} \bm{v}^2 + q,
    \label{eq:EulerEquation}
\end{equation}
where we introduced the superfluid velocity via
\begin{equation}
    \bm{v} = \frac{\hbar}{m} \bm{\nabla} \theta_0,
    \label{eq:SuperfluidVelocity}
\end{equation}
and consider the quantum pressure term
\begin{equation}
\label{eq:quantumPressure}
    q = - \frac{\hbar^2}{2 m} \frac{\bm{\nabla}^2 \sqrt{n_0}}{\sqrt{n_0}},
\end{equation}
in \eqref{eq:EulerEquation}, which allow us to be beyond the acoustic approximation. 
Note that $q$ in Eq.\ \eqref{eq:quantumPressure} is of second order in $\hbar$, as well as in spatial derivatives, such that it is expected to be subleading for sufficiently smooth density.

\subsection{Effective theory for fluctuations}
\label{subsec:EFT_Fluc}
Expanding the effective action $\Gamma[\Phi]$ up to second order in the fluctuating fields, $\Gamma[\Phi] = \Gamma_0[\phi_0] + \Gamma_2[\phi_1,\phi_2]$ (more details in \cite{Tolosa2022}), one finds 
\begin{equation}
    \begin{split}
        \Gamma_2 =& \int_{t,\bm{x}}\Bigg\{ \frac{1}{2} \phi_2 \left(\frac{\hbar^2}{2m} \bm \nabla^2+q\right) \phi_2 \\
        &+ \frac{1}{2} \phi_1 \left(-2 \lambda n_0 + \frac{\hbar^2}{2m} \bm \nabla^2 + q \right)  \phi_1 \\
        &- \hbar \phi_1\left[\partial_t \phi_2 + (\bm{v} \cdot \bm{\nabla}) \phi_2 +  \frac{1}{2}   (\bm{\nabla} \cdot \bm{v} )\phi_2\right] \Bigg\},
    \end{split}
\label{eq:FullFluctAction}    
\end{equation}
upon utilizing the Euler equation \eqref{eq:EulerEquation}.

Since we want to calculate the dispersive extension to the analog cosmological metric calculated in \cite{Tolosa2022}, we restrict to a stationary background flow, $\bm{v}=0$.
Furthermore, we assume a homogeneous background density, $n_0(r)=\bar{n}_0$.
This substantially simplifies the inclusion of dispersive effects since the quantum pressure still vanishes at the background level, $q = 0$.
Invoking these assumptions, the fluctuating fields $\phi_1$ and $\phi_2$ are described by the quadratic action 
\begin{equation}
    \begin{split}
        \Gamma_2 =& \int_{t,\bm{x}} \Big\{\phi_2 \hbar\partial_t \phi_1 +  \frac{\hbar^2}{4m} \phi_2 \bm{\nabla}^2 \phi_2   \\
        &  \qquad \qquad + \frac{1}{2} \phi_1 \left( \frac{\hbar^2}{2m} \bm{\nabla}^2 -2 \lambda n_0 \right) \phi_1  \Big\},
    \end{split}
\label{eq:FinalFluctAction}    
\end{equation}
and evolve according to the linearized equations of motion
\begin{equation}
2\lambda n_0 \phi_1 -\frac{\hbar^2}{2m} \bm{\nabla}^2 \phi_1  + \hbar\partial_t \phi_2 = 0,
    \label{eq:Phi1Phi2Rel}
\end{equation}
and
\begin{equation}
    \hbar\partial_t \phi_1  + \frac{\hbar^2}{2m} \bm{\nabla}^2 \phi_2=0.
\end{equation}

\subsection{Quantization and mode dynamics}
\label{subsec:QuantMode}
The conjugate momentum fields of $\phi_1$ and $\phi_2$ are obtained as
\begin{align}
    \pi_1(t,\bm{x}) &\equiv \frac{\delta \Gamma_2}{\delta \dot \phi_1(t,\bm{x})} = -\hbar \phi_2(t,\bm{x}), \\
    \pi_2(t,\bm{x}) &\equiv \frac{\delta \Gamma_2}{\delta \dot \phi_2(t,\bm{x})} = \hbar \phi_1(t,\bm{x}).
\end{align}
Hence, $\phi_1$ and $\phi_2$ are conjugate field variables.
With the splitting in \cref{eq:BackgroundSplit} and for $\phi_0 \in \mathbb{R}$, they can be interpreted as density- and phase-fluctuations (more details in \cref{app:BogTheoryComp}).
Therefore, we can quantize both fields in terms of a single set of annihilation and creation operators $\hat a_k$ and $\hat a_k^\dagger$.
To that end we employ the mode expansions
\begin{equation}
    \begin{aligned}
        \phi_1(t,\bm{x}) &= \int_{\bm{k}} [ \hat a_{\bm{k}} w_k(t) \mathrm{e}^{\mathrm{i} \bm{k} \cdot \bm{x}} + \hat a_{\bm{k}}^\dagger w_k^*(t) \mathrm{e}^{-\mathrm{i} \bm{k} \cdot \bm{x}}], \\
        \phi_2(t,\bm{x}) &= \int_{\bm{k}} [ \hat a_{\bm{k}} v_k(t) \mathrm{e}^{\mathrm{i} \bm{k} \cdot \bm{x}} + \hat a_{\bm{k}}^\dagger v_k^*(t) \mathrm{e}^{-\mathrm{i} \bm{k} \cdot \bm{x}}],
    \end{aligned}
    \label{eq:ModeExpansion}
\end{equation}
and demand the operators to obey the bosonic commutation relations
\begin{equation}
    \begin{aligned}
    \relax [\hat a_{\bm{k}}, \hat a_{\bm{k}}^\dagger] &= \delta(\bm{k}-\bm{k}'), \\
    \relax  [\hat a_{\bm{k}}, \hat a_{\bm{k}'}] &= [\hat a_{\bm{k}}^\dagger, \hat a_{\bm{k}'}^\dagger] = 0,
    \end{aligned}
\end{equation}
which in turn enable the equal-time canonical commutation relations 
\begin{align}
    [\phi_2(t,\bm{x}), \pi_2(t,\bm{y})] &= \mathrm{i} \hbar \delta(\bm{x} - \bm{y}), \\
    [\phi_2(t,\bm{x}), \phi_2(t,\bm{y})] &= [\pi_2(t,\bm{x}), \pi_2(t,\bm{y})] = 0,
\end{align}
if the normalization constraint,
\begin{equation}
    w_k(t) v_k^*(t) - v_k(t) w_k^*(t) = \mathrm{i},
    \label{eq:NormalizationConstraint}
\end{equation}
is fulfilled. This constraint is presered by the evolution equations
\begin{align}
    \hbar \partial_t w_k(t) &= E_k v_k(t), \label{eq:v1Dyn} \\
    - \hbar \partial_t v_k(t) &= \frac{\epsilon_k^2(t)}{E_k} w_k(t),  \label{eq:v2Dyn}
\end{align}
where we introduced
\begin{equation}
    \epsilon_k(t) \equiv \hbar \omega_k(t), \quad E_k \equiv \frac{\hbar^2 k^2}{2m},
\end{equation}
with the Bogoliubov dispersion relation
\begin{equation}
    \omega_k^2(t) \equiv c_\mathrm{s}^2(t) k^2 + \left(\frac{\hbar k^2}{2m}\right)^2,
    \label{eq:BogDispersion}
\end{equation}
and the squared speed of sound
\begin{equation}
    c_\mathrm{s}^2(t) \equiv \frac{\lambda(t) n_0}{m}.
    \label{eq:SoundSpeed}
\end{equation}
For our further discussion, the phase velocity of Bogoliubov modes
\begin{equation}
    c_\mathrm{ph}(t,k) = \frac{\omega_k(t)}{k} = c_\mathrm{s}(t) \sqrt{1 + \tfrac{1}{2} k^2 \xi^2(t)},
    \label{eq:phaseVel}
\end{equation}
and the healing length
\begin{equation}
    \xi(t) = \frac{\hbar}{\sqrt{2}m c_\mathrm{s}(t)},
    \label{eq:HealingLength}
\end{equation}
will play a central role. The latter is a characteristic length scale below which excitations lose their collective character and rather represent individual atoms.
Throughout all numerical evaluations of dispersive effects in this work, we assume the atomic mass in \cref{eq:HealingLength} to be given by $^{39}\mathrm{K}$ (corresponding to the BEC used in \cite{Sparn2024}).

\section{Dispersive effects in emergent spacetimes}
\label{sec:RainbowSpacetime}

The concepts developed so far provide a bona-fide framework to study the evolution and correlations of the non-relativistic quantum fields in the BEC. 
There exists, however, an alternative geometric formulation that bears a strong resemblance to relativistic quantum field theory which shall be layed out in the following.

\subsection{Dispersive metric}
\label{subsec:RainbowMetric}
In the dynamical system  \eqref{eq:v1Dyn} and \eqref{eq:v2Dyn} one can eliminate the mode $w_k(t)$ in favor of the time-derivative $\dot v_k(t)$
resulting in
\begin{equation}
    \ddot v_k(t) - 2 \frac{\dot c_\mathrm{ph}(t,k)}{c_\mathrm{ph}(t,k)} \dot v_k(t) + c_\mathrm{ph}^2(t,k) k^2 v_k(t) = 0.
    \label{eq:ModeEqCosmicTime}
\end{equation}
This mode elimination can be realized on the level of the effective action  \eqref{eq:FinalFluctAction} by integrating out $\phi_1$ in Fourier space.
One finds
\begin{equation}
    \begin{aligned}
        \Gamma_2 &= \frac{\hbar^2}{4m} \int \mathrm{d}t \int_{\bm{k}} \bigg \lbrace \frac{1}{c_\mathrm{ph}^2(t,k)} \partial_t \phi_2(t,-\bm{k}) \partial_t \phi_2(t,\bm{k}) \\
        &\hspace*{2.5cm}- k^2 \phi_2(t,-\bm{k}) \phi_2(t,\bm{k}) \bigg \rbrace,
    \end{aligned}
\end{equation}
as the effective action of phase-fluctuations in the BEC, whose velocity and momentum only depend on the absolute value $k = \lvert \bm{k} \rvert$ due to the spatially homogeneous background.

\begin{figure}
    \includegraphics[width=\columnwidth]{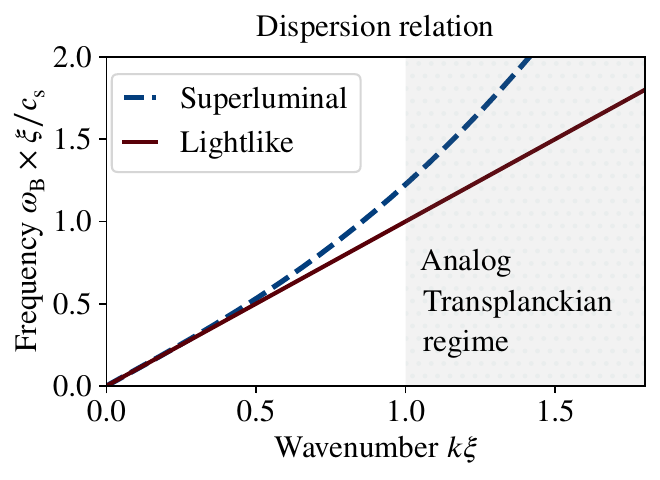}
    \caption{Modified dispersion relation in the quantum field simulator (dashed line) compared to the non-dispersive lightlike case (solid line).}
    \label{fig:MDR}
\end{figure}

Introducing the re-scaled field $\phi = \phi_2/\sqrt{2m}$ and the operator
\begin{equation}
    \tilde \partial_\mu = \mqty(\partial_t \\ - \mathrm{i} \bm{k})
\end{equation}
this effective action assumes a form resembeling a relativistic theory,
\begin{equation}
    \Gamma_2 = - \frac{\hbar^2}{2} \int \mathrm{d}t \int_{\bm{k}} \sqrt{g(t,k)} g^{\mu \nu}(t,k) \, \tilde \partial_\mu \phi(t,-\bm{k}) \, \tilde \partial_\nu \phi(t,\bm{k}),
    \label{eq:RainbowAction}
\end{equation} 
with the dispersive metric
\begin{equation}
    g_{\mu \nu}(t,k) = \mqty(-1 & 0 \\ 0 & c_\mathrm{ph}^{-2}(t,k) \mathds{1}_2), \quad \sqrt{g(t,k)} = c_\mathrm{ph}^2(t,k).
    \label{eq:RainbowMetric}
\end{equation}
Note that, due to the wavenumber dependence of the metric, the effective action \eqref{eq:RainbowAction} describes a non-local theory as one can show explicitly by transforming back to real space.
The wavenumber-dependent, geometric structure can be interpreted by tracing wavefronts of constant phase that propagate according to the Bogoliubov dispersion relation $\smash{\omega^2(t) = c_\mathrm{ph}^2(k,t) \, k^2}$
which translates into a null-condition,
\begin{equation}
    g_{\mu \nu}(t,k) k^\mu k^{\nu} = 0,
    \label{eq:RainbowNull}
\end{equation}
for the four-momentum $k^{\mu} \equiv (\omega,\bm{k})^\mathrm{T}$.
Note that \cref{eq:RainbowNull} is in general \emph{not} invariant under (analog) Lorentz transformations, $k^\mu \to \Lambda^{\mu}_{\, \nu} k^\nu$, such that the laboratory system defines a preferred frame of reference.
On scales where $k\xi/\sqrt{2} \ll 1$ however, we have $c_\mathrm{ph}(t,k) = c_\mathrm{s}(t) + \mathcal{O}(k^2\xi^2)$ such that \cref{eq:RainbowNull} becomes Lorentz-invariant on these scales.
In that sense, the healing length $\xi$ functions as an analog Planck length \cite{Volovik2009} below which Lorentz-invariance is broken and replaced by a non-local theory with a superluminal (or supersonic) dispersion relation (see \cref{fig:MDR} for a visualization).
The described analog spacetime structure is also known as rainbow metric \cite{Weinfurtner2009} and has been shown to admit a formulation in terms of a Finsler geometry \cite{Girelli2007b} (see \cite{Lobo2017} for a study on geodesics in such geometries).

\subsection{Quantization on dispersive, cosmological spacetime}

The metric \eqref{eq:RainbowMetric} can be brought into a form that is suitable to describe a cosmological spacetime (i.e.\@ a dynamic, homogenenous and isotropic space) but with a wavenumber-dependent scale-factor
\begin{equation}
    a_k(t) = c_\mathrm{ph}^{-1}(t,k),
    \label{eq:DispersiveScaleFactor}
\end{equation}
with which we will work in the following, bearing in mind that all statements can be equivalently made from (or translated to) the perspective of Bogoliubov excitations in a BEC with time-dependent contact interactions.

The form of \cref{eq:ModeEqCosmicTime} is equivalent to a mode equation of a massless Klein-Gordon field which is minimally coupled to a
such spacetime evolving dynamically with scale-factor $a_k$ where each mode experiences a different expansion (or contraction) rate.
In accordance with \cref{eq:DispersiveScaleFactor} we can write the dispersive scale-factor as
\begin{equation}
    a_k(t) = \frac{a(t)}{\sqrt{1 + \tfrac{1}{2}k^2 \xi^2(\eta)}},
    \label{eq:akaRel}
\end{equation}
where
\begin{equation}
     a(t) = \lim_{k \to 0} a_k(t) = c_\mathrm{s}^{-1}(t),
\end{equation}
is the scale factor of the acoustic metric derived in \cite{Tolosa2022}.

For completeness, let us also mention that the conjugate momentum of the relativistic field $\phi = \phi_2 / \sqrt{2m}$ is related to $\phi_1$ via
\begin{equation}
    \pi(t,\bm{k}) = \frac{\delta \Gamma_2}{\delta \dot \phi(t,\bm{k})} = \hbar^2 a_k^2 \, \dot \phi(t,-\bm{k}) = - \sqrt{2m} \hbar \, \phi_1(t,-\bm{k}),
    \label{eq:RelativisticMomentumField}
\end{equation}
as one would expect. 
The quantization of $\phi_1$ and $\phi_2$ is thus equivalent to quantizing $\phi$ and $\pi$ according to the momentum-space equal-time commutation relation
\begin{equation}
    [\phi(t,\bm{k}), \pi(t,\bm{k'})] = \mathrm{i} \hbar \delta(\bm{k}+\bm{k'}), \\[5pt]
    \label{eq:RelativisticCanonicalCommutationRelations}
\end{equation}
with the mode $v_k$ being subjected to the canonical normalization constraint
\begin{equation}
    \hbar a_k^2 \left( v_k \dot v_k^* - \dot v_k v_k^* \right) = \mathrm{i},
    \label{eq:WronskianNormalization}
\end{equation}
which is equivalent to the normalization condition \eqref{eq:NormalizationConstraint} upon mode-elimination and a generalization of the usual Wronskian determinant of cosmological scalar fields \cite{Tolosa2022} to the case of a dispersive spacetime described by the scale-factor $a_k(t)$. 

\subsection{Dispersive parametric oscillator}
\label{subsec:DispEffMass}
The goal of this section is to transform the mode equation \eqref{eq:ModeEqCosmicTime} into a parametric oscillator for convenience.
To that end we rescale,
\begin{equation}
    v_k(t) = [a(t) \mathcal{J}_k(t)]^{-1/2} \psi_k(t),
\end{equation}
where
\begin{equation}
    \mathcal{J}_k(t) \equiv \frac{c_\mathrm{s}^2(t)}{c_\mathrm{ph}^2(k,t)} = \left(1 + \frac{1}{2} k^2 \xi^2(t)\right)^{-1},
    \label{eq:CurlyJ}
\end{equation}
arises as a consequence of the modified canonical volume form in \cref{eq:RainbowAction} and captures the reduction in number of phase-space degrees of freedom \cite{Hassan2003}.

Furthermore, we introduce conformal time $\mathrm{d} \eta = \mathrm{d} t/a(t)$ with respect to the bare scale-factor $a(t)$
such that $\eta$ is still equivalent to the sound horizon in the BEC \cite{Schmidt2024}. 
The mode equation \eqref{eq:ModeEqCosmicTime} then takes the form
\begin{equation}
    \psi_k''(\eta) + \Omega_\mathrm{B}^2(k,\eta) \psi_k(\eta) = 0,
    \label{eq:ModeEquationFinal}
\end{equation}
with the effective Bogoliubov frequency 
\begin{equation}
   \Omega_\mathrm{B}^2(k,\eta) =  k^2/ \mathcal{J}_k(\eta) + m_\mathrm{eff}^2(k,\eta),
   \label{eq:EffBogFreq}
\end{equation}
and the effective mass
\begin{equation}
     m_\mathrm{eff}^2(k,\eta) = - \frac{1}{2} \frac{[a(\eta) \mathcal{J}_k(\eta)]''}{a(\eta) \mathcal{J}_k(\eta)} + \frac{1}{4} \left(\frac{[a(\eta) \mathcal{J}_k(\eta)]'}{a(\eta) \mathcal{J}_k(\eta)}\right)^2,
\end{equation}
(see \cref{app:DispEffMass} for a detailed derivation).

To clarify the notion of an effective mass, consider that \cref{eq:EffBogFreq} has the form of an energy-momentum relation, $E^2 = k^2 + m^2$, 
but with the effective mass being generated by the time-dependence of $\mathcal{J}_k a$ and one has $k^2 \to k^2/\mathcal{J}_k$ because of the dispersion.

The splitting of terms in the effective frequency \eqref{eq:EffBogFreq} is also motivated from the effective mass being responsible for non-adiabatic transitions \cite{MassarParentani1997}
whereas the modes can adiabatically adjust to $k^2 / \mathcal{J}_k(\eta)$. 
The magnitude of the former depends on the rate of change in spacetime (or the laboratory parameters in the simulator) and becomes adiabatically suppressed in the high-momentum regime $k^2 / \mathcal{J}_k(\eta) \gg m_\mathrm{eff}^2$ as we will show with explicit examples later-on.

Moreover, the magnitude of non-adiabatic transition is wavenumber-dependent in the first place since the cut-off scale (or comoving analog Planck length) $\xi(\eta)$ is time-dependent.
More explicitly, the effective mass has the form
\begin{equation}
\begin{aligned}
    m_\mathrm{eff}^2(k,\eta) &= -\frac{1}{4} \frac{3k^4 \xi^4(\eta) / 4 - 5 k^2 \xi^2(\eta) -1}{[1+ k^2 \xi^2(\eta)/2]^2}  \left(\frac{a'(\eta)}{a(\eta)}\right)^2  \\[5pt]
    &\hspace{0.5cm}- \frac{1}{2} \frac{1 - k^2 \xi^2(\eta)/2}{1 + k^2 \xi^2(\eta)/2} \left(\frac{a''(\eta)}{a(\eta)}\right),
    \label{eq:EffMass}
\end{aligned}
\end{equation}
where we have used the time-dependence of $\mathcal{J}_k(\eta)$.
Put differently, if $\xi$ would be time-independent, the effective mass $\eqref{eq:meffGeneral}$ would be wavenumber-independent such that the dispersive effects only influence the mode by its instantaneous eigenenergy in \cref{eq:ModeEquationFinal} as it is also the case for the ad-hoc modifications studied in references \@ \cite{MartinBrandenberger2000,MartinBrandenberger2001,MartinBrandenberger2001b,MartinBrandenberger2002,BrandenbergerJorasMartin2002,Niemeyer2001,NiemeyerParentani2001}.
The physical reason underlying the time-dependence of $\xi(\eta)$ (and thus $\mathcal{J}_k(\eta)$) are the time-dependent atomic interactions in the BEC. 
This places the BEC automatically in the frame comoving with the analog cosmos, which is also the preferred frame in which \cref{eq:RainbowNull} holds and where the Lorentz-invariance-violating physics need to be evaluated. 

As a consistency check consider the infrared and ultraviolet limit: 
In the infrared limit where $\mathcal{J}_k$ becomes time-independent, one finds the standard result \cite{Schmidt2024},
\begin{equation}
    \lim_{k \to 0} m_\mathrm{eff}^2(k,\eta) = - \frac{1}{2} \frac{a''(\eta)}{a(\eta)} + \frac{1}{4} \left(\frac{a'(\eta)}{a(\eta)}\right)^2.
    \label{eq:meffInfrared}
\end{equation}
In the ultraviolet limit, the mode equation is a harmonic oscillator with a quadratic, non-relativistic dispersion where the effective mass is adiabatically suppressed, as one would expect (details in \cref{app:DispEffMass}).

\subsection{Time-dependent dispersion and non-adiabaticity}
\label{subsec:Adiabaticity}
In this section we analyze the significance of the dispersive effects and contextualize with the literature \cite{MartinBrandenberger2000,MartinBrandenberger2001,MartinBrandenberger2001b,MartinBrandenberger2002,BrandenbergerJorasMartin2002,Niemeyer2001,NiemeyerParentani2001}.
Therein, it is common to write the effective frequency as 
\begin{equation}
    \Omega_F^2(k,\eta) = a^2(\eta) F^2(k/a(\eta))+ m_\mathrm{eff}^2(k = 0,\eta),
    \label{eq:EffGenFreq}
\end{equation}
where the form-function $F$ specifies the modification of the dispersion relation as a function of the physical wavenumber $k/a$ (i.e\@ the one that is subjected to redshift by an expanding space) and is required to become linear in the non-dispersive regime.
In fact, $F$ can be considered as an effective physical wavenumber, which is specified as a function of the bare physical wavenumber $k/a$ \footnote{Note that the additional factor of $a$ converts the effective physical wavenumber $F(k/a)$ into an effective comoving wavenumber $a(\eta) F(k/a)$ (i.e.\@ the wavennumber in the rest-frame of the cosmological background evolution)}
For example $F$ can be a quartic polynomial that is either curved subluminally or superluminally, as it is the case for the Corley-Jacobson dispersion relation \cite{CorleyJacobson1996,Corley1998}, or a hyperbolic tangent in the case of Unruh \cite{Unruh1995}.
In context of the mode equation in this work \eqref{eq:ModeEquationFinal} we would therefore introduce 
\begin{equation}
    F^2_\mathrm{B}\left(k, a(\eta)\right) = \left(\frac{k}{a(\eta)}\right)^2 + \frac{\xi_0^2}{2 a_0^2} k^4,
    \label{eq:FB}
\end{equation}
such that 
\begin{equation}
    \Omega_\mathrm{B}^2(k,\eta) = a^2(\eta) F^2_\mathrm{B}(k,a) + m_\mathrm{eff}^2(k,\eta).
\end{equation}
At this point we want to highlight two key-differences of the effective frequency \eqref{eq:EffBogFreq} relative to form \eqref{eq:EffGenFreq}
that stem from the time-dependence of $\xi(\eta)$:
\begin{enumerate}
    \item[(1)] The effective mass in \cref{eq:EffBogFreq} (and thus the probability for non-adiabatic transitions \cite{MassarParentani1997}) is wavenumber-dependent as we discussed in \cref{sec:RainbowSpacetime}.
    \item[(2)] The instantaneous mode eigenenergy $k^2 + \frac{1}{2} \xi^2(\eta) k^4$ is time-dependent since $\xi(\eta) =\xi_0 \, a(\eta)/a_0$. As an immediate consequence the function $F_\mathrm{B}$ defined in \cref{eq:FB} not be written as a function of the (bare) physical wavenumber $k/a$ alone.
\end{enumerate}
As a comparison consider the case where the healing length would be constant in time, $\xi = \xi_0$, leading to 
\begin{equation}
    F^2_{\mathrm{B,0}}\left(k/a(\eta)\right) = \left(\frac{k}{a(\eta)}\right)^2 + \frac{a_0^2 \xi_0^2}{2} \left(\frac{k}{a(\eta)}\right)^4,
    \label{eq:FCJ}
\end{equation}
which agrees with the superluminal Corley-Jacobson dispersion relation \cite{MartinBrandenberger2001,NiemeyerParentani2001,MartinBrandenberger2002} and can be written as a polynomial function solely depending on the physical momentum $k/a$.

For a quantitative analysis of these effects let us consider the non-adiabaticity parameter \cite{MartinBrandenberger2013}
\begin{equation}
    \alpha_F(k,\eta) = \bigg \lvert \frac{3}{4} \left(\frac{\Omega_F'(k,\eta)}{\Omega_F^2(k,\eta)} \right)^2 - \frac{1}{2} \frac{\Omega_F''(k,\eta)}{\Omega_F^3(k,\eta)} \bigg \rvert^{1/2},
    \label{eq:alphaF}
\end{equation}
which, when $\alpha_F \ll 1$, indicates that the mode function adiabatically adjusts to the effective frequency and remains in a WKB-form (which is also known as the adiabatic vacuum; see \cref{app:WKB} for details).
In contrast, when $\alpha_F \gg 1$, non-adiabatic transitions are highly probable and particle production is strong (consider \cite{MassarParentani1997} for a link of these concepts to the adiabatic theorem of quantum mechanics).

As an exemplary case study, let us consider a linearly expanding scenario (to be discussed in further detail in \cref{sec:analogParticleProduction}) and compute the non-adiabaticity parameter \eqref{eq:alphaF} for the time-dependent Bogoliubov dispersion \eqref{eq:EffBogFreq}.
As a proper reference, let us perform the same task for a time-independent Bogoliubov dispersion
\begin{equation}
    \Omega_\mathrm{B,0}^2(k,\eta) = a^2(\eta) F_\mathrm{B,0}^2(k/a) + m_\mathrm{eff}^2(k = 0,\eta),
    \label{eq:OmegaB0}
\end{equation}
where we impose the standard FLRW effective mass in $(2+1)$-dimensions, which is given by $\smash{m_\mathrm{eff}^2(k = 0,\eta)}$ (cf.\@ \cref{eq:EffMass}) and use a dispersion relation of the form \eqref{eq:FCJ}.
To have a non-dispersive reference, we also consider 
\begin{equation}
    \Omega_\mathrm{Ac}^2(k,\eta) = k^2 + m_\mathrm{eff}^2(k = 0,\eta),
    \label{eq:OmegaAc}
\end{equation}
which is the familiar non-dispersive FLRW-case.

\begin{figure}
    \includegraphics[scale = 0.42]{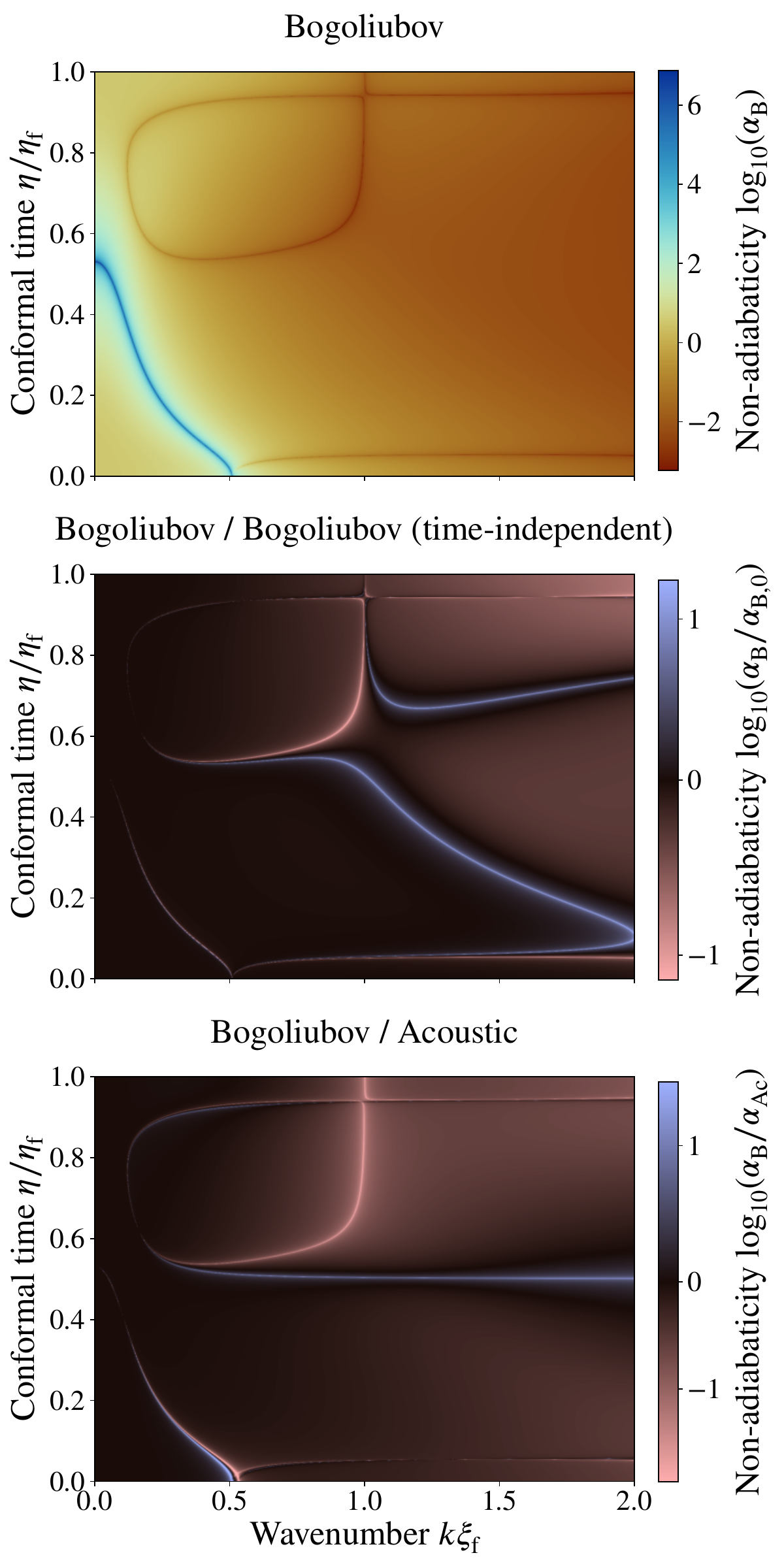}
    \caption{Non-adiabaticity for time-dependent Bogoliubov dispersion (absolute values, upper image) compared to the time-indepedent case  (relative values, central image) and the non-dispersive (acoustic) case (relative values, lower image) as a function of conformal time $\eta / \eta_\mathrm{f}$ and wavenumbers $k \xi_\mathrm{f}$ in case of the linearly expanding scenario discussed in \cref{sec:analogParticleProduction}.
    Blue regions in the upper image show non-adiabaticity to be highly localized in time along a single throat-like trajectory. As expected, one has an increasing adiabaticity towards the ultraviolet regime at all times.
    Dark regions in the central and lower image indicate similarity regarding (non)-adiabaticity whereas blue regions demonstrate a trend towards the Bogoliubov dispersion.
    To numerically evaluate $\alpha$ including the singular contributions to the effective mass (to be discussed in \cref{sec:analogParticleProduction}), we smoothed the delta-distributions via the Lorentzian profile $\delta_\epsilon(x) = (\epsilon/\pi)/(x^2 + \epsilon^2)$ where $\epsilon = 0.1$ models a relative switching sensitivity of ten percent.}
    \label{fig:Adiabaticity}
\end{figure}

In \cref{fig:Adiabaticity}, we show the non-adiabaticity parameter for the time-dependent Bogoliubov case (upper image) and take its ratio to the time-independent Bogoliubov case (center image) and acoustic case (lower image).
In the Bogoliubov case, we observe throat-like trajectories of high (low) non-adiabaticity in blue (red). 
Relative to the time-independent case, there are lines of high non-adiabaticity in the ultraviolet which move towards lower wavenumbers due to redshift, which can be explained via the dynamics of $\xi(\eta)$. 
At the critical (final) wavenumber $k \xi_\mathrm{f} = 1$, the Bogoliubov dispersion relation becomes adiabatic relative to the non-dispersive case as one can conclude from the central throat that appears at $\eta / \eta_\mathrm{f} \gtrsim 0.5$.

\section{Scale invariance}
\label{sec:ScalInv}
In this section we analyse the perspectives of approaching the Transplanckian problem of cosmology in analog BEC-Cosmology simulators with a time-dependent scattering length.
The following discussion is purely theoretical and can be understood as a first step, using the straight-forward scenarios of an exponential expansion and a power-law contraction, to gain intuition and experience for future scenarios where one has to stay within current experimental capabilities which requires innovative solutions.
In this section we set $\hbar = 1$ in accordance with the literature.

\subsection{Exponential expansion}
\label{subsec:ExpExp}

We first describe how a scale-invariant spectrum of quantum fluctuations emerges in a non-dispersive theory from an exponentially growing scale-factor, $\smash{a(t) = a_\mathrm{i} \mathrm{e}^{Ht}}$ which has to be identified with the reciprocal sound-speed in the analog simulator.
The passage of time is described by a growing (sound)-particle horizon (the sound-cone) $\eta$ which is defined as
\begin{equation}
    \eta(t) = \int_0^t \frac{\mathrm{dt}}{a(t)} =  \eta_0 (1 - \mathrm{e}^{-Ht}), \quad \eta_0 = \frac{1}{a_\mathrm{i}H},
    \label{eq:ConfTimeExpExp}
\end{equation}
and asymptotically grows from $\eta_\mathrm{i} = 0$ towards 
\begin{equation}
    \eta_\mathrm{f} = \eta_0 (1- \mathrm{e}^{-N}),
    \label{eq:finalTimesExpExp}
\end{equation}
where we introduced the number of e-foldings $N = \ln (a_\mathrm{f}/a_\mathrm{i})$.
Quantizing a non-dispersive, massless scalar field that is minimally coupled to a flat $(2+1)$-dimensional FLRW spacetime leads to the mode equation
\begin{equation}
    \psi_k''(\eta) + \left(k^2 - \frac{3}{4(\eta- \eta_0)^2}\right) \psi_k(\eta) = 0,
    \label{eq:ModeEquationExpExp}
\end{equation}
where the scale-factor reads 
\begin{equation}
    a(\eta) = a_\mathrm{i} (1 -\eta/\eta_0)^{-1},
    \label{eq:ScalfExpExp}
\end{equation}
as a function of conformal time.
The boundary condition is chosen as an instantaneous vacuum state at the initial time $\eta_\mathrm{i} = 0$ 
\footnote{This choice corresponds to the ground state in the interacting BEC.
In particular, the unique existence of that state places the initial superhorizon modes in a vacuum state.
Note that in the cosmological target system, the nature of the vacuum on initial superhorizon scales is generally unknown \cite{MukhanovWinitzki2007}.}
\begin{equation}
        \psi_k(\eta_\mathrm{i}) = 1 / \sqrt{2k}, \quad \psi_k'(\eta_\mathrm{i}) = -\mathrm{i}k \psi_k(\eta_\mathrm{i}),
        \label{eq:InitialInstVac}
\end{equation}
which leads to the canonically normalized mode solution
\begin{equation}
    \begin{aligned}
        \psi_k(\eta) = \frac{\mathrm{i} \pi}{4} &\frac{1}{\sqrt{2k}} \sqrt{\frac{\eta - \eta_0}{\eta_0}} \\
        &\times\big(r J_1[k  (\eta_0 - \eta)] + s Y_1[k  (\eta_0 - \eta)]  \big),
    \end{aligned}
    \label{eq:ModeSol2DExpExp}
\end{equation}
where $J_1, Y_1$ are Bessel-functions and the choice of the coefficients
\begin{align}
    r &= 2 k \eta_0  Y_0(k \eta_0) - (1 + 2 \mathrm{i} k \eta_0) Y_1(k\eta_0), \\
    s &= -2 k \eta_0 J_0(k \eta_0) + (1 + 2 \mathrm{i} k \eta_0) J_1(k \eta_0).
\end{align}
ensures an instantaneous vacuum at $\eta = 0$.
In the limit of many e-folds, $N \to \infty$, this corresponds to the Bunch-Davies state.

\begin{figure}
    \includegraphics[width=\columnwidth]{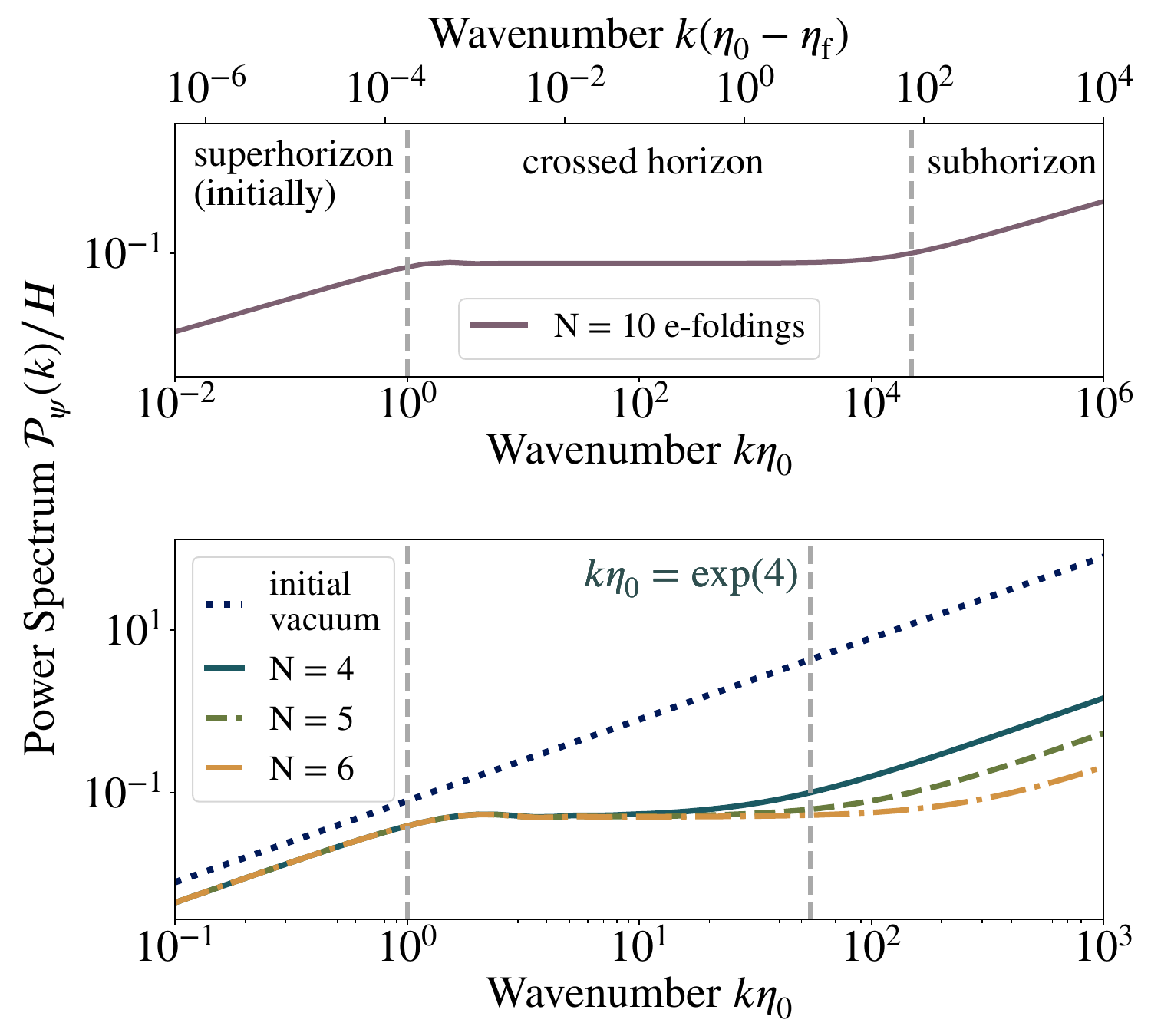}
    \caption{Upper image:
    Power spectrum $\mathcal{P}_\psi/H$ after an exponential expansion by $N = 10$ e-foldings compared to the initial vacuum spectrum.
    A constant (scale-invariant) value $\mathcal{P}_\psi / H = 1/2\pi^2$ occurs for modes $1 \ll k \eta_0 \ll \mathrm{e}^N$, where the boundaries become more accurate as $N$ increases. 
    In the limit $\eta_0 \to \infty$, the left boundary is shifted to arbitrary large scales, resulting in the well-known Bunch-Davies spectrum which would furthermore extend to arbitrary small scales in the limit $\eta_\mathrm{f} \to \eta_0$ (or $N \to \infty$).     
    Lower image: Time-dependence of the power spectrum as parametrized by the number of e-foldings $N$. The scale-invariant plateau grows with time towards smaller comoving scales.}
    \label{fig:QFSSpectraScalInv}
\end{figure}

Of central interest is the dimensionless power spectrum of field excitations which in two spatial dimensions is introduced as \footnote{where the pre-factor accounts for momentum space isotropy and that the conformal factor in $(2+1)$ dimensions is given by $\sqrt{a}$. }
\begin{equation}
    P_\psi(k,\eta) = \frac{1}{2\pi} \frac{k^2}{a(\eta)} \lvert \psi_k(\eta) \rvert^2.
    \label{eq:PowerSpecDef}
\end{equation}
At the final time, $\eta_\mathrm{f} = \eta_0 (1- \mathrm{e}^{-N})$, it takes the form 
\begin{equation}
    \frac{P_\psi(k,\eta_\mathrm{f})}{H k \eta_0}= \frac{\pi}{64} \mathrm{e}^{-2N}  \lvert r J_1(k\eta_0 \mathrm{e}^{-N}) + s Y_1(k\eta_0 \mathrm{e}^{-N}) \rvert^2,
    \label{eq:P_QFS_StartingFormula}
\end{equation}
which is a function of the dimensionless combination $k \eta_0$ only.
Now, for modes in the regime $1 \ll k \eta_0 \ll \mathrm{e}^N$ this spectrum becomes independent of $k$ (scale-invariance) as we show in \cref{fig:QFSSpectraScalInv}.
The physical reason is best understood in the comoving frame: 
Here one compares to mode wavelength $k^{-1}$ to the (comoving) Hubble horizon $\smash{(a H)^{-1} = (\eta_0 - \eta)}$ which indicates the distance beyond which two spatial positions recede faster than the speed of light (or sound in the simulator).
From the definition of the (sound)-particle horizon \eqref{eq:ConfTimeExpExp}, one finds that the comoving Hubble horizon exponentially shrinks from its initial value $\eta_0$ to its final value $\smash{\eta_0 \mathrm{e}^{-N}}$ as $\eta$ grows from $\eta_\mathrm{i} = 0$ towards $\smash{\eta_\mathrm{f} = \eta_0 (1- \mathrm{e}^{-N})}$.
Hence, the comoving wavelength $\smash{k^{-1}}$ of a mode transitions from being smaller than the Hubble horizon (subhorizon),
\begin{equation}
    k (\eta_0 - \eta) = k/aH > 1.
\end{equation}
to being larger than the Hubble horizon (superhorizon),
\begin{equation}
    k (\eta_0 - \eta) = k/aH < 1,
\end{equation}
where it freezes in because it is now too large for causal evolution.
In terms of the wavenumber in the physical (expanding) frame, $k_\mathrm{phys} = k a$, we can write 
\begin{equation}
    k (\eta_0 - \eta) = k_\mathrm{phys}/H,
\end{equation}
upon using \cref{eq:ConfTimeExpExp,eq:ScalfExpExp}. 
Thus, the time-evolution of the mode can be entirely absorbed into the physical scale $k_\mathrm{phys}$ \cite{MukhanovWinitzki2007}.
Put differently, $k (\eta_0 - \eta)$ is large at early times (or equivalently small scales) and small at late times (or equivalently large scales)
and the time evolution of a mode merely consists of transitioning from subhorizon to superhorizon scales around the turning point $k (\eta_0 - \eta) = k_\mathrm{phys} H = 1$.

Each mode crosses the Hubble horizon at a different value of the (sound)-particle horizon $\eta_k$ that fulfulls $k (\eta_0 - \eta_k) = 1$;
hence modes which have crossed the horizon within the time-window $\eta \in [\eta_\mathrm{i},\eta_\mathrm{f}]$ satisfy $k (\eta_0 - \eta_\mathrm{i}) = k \eta_0 > 1$ and
\begin{equation}
    k (\eta_0 - \eta_\mathrm{f}) = k \eta_0 \mathrm{e}^{-N} < 1.
\end{equation}
For modes which fall well into this regime, $k \eta_0 \mathrm{e}^{-N} \ll 1$, we find at leading order \footnote{We neglect the term involving $J_1$ in \cref{eq:P_QFS_StartingFormula} and approximate $Y_1(x) \approx -2 /\pi x$ for $x \ll 1$.}
\begin{equation}
    \begin{aligned}
        &P_\psi(k,\eta_\mathrm{f})  \\
        &= \frac{H}{16\pi k \eta_0} [ (J_1(k \eta_0) - 2 k \eta_0 J_0(k \eta_0))^2 + (2 k \eta_0)^2 J_1^2(k \eta_0)].
    \end{aligned}
    \label{eq:P_QFS_Intermediate}
\end{equation}
If we demand the modes to be significantly subhorizon initially, $k \eta_0 \gg 1$, we can use the large-argument asymptotics of the Bessel functions to find the constant value
\begin{equation}
    P_\psi(k,\eta_\mathrm{f}) = H/2\pi^2.
\end{equation}
If the modes are only weakly subhorizon initially, $\smash{k \eta_0 \gtrsim 1}$, one has a remnant oscilations in $k$ as is can be captured by \cref{eq:P_QFS_Intermediate} that is shown in \cref{fig:QFSSpectraScalInv}.
The magnitude of the described transition process and thus the size of the scale-invariant regime depends on the number of e-foldings $N$ (as we show in \cref{fig:QFSSpectraScalInv}).
In the limiting case $\eta_\mathrm{f} \to \eta_0$ (or $N \to \infty$), the scale-invariant regime would extend to arbitrary small scales.

\Cref{fig:QFSSpectraScalInv} suggests that e-foldings $N \in \lbrace 4,5,6 \rbrace$ suffice to realize scale-invariance within one order of magnitude in $k$.
We can constrain the remaining parameters such that this magnitude falls within the $\mu \mathrm{m}$-regime where the quantum field simulation takes place.
In terms of an initial sound-speed $c_\mathrm{s,i}$ and an expansion time $\Delta t$ we have
\begin{equation}
    \eta_0 = \frac{c_\mathrm{s,i}}{H} =  c_\mathrm{s,i} \Delta t / N,
\end{equation}
such that typical expansion times are required to be
\begin{equation}
    \Delta t = 20 \mathrm{ms} \,  \left(\frac{\eta_0}{10 \mu \mathrm{m}}\right) \, \left(\frac{2.5 \mu \mathrm{m}/ \mathrm{ms}}{c_\mathrm{s,i} } \right) \, \left(\frac{N}{5}\right),
    \label{eq:TypicalParamsScalInv}
\end{equation}
where the reference value of the initial sound-speed $c_\mathrm{s,i}$ represents relatively high initial scattering lengths \cite{Viermann2022,Sparn2024}.
The reference value of $\eta_0$ stems from demanding that scale-invariance should occur for $k \gtrsim  0.5 \mu \mathrm{m}^{-1}$ where an empirical factor of $5$ is read-off from \cref{fig:QFSSpectraScalInv} to effectively realize $k \gg 1/\eta_0  = 0.1 \mu \mathrm{m}^{-1}$.

\subsection{Contraction}
\label{subsec:Cont}
It is well-known that scale-invariant power spectra can also be achieved via contracting scenarios \cite{Wands1999}.
In two spatial dimensions, we find this to be realized for the power-law contraction 
\begin{equation}
    a(t) = a_\mathrm{i} \left(1 - t/\tau\right)^{3/4},
    \label{eq:ScalCont}
\end{equation}
that also been independently analyed in the context of dipolar BECs \cite{ChandranFischer2025}.
Here $\tau$ is a constant that corresponds to the time where $a(t)$ would reach zero.
In the contracting scenario, the Hubble parameter $H(t) = \dot a(t)/a(t)$ decreases according to
\begin{equation}
    H(t) = - \frac{3}{4 (\tau - t)},
    \label{eq:HubbleCont}
\end{equation}
instead of remaining constant as it was the case for the exponential expansion.
We again define the (sound)-particle horizon via the integral
\begin{equation}
    \eta(t) = \int_0^t \frac{\mathrm{d}s}{a(s)} = \eta_0 [1 - \left(1 - t/\tau\right)^{1/4}], \quad \eta_0 = \frac{4 \tau}{a_\mathrm{i}},
    \label{eq:ConfTimeCont}
\end{equation}
that asymptotes $\eta_0$.
As a function of $\eta$, one has
\begin{equation}
    a(\eta) = a_\mathrm{i} \left(1 - \frac{\eta}{\eta_0}\right)^{3}, \quad H(\eta) = - \frac{3}{4\tau} \left(1 - \frac{\eta}{\eta_0}\right)^{-4},
    \label{eq:ScalfCont}
\end{equation}
which generates the effective mass $\smash{3/4(\eta_0 - \eta)^{2}}$,
leading to the same mode equation \eqref{eq:ModeEquationExpExp} as an exponential expansion.
Choosing an instantaneous vacuum initially (given through \cref{eq:InitialInstVac}) we thus find the mode solution \eqref{eq:ModeSol2DExpExp}.
For the power spectrum \eqref{eq:PowerSpecDef} we have to consider that, in terms of the cosmological parameters given in \cref{eq:ScalfCont,eq:HubbleCont,eq:ConfTimeCont}, the condition where the modes freeze-out, 
$k (\eta_0 - \eta) = 1$, only translates into a horizon-crossing interpretation up to a factor of three, $3 k / aH = 1$.
Also, the comoving Hubble horizon $(aH)^{-1}$ does not shrink but rather grows according to 
\begin{equation}
    [a(t) H(t)]^{-1} = a_\mathrm{i}^{-1} (1-t/\tau)^{-3/4}.
\end{equation}
Nevertheless, we can compute the power-spectrum \eqref{eq:PowerSpecDef} at final times $\eta_\mathrm{f} = \eta_0 (1 - \mathrm{e}^{-N/3})$, leading to
\begin{equation}
    \begin{aligned}
    & P_\psi(k,\eta_\mathrm{f})\\
    &\quad = \frac{\pi}{64} \frac{k\eta_0}{4\tau} \mathrm{e}^{2N/3}  \lvert r J_1(k\eta_0 \mathrm{e}^{-N/3}) + s Y_1(k\eta_0 \mathrm{e}^{-N/3}) \rvert^2,
    \end{aligned}
    \label{eq:P_QFS_Cont}
\end{equation}
which is structurally similar to \eqref{eq:P_QFS_StartingFormula}. 
For $\smash{k \eta_0\ll \mathrm{e}^{N/3}}$, we find accordingly
\begin{equation}
    \begin{aligned}
    &P_\psi(k,\eta_\mathrm{f}) \\
    &= \frac{1}{4\tau} \frac{1}{16 \pi k \eta_0} \lvert r J_1(k\eta_0 \mathrm{e}^{-N/3}) + s Y_1(k\eta_0 \mathrm{e}^{-N/3}) \rvert^2,
    \end{aligned}
\end{equation}
which asymptotes to the constant value $P_\psi(k,\eta_\mathrm{f}) \approx (8\pi^2 \tau)^{-1}$
for $k \eta_0 \gg 1$. 
In summary, both the exponential expansion and the power-law contraction \eqref{eq:ScalCont} lead to a scale-invariant spectrum,
\begin{equation}
    P_\psi(k,\eta_\mathrm{f}) \approx (2 \pi^2 a_\mathrm{i} \eta_0)^{-1},
\end{equation}
where $\eta_0$ is given by \cref{eq:ConfTimeExpExp} (in case of an exponential expansion) or \cref{eq:ConfTimeCont} (in case of the power-law contraction).
However, the scale-invariant plateau is narrower than for an exponential expansion by the same strength, since the plateau is limited from above by the criterium $\smash{k \eta_0 \lesssim \mathrm{e}^{N/3}}$ (instead of $\smash{k \eta_0 \lesssim \mathrm{e}^{N}}$ in context of the exponential expansion).
Consequently, a scale-invariant spectrum within one order of magnitude requires $N \simeq 15$, which is significantly larger than in the expanding case.
Demanding the scale-invariant regime to be confined to micrometer scales requires the contraction time-scale to be of the order
\begin{equation}
    \tau = 2 \mathrm{ms} \left(\frac{\eta_0}{10 \mu \mathrm{m}}\right) \left(\frac{1.25 \mu \mathrm{m}/\mathrm{ms}}{c_\mathrm{s,i}}\right),
    \label{eq:TypicalParamsScalInvCont}
\end{equation}
which is much shorter than the time scale of the exponential expansion \eqref{eq:TypicalParamsScalInv}. 
Note that $N$ does not enter \cref{eq:TypicalParamsScalInvCont} but rather enters when computing the contraction duration $\smash{\Delta t = \tau / (1-\mathrm{e}^{-4N/3})}$ that basically equals $\tau$ for sufficiently large $N$.
Here we chose a smaller reference speed of sound in accordance with relatively low initial scattering lengths required for an analog contraction.

\subsection{Dispersive effects}
\label{subsec:DispEff}
Clearly, the presence of superluminal (or supersonic) modes can critically affect the horizon-crossing conditions and thus the emergence of scale-invariance.
In the following we analyse the mode equation \eqref{eq:ModeEquationFinal} for three classes of effective frequencies:
\begin{enumerate}
    \item[(i)]  The time-dependent Bogoliubov dispersion relation with $\Omega_\mathrm{B}^2$ defined by \cref{eq:EffBogFreq}.
    \item[(ii)] The time-independent Bogoliubov dispersion relation with $\Omega_{\mathrm{B,0}}^2$ defined by \cref{eq:OmegaB0}. 
    \item[(iii)] The non-dispersive case $\Omega_\mathrm{Ac}^2$ defined by \cref{eq:OmegaAc}.
\end{enumerate}
\subsubsection{Exponential expansion}
\begin{figure*}
    \includegraphics[scale=0.425]{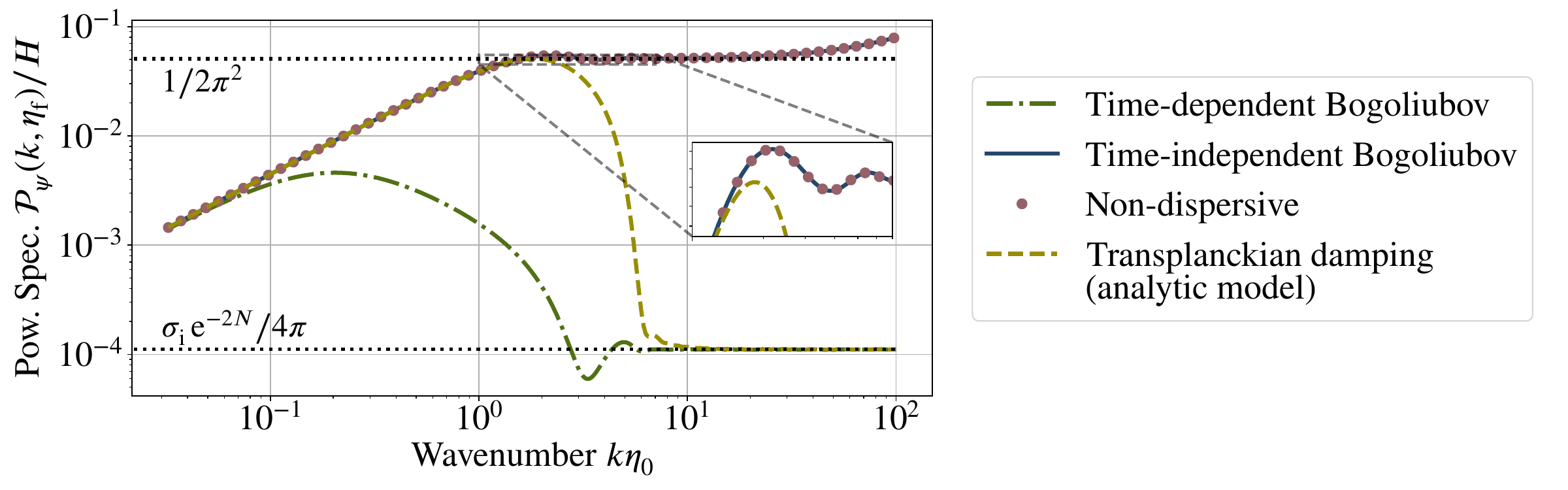}
    \caption{Sensitivity of scale-invariant power spectra produced by an exponential expansion to the dispersive effects described in cases (i) - (iii) using the reference values \eqref{eq:TypicalParamsScalInv} that can be combined into the initial scale-separation parameter $\sigma_\mathrm{i} = 30.68$ (more details in the main text and in context of \cref{fig:ScalSep}).
    The time-independent dispersion preserves the scale-invariant value $\mathcal{P}_\psi/H = 1/2\pi^2$ whereas the time-dependent case violates it on moderate ultraviolet scales due to Transplanckian damping but converges to another plateau in the far ultraviolet given by $\mathcal{P}_\psi / H = \sigma_\mathrm{i} \mathrm{e}^{-2N} / 2 \pi^2$. 
    } 
    \label{fig:ScalInvExpDisp}
\end{figure*}

In \cref{fig:ScalInvExpDisp} we show how these three cases modify the power spectrum created by an exponential expansion with reference values \eqref{eq:TypicalParamsScalInv} using numerical simulations of the mode evolution.
The time-independent Bogoliubov dispersion (which is equivalent to the superluminal Corley-Jacobson case \cite{Corley1998}) does respect scale-invariance as it is well-known \cite{MartinBrandenberger2001,MartinBrandenberger2001b,MartinBrandenberger2002,NiemeyerParentani2001}.
The time-dependent Bogoliubov case deviates from the non-dispersive case already in the mildly infrared regime $k \eta_0 \lesssim 0.1$, which we interprete as a dispersive reduction of non-adiabatic transitions (i.e.\@ the wavenumber-dependency of the effective mass).
The steep decline in power has been described as \emph{Transplanckian damping} in \cite{Hassan2003} (see also \cite{ChandranFischer2025} for a discussion in context of dipolar BECs) and was attributed to a reduced number of degrees of freedom at high energies that can be traced back to the canonical volume form in \cref{eq:RainbowAction} (and is captured by the function $\mathcal{J}_k$ defined in \cref{eq:CurlyJ}).

As an analytic model for the Transplanckian damping, we consult an auxiliary scenario where in the effective frequency \eqref{eq:EffBogFreq} the non-dispersive effective mass $m_\mathrm{eff}^2(k=0,\eta)$ is used  (cf.\@ \cref{eq:meffInfrared}) while keeping the time-dependence in $k^2/\mathcal{J}_k(\eta)$.
This leads to the mode equation
\begin{equation}
    \psi_k''(\eta) + \left(k^2 - \frac{\nu^2(k) -1/4}{(\eta - \eta_0)^2}\right) \psi_k(\eta) = 0,
    \label{eq:EffIndexModeEq}
\end{equation}
which can be solved analytically by Bessel-functions that have the wavenumber-dependent order
\begin{equation}
    \nu^2(k) = 1 - (k \eta_0)^4 / \sigma_\mathrm{i}^2, \quad \sigma_\mathrm{i} = \sqrt{2}\eta_0 / \xi_\mathrm{i},
    \label{eq:BesselOrderDisp}
\end{equation}
where we choose $\nu$ to have a positive real-part throughout the following.
Here we introduced the initial scale-separation parameter $\sigma_\mathrm{i}$ in terms of the initial healing length $\xi_\mathrm{i}$. 
It describes how small the initial (analog) Planck length $\xi_\mathrm{i}/\sqrt{2}$ is relative to the horizon-crossing scale $\eta_0$ 
such that the dispersive effects are expected to be weak if $\sigma_\mathrm{i} \gg 1$ (more details in the following subsection).
The auxiliary reference case can be classified into two regimes:
\paragraph{Transplanckian damping (mild ultraviolet):}
For $(k\eta_0)^4 \leq 1/\sigma_\mathrm{i}$, one has $\nu(k) \in \mathbb{R}$ and finds the power spectrum 
\begin{equation}
    \begin{aligned}
    P_\psi(k,\eta_\mathrm{f}) &= \frac{H}{32\pi} \Gamma^2(\nu(k)) \left(\frac{k \eta_0}{2}\right)^{1-2\nu(k)} \frac{ \mathrm{e}^{-2N(1 - \nu(k))} }{\sqrt{1 + (k \eta_0 \sigma_\mathrm{i})^2}}  \\
    \times \big \lbrace & \left[(-1 + 2 \nu(k)) J_{\nu(k)}(k \eta_0) - 2 k \eta_0 J_{\nu(k)-1}(k \eta_0) \right]^2 \\
    &+ (2 \omega_\mathrm{in} \eta_0)^2 J_{\nu(k)}(k\eta_0)^2 \big \rbrace,
    \end{aligned}
    \label{eq:MildUVSpec}
\end{equation}
for $k \eta_0 \ll \mathrm{e}^N$, which converges to the non-dispersive result \eqref{eq:P_QFS_Intermediate} in the infrared regime, $k \xi_\mathrm{i} \ll 1$, where $\nu(k)$ approaches unity.
Here, the Transplanckian damping arises from the factor $\smash{(k \eta_0)^{-2\nu(k)}}$ in \cref{eq:MildUVSpec}.

\paragraph{Transplanckian damping (far ultraviolet):}
In the far ultraviolet, $(k\eta_0)^4 \gg 1/\sigma_\mathrm{i}$, the power spectrum associated to both time-dependent and time-independent Bogoliubov dispersion remarkably converge to a plateau as well.
In this regime, non-adiabatic transitions do not occur anymore and the non-relativistic, instantaneous mode eigenenergy dominates the mode equation. 
This limit can be computed analytically resulting in
\begin{equation}
    P_\psi(k,\eta_\mathrm{f}) \approx \frac{H \sigma_\mathrm{i}}{4\pi} \mathrm{e}^{-2N},
\end{equation}
upon expressing the mode for $\nu(k) \in \mathrm{i}\mathbb{R}$ in terms of Dunsters Bessel-functions of purely imaginary order \cite{Dunster1990,Dunster2025} and performing a uniform asymptotic expansion in $\lvert \nu(k) \rvert \gg 1$ and $k \eta_0 \gg 1$ (details to be found in \cref{app:DispAnalSol}; note that \cref{eq:MildUVSpec} becomes numerically unstable for purely imaginary $\nu(k)$).

\subsubsection{Contraction}
\begin{figure}
    \includegraphics[width = \columnwidth]{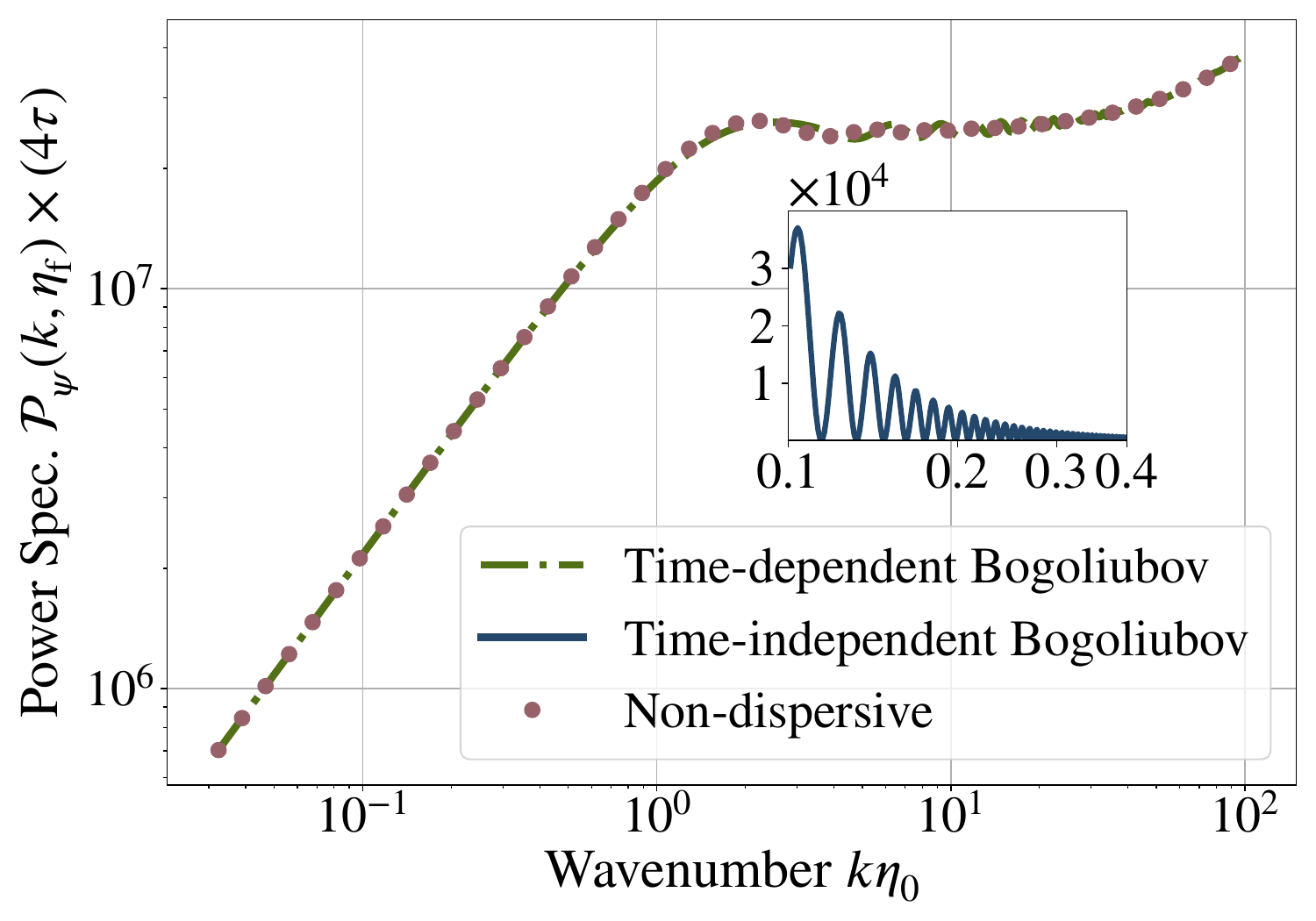}
    \caption{Sensitivity of scale-invariant power spectra produced by a power-law contraction to the dispersive effects described in cases (i) - (iii) using the reference values \eqref{eq:TypicalParamsScalInvCont}.
    The inset shows the time-indepedent Bogoliubov case (ii) where the spectrum is orders of magnitude lower and exhibits an oscillatory pattern.
    }
    \label{fig:ScalInvContDisp}
\end{figure}
In \cref{fig:ScalInvContDisp} we show the sensitivity of the power-law contraction with reference values \eqref{eq:TypicalParamsScalInvCont} to the dispersive effects described in cases (i) - (iii).
Therein, the situation is reversed in the sense that the time-dependent dispersion weakly alters the non-dispersive power spectrum by small oscillations whereas the time-independent case leads to a substantially damped signal of completely different form (as shown in the inset of \cref{fig:ScalInvContDisp}). 
\begin{figure}
    \includegraphics[width=\columnwidth]{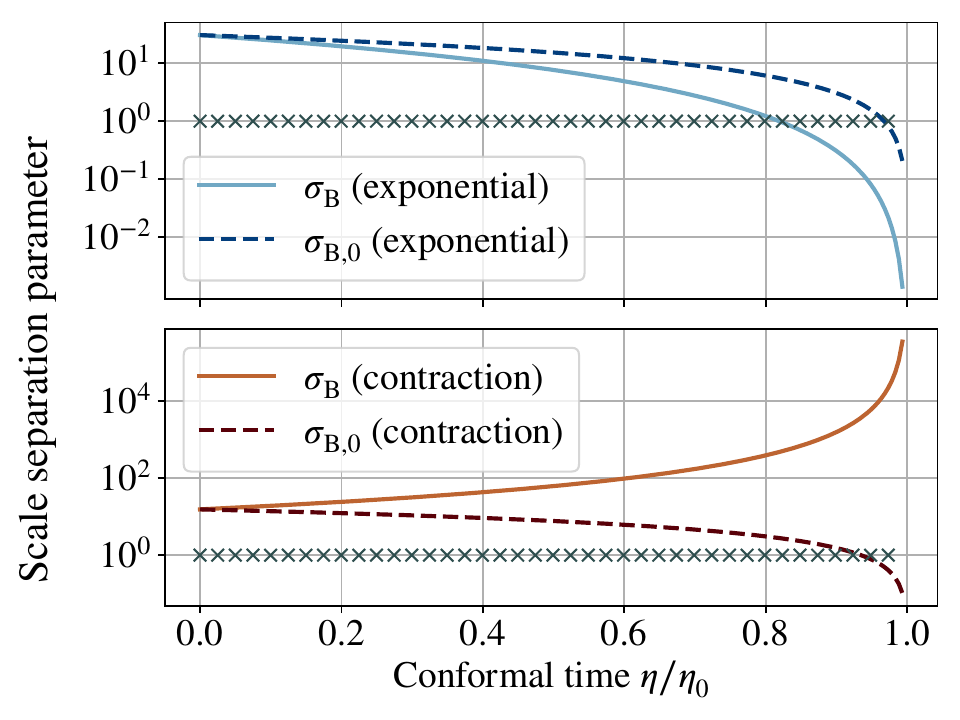}
    \caption{Scale separation parameters for the time-dependent and time-independent Bogoliubov dispersion relation in case of the exponential expansion ($N=5$, upper image) and the power-law contraction ($N=10$, lower image) 
    which are associated to \cref{fig:ScalInvExpDisp} and \cref{fig:ScalInvContDisp}, respectively.
    The initial scale-separation parameter is $\sigma_\mathrm{i} = 30.68$ for the exponential expansion and $\sigma_\mathrm{i} = 15.34$ for the power-law contraction.}
    \label{fig:ScalSep}
\end{figure}
This reversal can be understood as follows:
Per definition, the analog comoving Hubble-horizon $(a H)^{-1}$ indicates the boundary in space from which space recedes faster than the speed of sound (in comoving coordinates).
As discussed in \cite{NiemeyerParentani2001}, whether a violation of scale-invariance occurs can be efficiently reduced to whether the dispersion relation is significantly modified in the regime where the horizon-crossing occurs. 
More concretely, this can be be quantified via the so-called scale-separation parameter \cite{KempfNiemeyer2001,NiemeyerParentani2001,NiemeyerParentaniCampo2002} which we define as 
\begin{equation}
 \sigma =  \begin{cases}
 k_\mathrm{C} / aH & \mathrm{Expansion}, \\
 3 k_\mathrm{C} / a \lvert H\rvert & \mathrm{Contraction},
    \end{cases}
\end{equation}
where $k_\mathrm{C}$ is the comoving wavenumber well below which the dispersion relation is linear.
For instance for the effective comoving wavenumbers $F_\mathrm{B}$ and $F_\mathrm{CJ}$, we find 
\begin{equation}
    \sigma_\mathrm{B}(\eta) = \frac{\sqrt{2}}{\xi(\eta)} [a(\eta) H(\eta)]^{-1}, \, \, \sigma_\mathrm{B,0}(\eta) = \frac{\sqrt{2}}{\xi_\mathrm{i}}  [a(\eta) H(\eta)]^{-1}.
\end{equation}
Specifically for the expanding and contracting scenarios discussed in this section, one finds
\begin{equation}
    \sigma_\mathrm{B}(\eta) = \sigma_\mathrm{i}
    \begin{cases}
    (1 - \eta/\eta_0)^2 & \mathrm{Expansion} \\
    (1 - \eta/\eta_0)^{-2} & \mathrm{Contraction}    
    \end{cases},
\end{equation}
and 
\begin{equation}
    \sigma_\mathrm{B,0}(\eta) = \sigma_\mathrm{i}
    \begin{cases}
    (1 - \eta/\eta_0) & \mathrm{Expansion} \\
    (1 - \eta/\eta_0) & \mathrm{Contraction}
    \end{cases},
\end{equation}
where the initial scale-separation parameter $\sigma_\mathrm{i} = \sqrt{2} \eta_0 / \xi_\mathrm{i}$ was already used in context of \cref{eq:BesselOrderDisp} and is set by the chosen parameter configurations that are respectively given by \cref{eq:TypicalParamsScalInv} (expanding case) or \cref{eq:TypicalParamsScalInvCont} (contracting case).
The evolution in time of both scale-separation parameters is shown in \cref{fig:ScalSep} for both the expanding (cf.\@ \cref{fig:ScalInvExpDisp}) and the contracting case (cf.\@ \cref{fig:ScalInvContDisp}).
In the expanding case, the time-dependent Bogoliubov dispersion violates scale-separation at later times and therefore small scales, whereas the time-independent dispersion respects scale-separation for most times.
In the contracting case, this behavior is reversed such that the time-independent dispersion leads to a violation of scale-invariance.
In contrast, the time-dependent Bogoliubov case approaches scale invariance at late times.

\section{Influence of laboratory boundary conditions}
\label{sec:analogParticleProduction}
In the preceding discussion of scale-invariance we assumed an initial instantaneous vacuum state and traced the mode evolution up to some final instant in time where we evaluated the power spectrum.
Therein we did not account for boundary conditions that would be present in a laboratory setting such as an initial switch-on process or a final readout protocol of experimentally accessible field correlations.
The introduction and analysis of both is subject of this section.

\subsection{Non-equilibrium statistical functions}
\label{subsec:StatFunc}
\begin{figure}
    \includegraphics[scale=0.42]{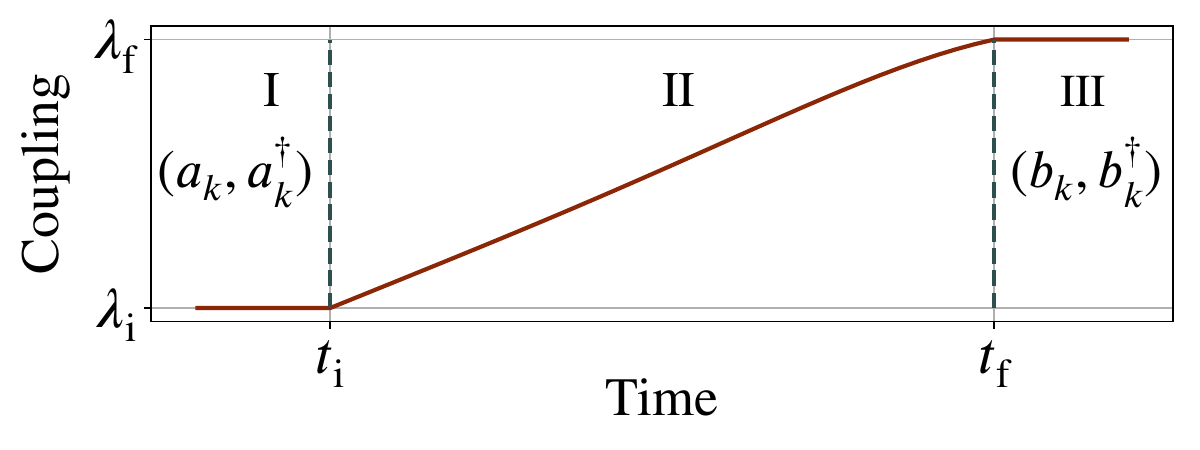}
    \caption{Non-equilibrium quantum dynamics via a temporal variation of the interatomic coupling constant $\lambda(t)$. 
    In the Heisenberg picture the process is understood as follows: A vacuum state is defined initially as the state that is annihilated by $\hat a_k$; 
    the two-point correlations are read out finally where the state can be described by the Bogoliubov-transformed operator $b_k = \alpha_k^* a_k - \beta_k^* a_{-k}^\dagger$. }
    \label{fig:ramp}
\end{figure}

In the quantum field simulator, the non-equilibrium dynamics are initiated through a specific temporal profile of $\lambda(t)$ for $t \in [t_\mathrm{i},t_\mathrm{f}$],
whereas $\lambda$ is static initially ($t < t_\mathrm{i}$) and finally ($t > t_\mathrm{f}$) (consider \cref{fig:ramp} for a schematic visualization). 
This protocol leads to pair-production of Bogoliubov excitations which are associated to well defined vacuum states for $t < t_\mathrm{i}$ and $t > t_\mathrm{f}$.
More details on this process can be found in \cite{Tolosa2022,Viermann2022,Schmidt2024,Sparn2024}.
Evaluated in the initial vacuum state, the equal-time statistical functions of the fluctuating fields $\phi_1$ and $\phi_2$ take the form
\begin{align}
    \frac{1}{2} \langle \lbrace \phi_1(t,\bm{x}), \phi_1(t,\bm{x}') \rbrace \rangle_c &= \int_k J_0(kL) \, \lvert w_k(t)\rvert^2, \label{eq:phi1corr_mode} \\[5pt]
    \frac{1}{2} \langle \lbrace \phi_2(t,\bm{x}), \phi_2(t,\bm{x}') \rbrace \rangle_c &= \int_k J_0(kL) \, \lvert v_k(t)\rvert^2, \label{eq:phi2corr_mode} 
\end{align}
where we already carried out the angular integral in Fourier space, due to the isotropy of the correlations. 
Due to homogeneity, the geometric kernel, given in terms of the cylindrical Bessel function $J_0(kL)$ only depends on the distance $L = \lvert \bm{x} - \bm{x}' \rvert$.
In the Schrödinger picture, the time-dependence of the state is encoded in the mode functions.
In particular, the pair-production process resulting from the out-of-equilibrium quantum dynamics can be captured by means of the Bogoliubov transformation
\begin{equation}
    v_k(t) = \alpha_k^* u_k(t) - \beta_k u_k^*(t),
    \label{eq:BogoliubovMode}   
\end{equation}
where $u_k$ is a positive frequency plane-wave in the final region,
\begin{equation}
     u_k(t \geq t_\mathrm{f}) = \sqrt{\frac{m c_k(t_\mathrm{f})}{\hbar k}} \mathrm{e}^{- \mathrm{i} \omega_k(t_\mathrm{f}) t}.
     \label{eq:PositiveFrequencyMode}
\end{equation} 
According to \cref{eq:v2Dyn} which relates $w_k$ and $v_k$, one consequently has 
\begin{equation}
    w_k(t) = -\mathrm{i} \frac{E_k}{\epsilon_k(t_\mathrm{f})} \left[ \alpha_k^* u_k(t) + \beta_k u_k^*(t) \right],
\end{equation}
where the pre-factor in \cref{eq:PositiveFrequencyMode} ensures the canonical normalization constraint \eqref{eq:NormalizationConstraint} to be fulfilled.
Evaluating the mode functions in the final region, one finds 
\begin{align}
    &\frac{\lvert w_k(t) \rvert^2}{k} = \frac{E_k}{\epsilon_k(t_\mathrm{f})} \left(\frac{1}{2} + \lvert \beta_k \rvert^2 - \Re[\alpha_k \beta_k \mathrm{e}^{2\mathrm{i} \hbar \omega_k(t_\mathrm{f}) t}]\right), \label{eq:wksq} \\
    &\frac{\lvert v_k(t) \rvert^2}{k} = \frac{\epsilon_k(t_\mathrm{f})}{E_k} \left(\frac{1}{2} + \lvert \beta_k \rvert^2 + \Re[\alpha_k \beta_k \mathrm{e}^{2\mathrm{i} \hbar \omega_k(t_\mathrm{f}) t}]\right), \label{eq:vksq}
\end{align}
with 
\begin{equation}
 \frac{\epsilon_k}{E_k} = \frac{\sqrt{2}}{k\xi} \sqrt{1 + \tfrac{1}{2} k^2 \xi^2}.
\end{equation} 
The phase-factor in \cref{eq:wksq,eq:vksq} indicates coherent oscillations in the final region $t \geq t_\mathrm{f}$ \cite{Robertson2017,Chen2021,Viermann2022,Sparn2024}, which stems from interference effects within a produced particle pair.
In the literature \cite{Chen2021,Duval2023}, the left-hand-side of \cref{eq:wksq,eq:vksq} is also known as the structure factor. Indeed, we may equivalently write
\begin{equation}
    \frac{1}{2} \langle \lbrace \phi_a(t,\bm{x}), \phi_b(t,\bm{x}') \rbrace \rangle_c = \int_k J_0(kL) \mathcal{S}_{ab}(t,k),
\end{equation}
with $\mathcal{S}_{ab}(t,k)$ as the structure factor.
If the quasi-particles are in thermal equilibrium, the structure factors becomes independent of time and take the form
\begin{equation}
    \mathcal{S}_{ab}(k) = \frac{1}{2} \coth(\frac{1}{2} \frac{\hbar \omega(k)}{k_B \,T})
    \begin{pmatrix}
   E_k/\epsilon_k & 0 \\
    0 & \epsilon_k/E_k  
    \end{pmatrix},
    \label{eq:StructureFactor_Thermal}
\end{equation}
as we show in \cref{app:ThermalStructureFactor} (see also \cite{LifshitzPitaevskii2013} and supplementary material of \cite{Hung2013}).
Therefore, if the initial state is thermal, the prefactors of \cref{eq:wksq,eq:vksq} have to be adjusted according to the thernal structure factor \eqref{eq:StructureFactor_Thermal}.

\subsection{Density correlations}
\label{subsec:DensCorr}
Let us consider the density-contrast relative to a homogeneous background, which is defined as
\begin{equation}
    \delta_c(t,\bm{x}) = \frac{n(t,\bm{x}) - n_0}{n_0}.
\end{equation}
Its equal-time two-point correlation function is related to the statistical function of $\phi_1$ via
\begin{equation}
    \begin{aligned}
        \mathcal{G}_{nn}(t,L) &\equiv \langle \delta_c(t,\bm{x}) \delta_c(t,\bm{x'}) \rangle, \\
        &= \frac{1}{n_0} \langle \lbrace \phi_1(t,\bm{x}), \phi_1(t,\bm{x}') \rbrace \rangle_c + \mathcal{O}(\phi^3).
    \end{aligned}
    \label{eq:GnnPhi1Phi1}
\end{equation}
In terms of a Fourier-decomposition, one thus has
\begin{equation}
    \mathcal{G}_{nn}(t,L) = \frac{2}{n_0} \int_k J_0(kL) \lvert w_k(t) \rvert^2.
\end{equation}
Employing \cref{eq:v2Dyn}, we find
\begin{equation}
    \begin{aligned}
        \mathcal{G}_{nn}(t,L) &=  \frac{2}{n_0} \int_k J_0(kL) \frac{\hbar^2}{(2m)^2} a_k^4 \lvert \dot v_k(t) \rvert^2 \\
        &= \frac{\hbar a_\mathrm{f}}{m n_0} \int_k J_0(kL) \frac{k \, S_k}{\sqrt{1 + \tfrac{1}{2}k^2 \xi_\mathrm{f}^2}},
    \end{aligned}
    \label{eq:GnnSk}
\end{equation}
where we used that $\lvert \dot v_k \rvert^2 \simeq \omega_k^2(t_\mathrm{f}) \lvert v_k \rvert^2$ up to a $\pi$-phase-shift in the coherent term.
Here we introduced the spectrum
\begin{equation}
    S_k(t) = \frac{\hbar}{2m} \frac{k}{c_k} \lvert v_k(t) \rvert^2,
    \label{eq:SkDef}
\end{equation}
as a generalization to the discussion in \cite{Tolosa2022} to which \cref{eq:GnnSk} is consistent for $k\xi_\mathrm{f}\ll 1$ \footnote{Note that \cref{eq:GnnSk} corrects a typo in equation (S2) of the supplementary material of \cite{Sparn2024}.}.
Moreover, \cref{eq:GnnSk} is consistent with Bogoliubov theory \cite{Esteve2006,Robertson2017,Chen2021} where one finds a well-known shot noise spectrum in the ultraviolet regime $k\xi_\mathrm{f}
\gg 1$ (consider \cref{app:LimitingCases} for a derivation of the shot-noise).

\subsection{Singular contributions}
\label{subsec:SingularCont}
In the specific case of the non-equilibrium setup shown in \cref{fig:ramp}, the analog scale-factor is kept static initially and finally,
\begin{align}
  &a(t) = a_\mathrm{i} &\hspace{-2cm} \mathrm{for} \, \, t \leq t_\mathrm{i},  \\ 
  &a(t) = a_\mathrm{f}  &\hspace{-2cm} \mathrm{for} \, \, t \geq t_\mathrm{f},
\end{align}
and switched on (and off) at $t_\mathrm{i}$ (and $t_\mathrm{f}$) leading to singular contribution to the effective mass, as we previously discussed in the acoustic approximation in \cite{Schmidt2024}.
In order to generalize this discussion to the dispersive case, we insert the laboratory scale-factor into \cref{eq:EffMass} and transforming back to laboratory time, resulting in
\begin{equation}
    m_\mathrm{eff}^2(k,t) = m_\mathrm{eff;reg}^2(k,t) + m_\mathrm{eff;sing}^2(k,t),
\end{equation}
with the regular terms 
\begin{equation}
    \begin{aligned}
        m_\mathrm{eff;reg}^2(k,t)= &- \frac{1}{4}  \frac{1 - 5 k^2 \xi^2(\eta) + 4 k^4 \xi^4(\eta)}{[1+ \tfrac{1}{2} k^2 \xi^2(\eta)]^2} \dot a^2(t) \\[5pt]
        &- \frac{1}{2}  \frac{1 - \tfrac{1}{2} k^2 \xi^2(\eta) }{1 + \tfrac{1}{2} k^2 \xi^2(\eta)}  \ddot a(t) a(t),
    \end{aligned}
\end{equation}
and the singular contributions
\begin{equation}
        \begin{aligned}
            m_\mathrm{eff;sing}^2(k,t) &= - \frac{\mathcal{H}(\eta)}{2} \frac{1- \tfrac{1}{2} k^2 \xi^2(\eta)}{1+ \tfrac{1}{2} k^2 \xi^2(\eta)} \\
            &\quad \times \left[\delta(t(\eta) - t_\mathrm{i}) - \delta(t(\eta) - t_\mathrm{f}) \right],
        \end{aligned}
\label{eq:EffectiveMassSingular}
\end{equation}
where we introduced the conformal Hubble rate $\mathcal{H}(\eta) = a'(\eta) / a(\eta)$.
The expression \eqref{eq:EffectiveMassSingular} is consistent with our previous result given in \cite{Schmidt2024} in the acoustic limit.
In the presence of dispersion however, these distributional terms differ for each mode and cross through zero at $k = \sqrt{2}/\xi$. 

\subsection{Boundary conditions}
\label{subsec:BoundaryConditions}
Due to the initial and final stasis of space, the canonically normalized mode function obeys the boundary conditions
\begin{align}
  &\psi_k(\eta \leq \eta_\mathrm{i}) = \frac{1}{\sqrt{2 \omega_\mathrm{in}(k)}} \mathrm{e}^{-\mathrm{i} \omega_\mathrm{in}(k)\eta}  \label{eq:PsiI} \\ 
  &\psi_k(\eta \geq \eta_\mathrm{f}) = \frac{1}{\sqrt{2 \omega_\mathrm{out}(k)}} \left( \alpha_k^* \mathrm{e}^{-\mathrm{i} \omega_\mathrm{out}(k)\eta} - \beta_k \mathrm{e}^{\mathrm{i} \omega_\mathrm{out}(k)\eta} \right) \label{eq:PsiIII}
\end{align}
with the Bogoliubov coefficients $\alpha_k, \beta_k$ as well as the incoming and outgoing frequencies,
\begin{equation}
\omega_\mathrm{in}(k) = k \sqrt{1 + \tfrac{1}{2} k^2 \xi^2_\mathrm{i}}, \quad \omega_\mathrm{out}(k) = k \sqrt{1 + \tfrac{1}{2} k^2 \xi^2_\mathrm{f}},
\end{equation}
where $\xi_\mathrm{i}$ and $\xi_\mathrm{f}$ are the initial and final values of the healing length, respectively.
In general the choice of the initial vacuum is ambiguous in dispersive theories, which has been discussed extensively \cite{MartinBrandenberger2001}.
However, in the BEC-cosmology simulator the mode $\psi_k(\eta)$ for $\eta \leq \eta_\mathrm{i}$ is associated to a unique quantum state that is determined by the initial stasis. 
These boundary conditions affect the mode function via matching conditions that read
\begin{equation}
\begin{aligned}
    \lim_{\eta \searrow \eta_\mathrm{i}} \psi_k(\eta) &= \lim_{\eta \nearrow \eta_\mathrm{i}} \psi_k(\eta), \\  
    \lim_{\eta \searrow \eta_\mathrm{i}} \psi_k'(\eta) &= \left(- \mathrm{i} \omega_\mathrm{i}(k) + \frac{\mathcal{H}(\eta_\mathrm{i})}{2} \frac{1-k^2 \xi^2_\mathrm{i}/2}{1+k^2 \xi^2_\mathrm{i}/2} \right) \lim_{\eta \nearrow \eta_\mathrm{i}} \psi_k(\eta),
\end{aligned}
\label{eq:MatchingLeft}
\end{equation}
at $\eta = \eta_\mathrm{i}$ and
\begin{equation}
\begin{aligned}
    \lim_{\eta \nearrow \eta_\mathrm{f}} \psi_k(\eta) &= \lim_{\eta \searrow \eta_\mathrm{f}} \psi_k(\eta), \\  
    \lim_{\eta \nearrow \eta_\mathrm{f}} \psi_k'(\eta) &= \lim_{\eta \searrow \eta_\mathrm{f}} \psi_k'(\eta) + \frac{\mathcal{H}(\eta_\mathrm{f})}{2} \frac{1-k^2 \xi^2_\mathrm{i}/2}{1+k^2 \xi^2_\mathrm{i}/2} \lim_{\eta \searrow \eta_\mathrm{f}} \psi_k(\eta),
\end{aligned}
\label{eq:MatchingRight}
\end{equation}
at $\eta = \eta_\mathrm{f}$, where the extra terms proportional to the conformal Hubble rate $\mathcal{H}$ stem from the singular contributions to the effective mass. 
The coherent oscillations discussed in context of \cref{eq:wksq,eq:vksq} manifest themselves in the relativistic particle spectrum $S(k)$ introduced in \cref{eq:SkDef} via the form
\begin{equation}
    S_k(\eta) = \frac{1}{2} + N_k + \Delta N_k^0 \cos(2 \omega_\mathrm{f} (\eta - \eta_\mathrm{f}) + \vartheta_k),
    \label{eq:SkOscillationForm}
\end{equation}
where the particle number,
\begin{equation}
    N_k = \lvert \beta_k \rvert^2,
    \label{eq:NkDef}
\end{equation}
the oscillation amplitude,
\begin{equation}
    \Delta N_k^0 = \lvert \alpha_k \beta_k \rvert,
    \label{eq:DeltaNk0Def}
\end{equation}
and the phase,
\begin{equation}
    \vartheta_k = \mathrm{arg}( \alpha_k \beta_k \mathrm{e}^{2 \mathrm{i} \omega_\mathrm{f} \eta_\mathrm{f}}),
    \label{eq:ThetakDef}
\end{equation}
follow from the matching conditions \eqref{eq:MatchingLeft} and \eqref{eq:MatchingRight}, as it is extensively discussed for example in \cite{Tolosa2022,Sanchez2022,Schmidt2024}.

\subsection{Influence of switching effects on scale-invariance}
\label{subsec:SwitchingEff}
\begin{figure}
    \includegraphics[width=\columnwidth]{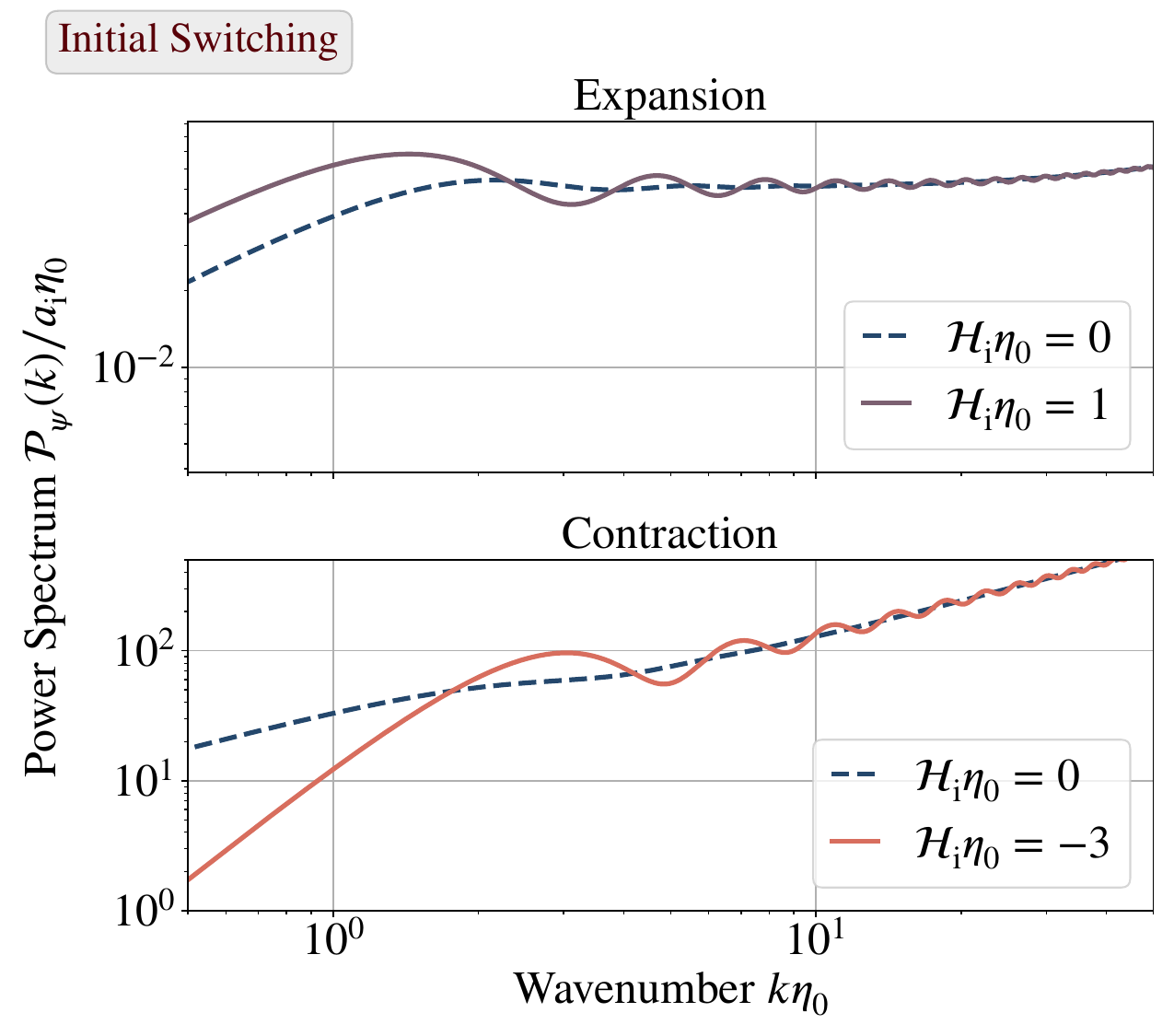}
    \includegraphics[width=\columnwidth]{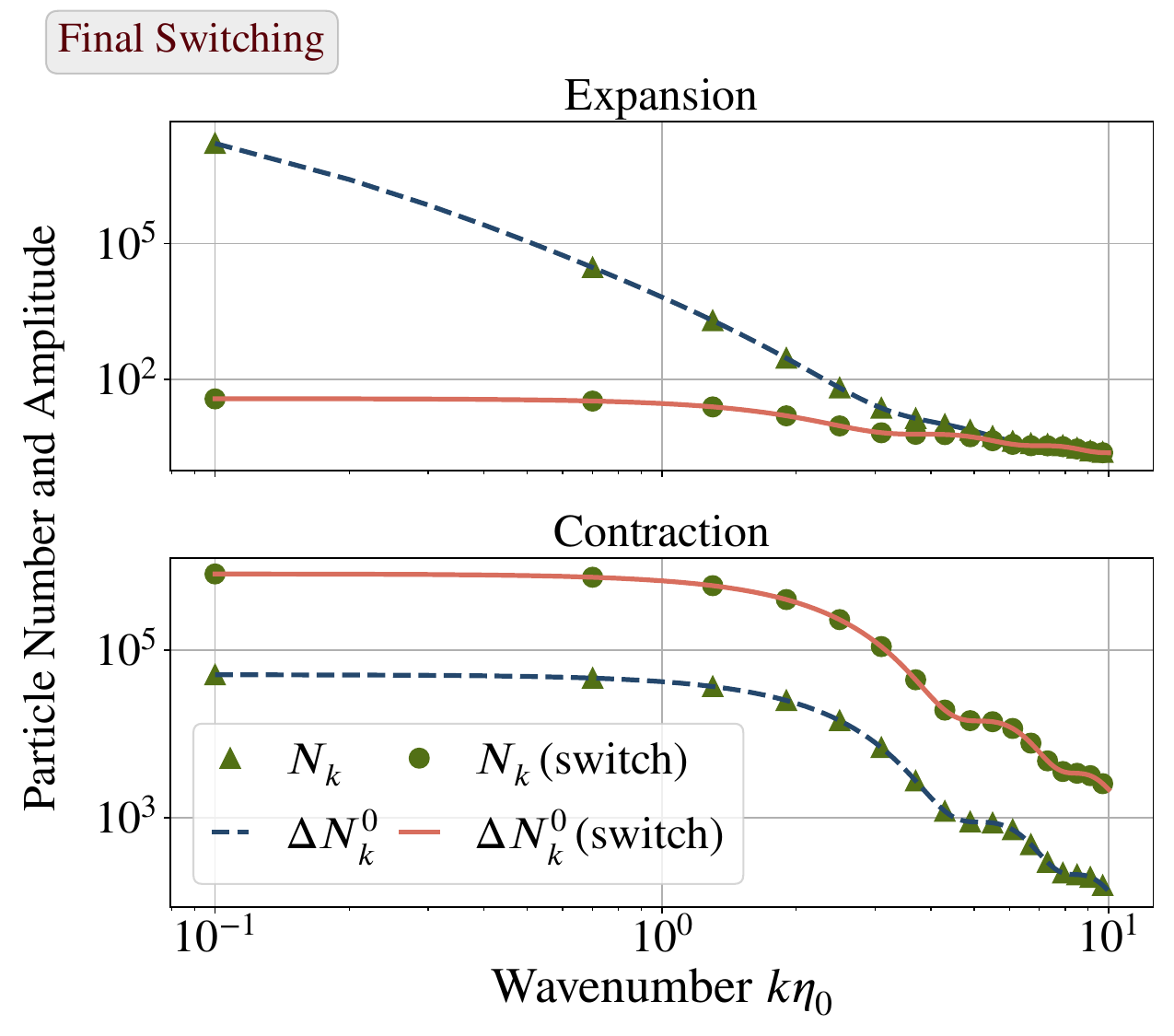}
    \caption{Influence of an initial switching (upper two panels) and a final switching (lower two panels) on the power spectrum (upper two panels) or the coherent oscillation amplitude (lower two panels) produced by either an exponential expansion and power-law contraction for $N=5$ according to non-dispersive theory.}
    \label{fig:Switching}
\end{figure}

Let us now consider the exponential expansion or power-law contraction to be embedded as a dynamical region II in the sense of \cref{fig:ramp}.
To separate these switching effects from the dispersive effects, we restrict the following analysis to the acoustic (non-dispersive) limit.
Accounting for an initial switch-on at a conformal Hubble rate $\mathcal{H}_\mathrm{i} \equiv \mathcal{H}(\eta_\mathrm{i})$ via invoking the boundary condition \eqref{eq:MatchingLeft} changes the power spectrum on final superhorizon scales into 
\begin{equation}
    \begin{aligned}
        &\frac{P_\psi(k,\eta_\mathrm{f})}{a_\mathrm{i}\eta_0}  = \frac{1}{16\pi k \eta_0} \\
        &\times [ (J_1(k \eta_0) - 2 k \eta_0 J_0(k \eta_0))^2 + (2 k \eta_0 - \mathcal{H}_\mathrm{i} \eta_0)^2 J_1^2(k \eta_0)],
    \end{aligned}
    \label{eq:PowerSpecSwitch}
\end{equation} 
where $\mathcal{H}_\mathrm{i} \eta_0 = 1$ (in the expanding case) and $\mathcal{H}_\mathrm{i} \eta_0 = -3$ (in the contracting case).
The contributions of these finite values to \cref{eq:PowerSpecSwitch} leads to oscillations instead of a scale-invariant plateau as shown in the upper two panels of \cref{fig:Switching}.
Interestingly, the initial switching reduces power in the infrared regime for a contracting scenario (or desqueezes the state \cite{Schmidt2024}) since $\mathcal{H}_\mathrm{i} \eta_0 < 0$ as opposed to the expanding case.

Considering a final switch-off at a conformal Hubble rate $\mathcal{H}_\mathrm{f} \equiv \mathcal{H}(\eta_\mathrm{f})$ and computing the particle number $N_k$ as well as the oscillation amplitude $\Delta N_k^0$ one finds the results shown in the lower two panels of \cref{fig:Switching}
where $\smash{\mathcal{H}_\mathrm{f} \equiv \mathrm{e}^N \mathcal{H}_\mathrm{i}}$ (in the expanding case) and $\smash{\mathcal{H}_\mathrm{f} = \mathrm{e}^{N/3} \mathcal{H}_\mathrm{i}}$ (in the contracting case).
Therein one sees a desqueezing of the state after the expanding case because $\mathcal{H}_\mathrm{f}$ enters with a relative minus sign to $\mathcal{H}_\mathrm{i}$ in \cref{eq:MatchingLeft,eq:MatchingRight} as opposed to the contracting case; 
note that the switch-off effect is orders of magnitude stronger than the switch-on effect due to the strong blueshift, $\mathcal{H}_\mathrm{f} \gg \mathcal{H}_\mathrm{i}$.

To realize why the particle number $N_k$ agrees with the amplitude $\Delta N_k^0$, consider that for a strong particle production process where $\lvert \beta_k \rvert \gg 1$ one can use the
the symplectic property of the Bogoliubov transformation, $\lvert \alpha_k \rvert^2 - \lvert \beta_k \rvert^2 = 1$, to infer that 
\begin{equation}
   N_k = \Delta N_k^0 \, [1 - \tfrac{1}{2} \lvert \beta_k \rvert^{-2} + \mathcal{O}(\lvert \beta_k \rvert^{-4})].
\end{equation}

\subsection{Deeper dive into experimental realizations}
In this final subsection, we utilize the concepts developed in this work to provide a further detailed theoretical description of scenarios that were experimentally realized in \cite{Sparn2024}.

\subsubsection{Linear expansion}

\begin{figure*}
    \includegraphics[width=\textwidth]{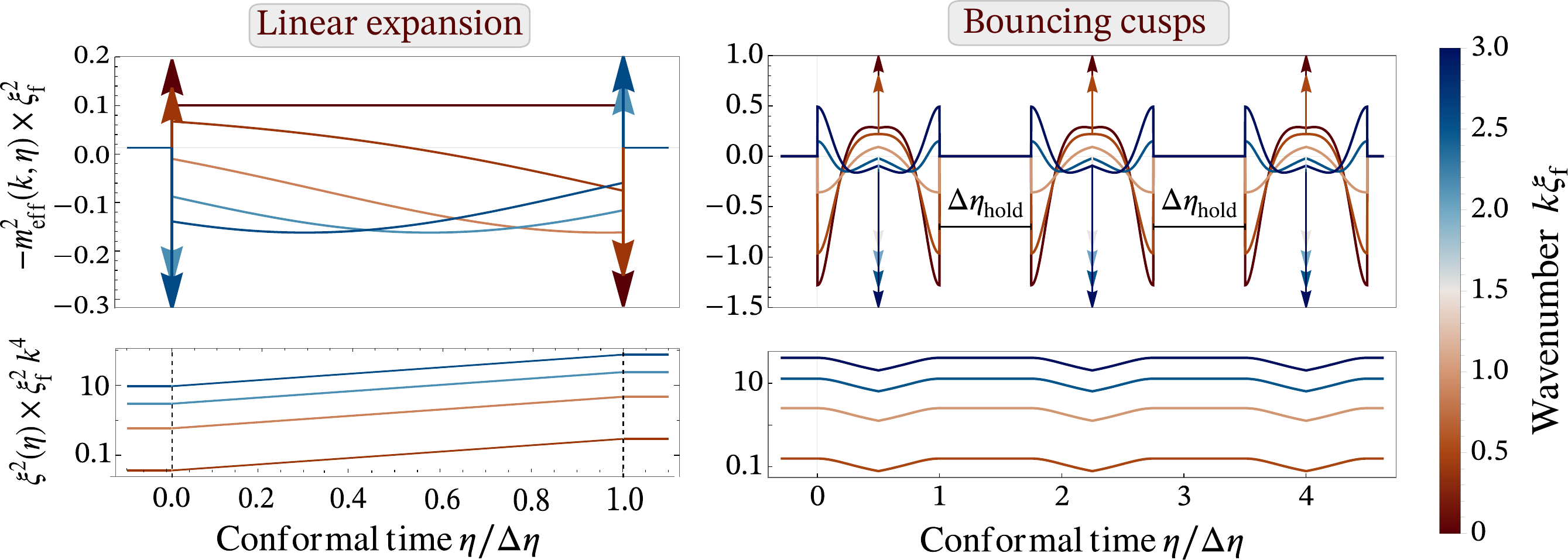}
    \caption{Effective mass $m_\mathrm{eff}^2(k,\eta)$ (top) and the purely dispersive factor $k^4 \xi(\eta)^2$ (bottom) for the linear expansion (left panels) and the bouncing cusp-scenario (right panels).
    Open parameters are set in accordance with the experiment \cite{Sparn2024}:
    In the former case, space expands by a ratio of $a_\mathrm{f}/a_\mathrm{i}=\sqrt{8}$ whereas it bounces by $a_\mathrm{b}/a_\mathrm{i} = \sqrt{2}$ in the latter case.
    The initial healing length is set to $\xi_\mathrm{i} =  0.12 \,  c_\mathrm{s}^\mathrm{i} \Delta t$ (for the expanding scenario) and to $\xi_\mathrm{i} = 1.6 \, c_\mathrm{s}^\mathrm{i}/\omega$ (for the bouncing scenario), 
    where $c_\mathrm{s}^\mathrm{i} \Delta t$ and $c_\mathrm{s}^\mathrm{i}/\omega$ represent characteristic sound (or particle) horizon scales in terms of the initial sound-speed $c_\mathrm{s}^\mathrm{i}$, the expansion duration $\Delta t$ and the bouncing frequency $\omega$.
    Distributional contributions are indicated via arrows.}
    \label{fig:EffectiveMass}
\end{figure*}

\begin{figure}
    \includegraphics[width=\columnwidth]{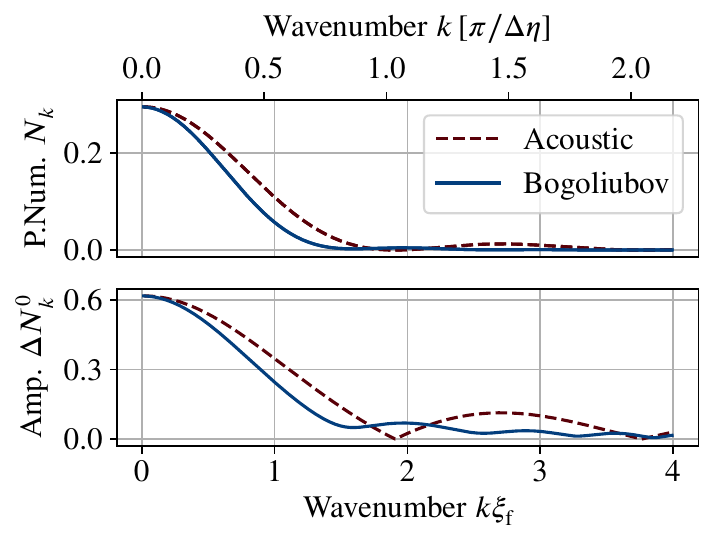}
    \caption{Particle number (upper image) and oscillation amplitude (lower image) after a linear expansion scenario.}
    \label{fig:RainbowRamp_Solution}
\end{figure}

A linear expansion between initial and final scale-factor values $a_\mathrm{i,f}$ corresponds to $a(t)/a_\mathrm{i} = 1 + H_0 t$ 
where the constant Hubble rate is $H_0 = (a_\mathrm{f}/a_\mathrm{i} - 1)/(c_\mathrm{s}^\mathrm{i} \Delta t)$ in terms of the expansion duration $\Delta t$.
The resulting effective mass and the purely dispersive term are shown in the left panel of \cref{fig:EffectiveMass}, where one sees that the latter dominates in the dispersive regime whereas the former is approximately constant in the infrared and switches sign towards the ultraviolet.  
\Cref{fig:RainbowRamp_Solution} shows the particle number $N_k$ and the oscillation amplitude $\Delta N_k^0$ obtained from a numerical integration of the mode equation \eqref{eq:ModeEquationFinal} under consideration of the matching prodecure described in \cref{subsec:BoundaryConditions}.
\label{subsec:LinExp}
In the acoustic approximation, zero-crossings of the amplitude occur for modes which satisfy $\sqrt{k^2 - H_0^2/4} = n \pi /\Delta \eta$ where $n$ is a positive integer and $\smash{\Delta \eta = (c_\mathrm{s}^\mathrm{i} \Delta t) \,  \ln(a_\mathrm{f}/a_\mathrm{i}) / (a_\mathrm{f}/a_\mathrm{i}-1)}$ is the final particle horizon.
This was explained in terms of a scattering picture in \cite{Schmidt2024,Sparn2024}. 
Taking into account the Bogoliubov dispersion, these quantum recurrences of the initial state become minima which are narrower spaced since the phase-velocity $c_k$ is larger than the speed of sound $c_\mathrm{s}$, which also leads to a weaker signal in total.   

\subsubsection{Bouncing cusps}
\label{subsec:CuspModul}
\begin{figure}
    \includegraphics[width = \columnwidth]{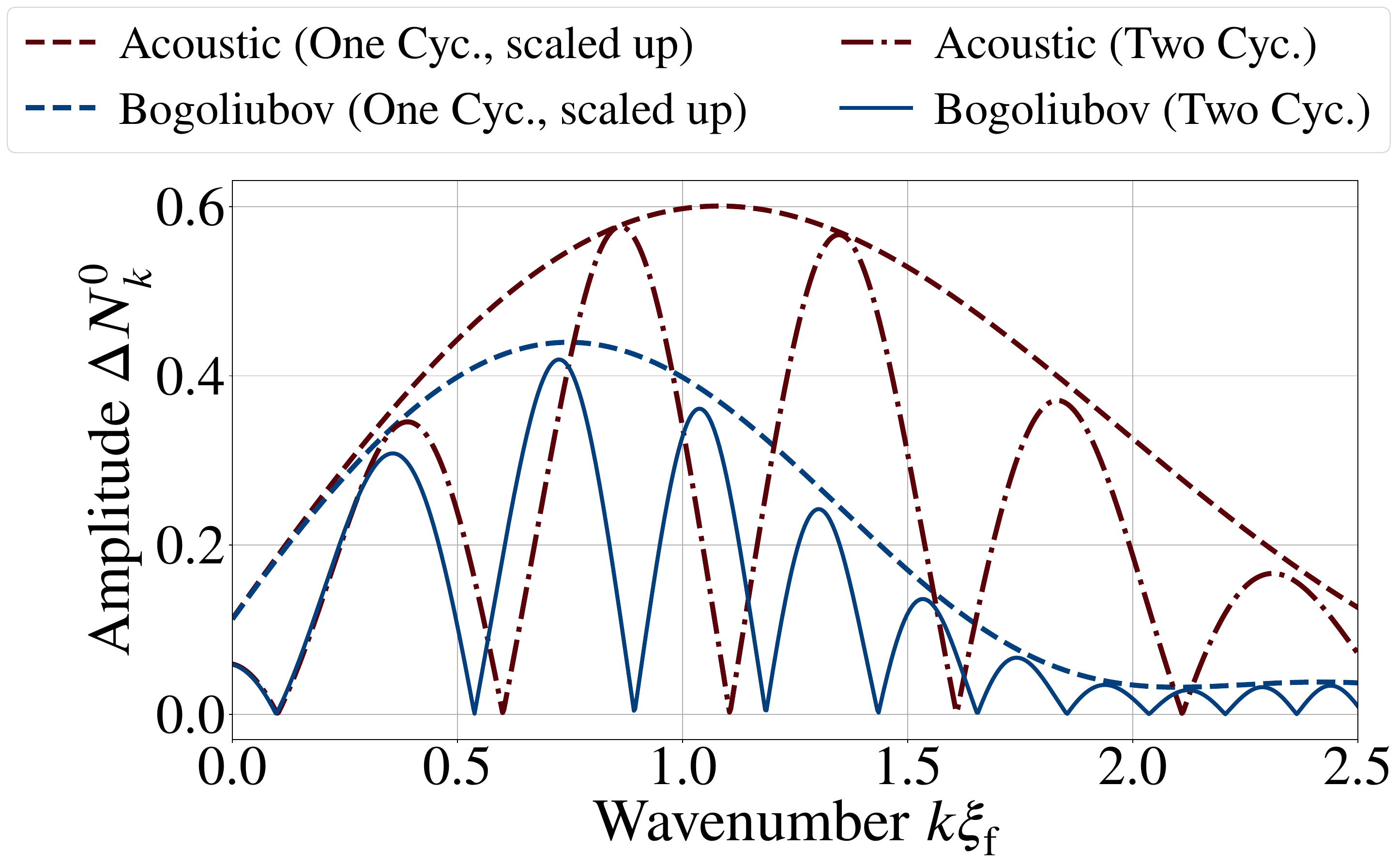}
    \caption{Coherent oscillation amplitude after a single (solid line) and two cycles (dashed line) in both acoustic (green curves) and Bogoliubov theory (red curves).
    Note that the signal actually grows with the amount of cycles, we merely scaled up the single cycle amplitudes by a factor of $2.5$ for better visibility of the diffractive pattern.}
    \label{fig:RainbowCuspsEnvelope}
\end{figure}

To increase the magnitude of the particle production signal, one can consider a cosmological contraction followed by an expansion by a ratio $a_\mathrm{b}/a_\mathrm{i}<1$  via the scale-factor 
\begin{equation}
    a_\mathrm{cusp}(t) = \left[(a_\mathrm{b}/a_\mathrm{i})^2 - \lvert \cos(\omega_0 t) \rvert \right]^{-1/2},
\end{equation}
with a cusp-singularity at the turning point.
This has the advantage that the probability of non-adiabatic transitions is boosted by the cusp.
\begin{figure}[h]
    \centering
    \includegraphics[{width=\columnwidth}]{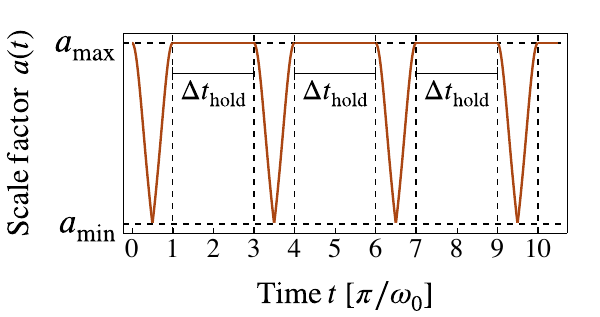}
    \caption{Temporal profile of a cusp-modulated scale-factor with $a_\mathrm{b}/a_\mathrm{i} = \sqrt{2}$ for $\Delta t_\mathrm{hold} = 2 \Delta t_\mathrm{cusp}$.}
    \label{fig:BouncingCusps}
\end{figure}
Periodically continuing such bouncing cusps with intervals of constant sound-speed $c_\mathrm{s,i}= 1/a_\mathrm{i}$ in between (as displayed in \cref{fig:BouncingCusps})
creates a periodic pattern in the effective mass with periodicity $\Delta \eta_\mathrm{hold} = c_\mathrm{s,i} \Delta t_\mathrm{hold}$ where $\Delta t_\mathrm{hold}$ is the duration of the holding intervals.
The dynamical evolution of modes is then equivalent to a propagation of waves through a dispersive medium 
whose elementary cells have substructure of size $\Delta \eta = \int_0^{\pi/\omega_0} \mathrm{d}t/a_\mathrm{cusp}(t)$, as it is shown in the right panel of \cref{fig:EffectiveMass}. 
Since particle production can be described as reflection in that medium \cite{Schmidt2024,Sparn2024}, a pattern of minima and maxima developes as a result of Bragg-diffraction of reflected waves
which is visualized in \cref{fig:RainbowCuspsEnvelope}.
Therein it is clearly visible that the single-cycle amplitude sets an envelope for the multi-cycle amplitudes which is related to the Fourier convolution theorem and a characteristic feature of Bragg-diffraction.

\section{Conclusion and outlook}
\label{sec:Conclusion}

We have derived an effective field theory for the U(1)-Goldstone boson of a scalar Bose-Einstein condensate with time-dependent contact interactions that is consistent with Bogoliubov theory and mapped it to a superluminal, non-local quantum field theory on a dispersive, curved spacetime whose geometry is set by the phase-velocity of Bogoliubov excitations and can be classified as a rainbow geometry.
Assuming a spatially homogeneous background density profile enabled us to neglect quantum pressure terms, thereby considerably facilitating the mapping at the expense of a spatially flat emergent spacetime.
Nevertheless these developments could also be used as an approximation to the spatially curved case (with an inhogeneous background density) for modes whose wavelength is smaller than the curvature radius (which is also known as the eikonal approximation \cite{Weinfurtner2009} in analogy to the small-wavelength-limit of geometric optics).

In the developed theory, effective Lorentz-invariance is restored on scales larger than the healing length which serves as an analog of the Planck length, with a modified dispersion relation assumed for the Transplanckian regime.
We continued by exploring this framework in a cosmological setup where the (dispersive) scale-factor is identified with the inverse phase-velocity and the comoving frame is the laboratory frame.
The mode equation was transformed into a parametric oscillator form that features wavenumber-dependent non-adiabatic transitions.
The latter originate from a dynamical, analog Planck length in the comoving frame or, equivalently, time-dependent interatomic interactions.
This scenario is different from ad-hoc approaches to Transplanckian regimes that only modify the instantaneous eigenenergy of a mode.

Subsequently, we discussed aspects of the cosmological Transplanckian problem, approached from the perspective of the BEC-analog simulator.
To prepare the discussion, we compared the time-dependent dispersion (with the dynamical analog Planck length) to the time-independent and the non-dispersive case regarding non-adiabaticity. 
We continued by describing how a scale-invariant power spectrum of fluctuations can emerge from an exponential expansion or power-law contraction in the BEC and estimated the required parameters to produce such a spectrum within one order of magnitude in the micrometer regime, which resulted in a required number of four to six e-foldings for an expansion time of about $20 \mathrm{ms}$ (in the expanding case)
whereas the contracting case demands an equivalent of $15$ e-foldings and a contraction time scale on the order of $2 \mathrm{ms}$; both scenarios clearly lie beyond current experimental capabilities.
It followed an investigation of the sensitivity of scale-invariance under the time-dependent Bogoliubov dispersion with a comparison to the time-independent case (that has been discussed extensively in the literature \cite{MartinBrandenberger2001,MartinBrandenberger2001b,MartinBrandenberger2002,NiemeyerParentani2001}).
The time-dependent case violates scale-invariance in the exponentially expanding scenario and showcases Transplanckian damping \cite{Hassan2003} (see also \cite{ChandranFischer2025} for a discussion in context of a dipolar BEC) that can be traced back to a reduced density of analog relativistic states at high energies. 
The time-indepedent case on the other hand preserves scale-invariance since the analog Planck scale and the horizon scale are well separated \cite{KempfNiemeyer2001,NiemeyerParentani2001,NiemeyerParentaniCampo2002} at all times (in contrast to the time-dependent case).
Similar arguments explain the reversed behavior in the contracting case.

In a final section, we introduced laboratory boundary conditions in time as typical for a BEC-experiment.
Relationships between the non-equilibrium statistical two-point functions of the canonically normalized, relativistic field variables and experimentally observarble \emph{in-situ} density-correlations were obtained including Bogoliubov dispersion.
We furthermore studied the impact of switching protocols on the scale-invariant power spectra, where switch-on processes lead to residual acoustic oscillations in the power spectrum whose frequency is determined by the switch-on-rate. 
Depending on a contracting or an expanding scenario, the switch-off process can lead to desqueezing and typically is more influential than the switch-on process due to the strong blueshift.

Finally, we applied the developed concepts to previously realized experimental scenarios \cite{Sparn2024}, thereby showcasing the occurence of Bragg-reflection and a suppression of the particle-production-signal relative to the acoustic case which surprisingly occurs already at scales well above the healing length
which needs to be taken into account when, for example, making statements about two-mode quantum entanglement close to the dispersive regime. 

With the present work we have contributed to the general discussion of the Transplanckian problem of cosmology via an analysis
of dispersive effects with a time-dependent UV-cutoff in the comoving frame that is motivated from the perspective of a BEC-cosmology simulator with time-dependent contact interactions.
This type of time-dependence dynamically disfavors scale-separation between the Hubble horizon and the cutoff-scale in the expanding case but dynamically favors it in the contracting case, such that the scale-invariance of only the latter is robust to such dispersive effects.
We furthermore conclude that Transplanckian damping effects induced by this time-dependence can settle to a another scale-invariant regime at much lower power in the far UV spectrum.
We also would like to note that Transplanckian damping occured despite the absence of a minimal length principle (in context of which it is studied for example in references \cite{Hassan2003,ChandranFischer2025}).

The extended framework provides quantitative access to stronger particle production scenarios whose spectrum can be measured deeper into the ultraviolet regime which enables a path toward constructing innovative scenarios that lie within experimental capabilities and overcome the described effects that obstruct scale-invariance.
This may ultimately lead to an experimental observation of a scale-invariant power spectrum of an analog relativistic field.

\section*{Acknowledgements}
The authors thank M.\@ Tolosa-Simeón, M.\@ Sparn, E.\@ Kath, N.\@ Liebster, J.\@ Duchene, H.\@ Strobel and M.\@ Oberthaler for valuable discussions.
C.F.S acknowledges support by the Studienstiftung des Deutschen Volkes and the Deutsche Forschungsgemeinschaft (DFG) under Grant No 406116891 within the Research Training Group RTG 2522/1.
The graphics in this work were made using perceptually uniform colormaps \cite{Crameri2021}.

\appendix 
\section{Comparison to Bogoliubov theory}
\label{app:BogTheoryComp}
In this section we compare the field-theory formalism of \cref{sec:BeyondAcoustic} to the more common Bogoliubov theory in the operator formalism.
First, it will be important to realize that the fluctuating fields $\phi_1,\phi_2$ are related to density and phase fluctuations $\delta n, \delta \theta$ which are the variables commonly used in the literature (a non-exhaustive list is \cite{Weinfurtner2006,Jain2007,Weinfurtner2009,Robertson2017}) and can be introduced here via 
\begin{equation}
    \Phi(t,\bm{r}) = [n_0(t,\bm{r}) + \delta n(t,\bm{r})] \, e^{\mathrm{i} [\theta_0(t,\bm{r}) + \delta S(t,\bm{r})]}.
\end{equation}
At leading order in both fluctuations, one finds  
\begin{align}
    \phi_1(t,\bm{r}) &= \delta n(t,\bm{r}) /\sqrt{2 n_0(t,\bm{r})} \label{eq:phi1_density}, \\ 
    \phi_2(t,\bm{r}) &= \delta \theta(t,\bm{r}) \, \sqrt{2 n_0(t,\bm{r})} \label{eq:phi2_phase},
\end{align}
after a local gauge transformation $\Phi \to \mathrm{e}^{-\mathrm{i}\theta_0(t,\bm{r})} \Phi$.
However, since phase fluctuations become ill-defined in the limit $n_0 \to 0$, where \cref{eq:phi2_phase} becomes singular, we work with the fields $\phi_1$ and $\phi_2$ in the main text.
In the following we closely follow \cite{Robertson2017}.

\subsection{Atomic density correlations}
At linear order, the total atomic density operator is expanded into a background- and fluctuating part
\begin{equation}
    \hat \rho(t,\bm{x}) = \rho_0 + \sqrt{\rho_0} \, (\delta \hat \phi + \delta \hat \phi^\dagger).
\end{equation}
The connected equal-time density correlator,
\begin{equation}
    \begin{aligned}
        &G^{\rho \rho}(t,\bm{x};t',\bm{x}') = \langle \hat \rho(t,\bm{x}) \hat \rho(t,\bm{x}') \rangle - \langle \hat \rho(t,\bm{x}) \rangle \langle \hat \rho(t,\bm{x}') \rangle
    \end{aligned}
\end{equation}
can be expanded in Fourier space,
\begin{equation}
    \begin{aligned}
        G^{\rho \rho}(t,\bm{k};t',\bm{k}') &= \int \mathrm{d}\bm{x} \, \mathrm{d}\bm{x'} \mathrm{e}^{-\mathrm{i}(\bm{k}\bm{x} + \bm{k}'\bm{x}')} G^{\rho \rho}(t,\bm{x};t',\bm{x}') \\
        &= \langle \delta \hat \rho_{\bm{k}}(t) \delta \hat \rho^\dagger_{\bm{k}'}(t) \rangle,
    \end{aligned}
\end{equation}
where 
\begin{equation}
   \delta  \hat \rho_{\bm{k}}(t) = \int \mathrm{d}\bm{x} \, \mathrm{e}^{-\mathrm{i} \bm{k} \bm{x}} \delta \hat \rho(t,\bm{x}).
\end{equation}
Since $\hat \rho(t,\bm{x})$ is a Hermitian operator, one has
\begin{equation}
\delta \hat \rho_{\bm{k}}(t) = \delta \hat \rho_{-\bm{k}}^\dagger(t).
\end{equation}
In terms of perturbative operators we have
\begin{equation}
    \delta \hat \rho_{\bm{k}}(t) = \sqrt{N} (\delta \hat \phi_{\bm{k}}(t) + \delta \hat \phi_{\bm{-k}}^\dagger(t)),
\end{equation}
where $\sqrt{N}$ ensures the fulfillment of the equal-time commutation relations
\begin{equation}
    [\delta \hat \phi_{\bm{k}}(t), \delta \hat \phi_{\bm{k'}}^\dagger(t)] = \delta_{\bm{k}\bm{k'}}.
\end{equation}
As a result of both $k$ and $-k$ excitations appearing in $\delta \hat \rho(\bm{k})$, it is indistinguishable to destroy a phonon of momentum $\bm{k}$ or to create one with momentum $\bm{-k}$. 
Furthermore, these two excitations necessarily interfere when performing density measurements. 

\subsection{Collective excitations} 
The Hamiltonian of the linear perturbations is diagonalized by the Bogoliubov transformation
\begin{equation}
    \delta \phi_{\bm{k}} = u_{\bm{k}} \hat \varphi_{\bm{k}} + v_{-\bm{k}}^* \hat \varphi_{-\bm{k}}^\dagger,
\end{equation}
where we introduce the quasi-particle operators $\hat \varphi_{\bm{k}}, \hat \varphi_{\bm{k}}^\dagger$.
The Bogoliubov coefficients can be parametrized as
\begin{equation}
        u_k = \cosh \chi_k, \quad v_k = \sinh \chi_k,
\end{equation}
with 
\begin{equation}
    \coth(2 \chi_k) = - 1 - k^2 \xi^2,
\end{equation}
where 
\begin{equation}
    \frac{\hbar^2}{2m\xi^2} = \lambda n_0 = m c_\mathrm{s}^2.
\end{equation}
In the following, we restrict the analysis to a stationary and homogeneous system.
Additionally we assume that the phonon states are statistically homogeneous. 
As a consequence, the correlations only depend the spatial distance $\lvert \bm{x - x'} \rvert$ or absolute values $k = \lvert \bm{k} \rvert$ in Fourier space.
Moreover, one only has correlations between $k$ and $-k$. 

\subsubsection{Density fluctuations in the phonon basis}
In terms of the quasi-particle operators, the density-fluctuation operator reads 
\begin{equation}
    \delta \hat \rho_{\bm{k}} = \sqrt{N} (u_k  + v_k) \left(\hat \varphi_{\bm{k}} + \hat \varphi_{-\bm{k}}^\dagger \right).
\end{equation}
In the quasi-particle ground-state, the density correlations are
\begin{equation}
    G_{\mathrm{vac}}^{\rho \rho}(k) = N (u_k + v_k)^2, 
    \label{eq:VacuumDensityCorrelator}
\end{equation}
whereas in an excited phonon state one has 
\begin{equation}
    G_{\mathrm{exc}}^{\rho \rho}(k) = G_{\mathrm{vac}}^{(2)}(k) (1 + n_k + n_{-k} + 2 \mathrm{Re}[c_k \mathrm{e}^{-\mathrm{i}\Omega_k t}]),
\end{equation}
where
\begin{equation}
    \begin{aligned}
        n_{\pm k} &= \langle \hat \varphi_{\pm \bm{k}}^\dagger \hat \varphi_{\pm \bm{k}} \rangle, \qquad c_k = \langle \hat \varphi_{\bm{k}} \hat \varphi_{-\bm{k}} \rangle, \\
    \end{aligned}
\end{equation}
represent the occupation numbers of and the coherence between $k$ and $-k$ excitations.

\subsubsection{Phase fluctuations in the phonon basis}
Phase fluctuations can be introduced in the atomic basis via the operator
\begin{equation}
    \delta \hat \phi(t,\bm{x}) = \frac{\delta \hat \rho}{2 \sqrt{\rho_0}} + \mathrm{i} \sqrt{\rho_0} \delta \hat \theta.
\end{equation}
In terms of the quasi-particle operators, one has
\begin{equation}
    \delta \hat \theta_k = \frac{\sqrt{N}}{2 \mathrm{i} \rho_0} (u_k - v_k) \left(\hat \varphi_{\bm{k}} - \hat \varphi_{-\bm{k}}^\dagger \right),
\end{equation}
such that $\delta \hat \phi$ is the proper conjugate variable to the density fluctuation $\delta \hat n_k$.
In the quasi-particle ground state, the phase correlations are 
\begin{equation}
    G^{\theta \theta}_{\mathrm{vac}}(k) = \frac{N}{4 \rho_0^2} (u_k - v_k)^2,
    \label{eq:VacuumPhaseCorrelator}
\end{equation}
whereas for an excited state 
\begin{equation}
    G^{\theta \theta}_{\mathrm{exc}}(k) = G^{\theta \theta}_{\mathrm{vac}}(k) (1 + n_k + n_{-k} - 2 \mathrm{Re}[c_k \mathrm{e}^{-\mathrm{i}\Omega_k t}]).
\end{equation}
This correlator contains no additional information on the phonon pair state as the coherent term is merely $\pi$-shifted.
The vacuum normalization is the respective inverse of the density normalization factor as it should be for proper conjugate variables. 
These considerations follow from the useful identities
\begin{equation}
   \begin{aligned}
    (u_k + v_k)^2 &= \frac{k \xi}{\sqrt{2}} \frac{1}{\sqrt{1 + \tfrac{1}{2}(k\xi)^2}} = \frac{E_k}{\epsilon(k)}, \\
    (u_k - v_k)^2 &= \frac{\sqrt{2}}{k \xi} \sqrt{1 + \tfrac{1}{2}(k\xi)^2} = \frac{\epsilon(k)}{E_k}, \\
   \end{aligned} 
   \label{eq:SqueezingFactors}
\end{equation}
where 
\begin{equation}
    \begin{aligned}
        \epsilon(k) &= c_\mathrm{s} k \sqrt{1 + \tfrac{1}{2} k^2 \xi^2}, \quad E_k = \frac{\hbar^2 k^2}{2m}.
    \end{aligned}
\end{equation}

\section{Derivation of dispersive effective mass}
\label{app:DispEffMass}

The goal of this section is to absorb the influence of the time-dependent and dispersive cosmological background into an effective mass, by transforming away the Hubble friction term in \cref{eq:ModeEqCosmicTime}.
To that end we rescale
\begin{equation}
    v_k(t) = f_k(t) \psi_k(t),
\end{equation}
and introduce conformal time $\mathrm{d} \eta = \mathrm{d} t/a(t)$ with respect to the bare scale-factor $a(t) = 1/c_\mathrm{s}(t)$
such that $\eta$ is still equivalent to the sound horizon in the BEC \cite{Schmidt2024}. 
The mode equation \eqref{eq:ModeEqCosmicTime} then takes the form
\begin{equation}
    \begin{aligned}
    &\psi_k'' + 2 \left( \frac{f_k'}{f_k} + \frac{a_k'}{a_k} - \frac{1}{2} \frac{a'}{a} \right) \psi_k' \\
    &\hspace*{1cm}+ \left[ \frac{f_k''}{f_k}  -2\left(\frac{1}{2} \frac{a'}{a}-\frac{a_k'}{a_k} \right) \frac{f_k'}{f_k}\right] \psi_k = - k^2 \frac{a^2}{a_k^2} \psi_k
    \end{aligned}
    \label{eq:ModeEquationConfIntermediate}
\end{equation}
To have a vanishing Hubble friction term, the scaling function is specified to fulfill
\begin{equation}
\frac{f_k'(\eta)}{f_k(\eta)} + \frac{a_k'(\eta)}{a_k(\eta)} =  \frac{1}{2} \frac{a'(\eta)}{a(\eta)},
\label{eq:gkRequirement}
\end{equation} 
such that the mode equation becomes 
\begin{equation}
    \psi_k''(\eta) + \left[k^2 + m_\mathrm{eff}^2(k,\eta)+ \frac{1}{2} k^4 \xi^2(\eta) \right] \psi_k(\eta) = 0,
\end{equation}
with the effective mass
\begin{equation}
    m_\mathrm{eff}^2(k,\eta) = \frac{\mathrm{d}}{\mathrm{d}\eta} \frac{f_k'(\eta)}{f_k(\eta)} - \left(\frac{f_k'(\eta)}{f_k(\eta)}\right)^2.
    \label{eq:meffGeneral}
\end{equation}
The specific form of the function $f_k$ can be obtained by considering \cref{eq:akaRel}
We can then rewrite \cref{eq:gkRequirement} as
\begin{equation}
    \frac{f_k'(\eta)}{f_k(\eta)} = \left(\frac{1}{2} - \mathcal{J}_k(\eta) \right) \frac{a'(\eta)}{a(\eta)},
\end{equation}
which can be integrated to 
\begin{equation}
    f_k(\eta) = [a(\eta) \mathcal{J}_k(\eta)]^{-1/2},
    \label{eq:ScalingFunction}
\end{equation}
and is the $(2+1)$-dimensional version of the mode rescaling employed in \cite{Hassan2003}.

\paragraph*{Acoustic limit.}
As a consistency check, we find in the infrared regime, $k \ll k_\xi$, where the dispersion relation \eqref{eq:BogDispersion}  is linear and thus $a_k = a + \mathcal{O}((k/k_\xi)^2)$, that
\begin{equation}
    m_\mathrm{eff}^2(k,\eta)  =  \frac{1}{4} \left(\frac{a'(\eta)}{a(\eta)}\right)^2 - \frac{1}{2} \frac{a''(\eta)}{a(\eta)} + \mathcal{O}\left( \frac{k^2}{k_\xi^2}\right),
\end{equation}
which coincides with the analysis \cite{Schmidt2024}. 

\paragraph*{Ultraviolet limit.}
In the ultraviolet limit, $k \gg \sqrt{2}/\xi$, the elementary excitations obey a non-relativistic, quadratic dispersion relation and represent individual atoms that are resolved on small scales $k^{-1} \ll \xi/\sqrt{2}$. 
As it is not possible anymore to project phonons out of the condensate basis, analog particle production should ceise. 
Indeed, the phase velocity becomes 
\begin{equation}
    c_k = \frac{\hbar k}{2m} + \mathcal{O}(\hbar^2k^2/4m^2c_\mathrm{s}^2(t)).
\end{equation}
Then, the mode equation \eqref{eq:ModeEqCosmicTime} is a time-independent harmonic oscillator
\begin{equation}
    \ddot v_k(t) + \frac{\hbar^2 k^4}{(2m)^2} v_k(t)  = 0,
\end{equation}
with non-relativistic dispersion and no time-dependent frequency, such that quanta can not be excited.
The ultraviolet limit of the effective mass is 
\begin{equation}
    \lim_{k \to \infty} m_\mathrm{eff}^2(k,\eta) = - \frac{3}{4} \left(\frac{a'(\eta)}{a(\eta)}\right)^2 + \frac{1}{2} \frac{a''(\eta)}{a(\eta)},
\end{equation}
which is adiabatically suppressed by the purely dispersive term $k^4/k_\xi^2(\eta)$ in the ultraviolet limit (as visible in \cref{fig:EffectiveMass}).

\section{Adiabatic vacuum via WKB-approximation}
\label{app:WKB}
We recall the well-known WKB-solution of a parametric oscillator with frequency $\Omega(k,\eta)$ that reads
\begin{equation}
    \psi_k^\mathrm{WKB}(\eta) = \frac{1}{\sqrt{2 W_k(\eta)}} \exp[-\mathrm{i} \int_{\eta_\mathrm{i}}^\eta W_k(\eta') \mathrm{d}\eta'],
\end{equation}
where the pre-factor ensures canonical normalization and the quantity $W_k$ satisfies the non-linear differential equation \cite{birrell_davies_1982}
\begin{equation}
    W_k^2(\eta) = \Omega^2(k,\eta) - \frac{1}{2} \left[\frac{W_k''(\eta)}{W_k(\eta)} - \frac{3}{2} \left(\frac{W_k'(\eta)}{W_k(\eta)}\right)^2 \right].
\end{equation}
This equation can be solved iteratively, resulting in an adiabatic expansion of the mode function where each iteration is referred to as an \emph{adiabatic order} \cite{birrell_davies_1982}.
The vacuum associated to a mode function at a certain adiabatic order is called the \emph{adiabatic vacuum}.
At zeroth adiabatic order one has $W_k^2(\eta) =  \Omega^2(k,\eta)$ and the mode \emph{adiabatically adjusts} to the time-dependent background (an adjustment local in time).
The zeroth-order adiabatic approximation is valid as long as both the conditions 
\begin{equation}
    \bigg \lvert \frac{\Omega'(k,\eta)}{\Omega(k,\eta)} \bigg \rvert \ll \lvert \Omega (k,\eta) \rvert, \quad \bigg \lvert  \frac{\Omega''(k,\eta)}{\Omega(k,\eta)} \bigg \rvert \ll \Omega^2(k,\eta),
    \label{eq:AdiabaticityCondition}
\end{equation}
are fulfilled \cite{Ford2021}.

\section{Analytic model for Transplanckian damping}
\label{app:DispAnalSol}

Neglecting the wavenumber dependence of the effective mass but keeping the time-dependence of the healing length,
the mode equation induced by an exponentially expanding spacetime reads
\begin{equation}
    \psi_k''(\eta) + \left(k^2 - \frac{\nu(k)^2 -1/4}{(\eta - \eta_0)^2}\right) \psi_k(\eta) = 0
\end{equation}
with 
\begin{equation}
    \nu(k)^2 = 1 - (k \eta_0)^4 / \sigma_\mathrm{i}^2 ,
\end{equation}
whose square-root we take to have a positive real part in the following.
We also did introduce the initial scale-separation parameter
\begin{equation}
    \sigma_\mathrm{i} = \sqrt{2}\eta_0 / \xi_\mathrm{i},
\end{equation}
to conveniently analyze the spectrum as a function of $k \eta_0$. 
The general solution to the mode equation can be written as
\begin{equation}
    \begin{aligned}
    \psi_k(\eta) &= \frac{\mathrm{i} \pi}{4} \frac{1}{\sqrt{2\omega_\mathrm{i}}} \sqrt{\frac{\eta_0 - \eta}{\eta_0}}  \\
    &\quad \times \left( r J_{\nu(k)}[k  (\eta_0 - \eta)] + s Y_{\nu(k)}[k  (\eta_0 - \eta)]\right)
    \end{aligned}
\end{equation}
The boundary conditions 
\begin{equation}
    \psi_k(0) = 1/\sqrt{2\omega_\mathrm{in}}, \quad \psi_k'(0) = - \mathrm{i} \omega_\mathrm{in} \psi_k(0),
\end{equation}
are solved by 
\begin{equation}
    r =  (1 - 2 \nu(k) - 2 \mathrm{i} \omega_\mathrm{i} \eta_0)  Y_{\nu(k)}(k \eta_0) + 2 k \eta_0 Y_{\nu(k)- 1}(k \eta_0) \\
\end{equation}
and 
\begin{equation}
    s = (-1 + 2 \nu(k) + 2 \mathrm{i} \omega_\mathrm{i} \eta_0) J_{\nu(k)}(k \eta_0) - 2 k \eta_0 J_{\nu(k)- 1}(k \eta_0).
\end{equation}
The power spectrum is 
\begin{align}
    \frac{P_\psi(k,\eta)}{H k \eta_0} &= \frac{\pi}{64} \frac{\mathrm{e}^{-2N}}{\sqrt{1 + \tfrac{1}{2} (k \xi_\mathrm{i})^2}}  \\
    & \times \lvert r J_{\nu(k)}[k(\eta_0 - \eta)] + s Y_{\nu(k)}[k(\eta_0 - \eta)] \rvert^2.
\end{align}
Let us again evaluate it at final times $\eta_\mathrm{f} = \eta_0 (1-\mathrm{e}^{-N})$ for modes with $1 \ll k \eta_0 \ll \mathrm{e}^{N}$. 

\subsection{Mild ultraviolet regime}
As long as $(k\eta_0)^4 < 1/\sigma$ we have $\nu(k) \in \mathbb{R}$ and therefore can approximate
\begin{equation}
    Y_{\nu(k)}(k \eta_0 \mathrm{e}^{-N}) \approx - \frac{\Gamma(\nu(k))}{\pi} \left(\frac{k \eta_0 \mathrm{e}^{-N}}{2}\right)^{-\nu(k)},
\end{equation}
to find 
\begin{widetext}
\begin{equation}
    \begin{aligned}
            P_\psi(k,\eta_\mathrm{f}) = \frac{H}{32\pi} \Gamma^2(\nu(k)) \left(\frac{k \eta_0}{2}\right)^{1-2\nu(k)} \frac{ \mathrm{e}^{-2N(1 - \nu(k))} }{\sqrt{1 + (k \eta_0 \sigma)^2}}
    \big \lbrace & \left[(-1 + 2 \nu(k)) J_{\nu(k)}(k \eta_0) - 2 k \eta_0 J_{\nu(k)-1}(k \eta_0) \right]^2  \\
    &\qquad + (2 \omega_\mathrm{i} \eta_0)^2 J_{\nu(k)}(k\eta_0)^2 \big \rbrace,
    \end{aligned}
\end{equation}
\end{widetext}
which converges to the non-dispersive result \eqref{eq:P_QFS_Intermediate} in the infrared regime, $k \xi_\mathrm{i} \ll 1$, where $\nu(k) \to 1$.

\subsection{Far ultraviolet regime}
For $(k\eta_0)^4 > 1/\sigma_\mathrm{i}$, the order $\nu(k)$ becomes purely imaginary, $\nu(k) \in \mathrm{i}\mathbb{R}$.
This requires to write the mode solution in a different basis,
\begin{equation}
    \begin{aligned}
            \psi_k(\eta) &= \frac{\mathrm{i} \pi}{4} \frac{1}{\sqrt{2\omega_\mathrm{i}}} \sqrt{\frac{\eta_0 - \eta}{\eta_0}}  \\
            &\qquad \times \left(r F_{\lvert \nu(k) \rvert}[k(\eta_0 - \eta)] + s G_{\lvert \nu(k) \rvert}[k(\eta_0 - \eta)] \right),
    \end{aligned}
\end{equation}
with the coefficients
\begin{align}
    r &= -(1 + 2 \mathrm{i} \omega_\mathrm{i} \eta_0) G_{\lvert \nu(k) \rvert} (k \eta_0) - 2 k \eta_0 \partial_x G_{\lvert \nu(k) \rvert}(x) \lvert_{x = k \eta_0}, \\
    s &= (1 + 2 \mathrm{i} \omega_\mathrm{i} \eta_0) F_{\lvert \nu(k) \rvert} (k \eta_0) + 2  k \eta_0 \partial_x F_{\lvert \nu(k) \rvert}(x) \lvert_{x = k \eta_0},
\end{align}
and the basis-functions
\begin{equation}
    F_\nu(x) = \frac{1}{2 \cosh(\pi \nu/2)} \left(J_{\mathrm{i}\nu}(x) + J_{-\mathrm{i}\nu}(x)\right), \label{eq:FnuDef} \\
\end{equation}
and 
\begin{equation}
    \begin{aligned}
            G_\nu(x) &= \frac{1}{2\mathrm{i} \sinh(\pi \nu/2)} \left(J_{\mathrm{i}\nu}(x) - J_{-\mathrm{i}\nu}(x)\right) \\
             &=  \frac{1}{2 \cosh(\pi \nu/2)} \left(Y_{\mathrm{i}\nu}(x) + Y_{-\mathrm{i}\nu}(x)\right),
    \end{aligned}
    \label{eq:GnuDef}
\end{equation}
which are numerically stable alternatives to the unmodified Bessel-functions of purely imaginary order (see \cite{Dunster1990} for the construction by Dunster). 
They are real for positive arguments, have the similar small-argument asymptotics 
\begin{align}
    F_\nu(x) &= \sqrt{\frac{2 \tanh(\pi \nu/2)}{\pi \nu}} \cos\left(\nu \ln(x/2) - \mathrm{arg} \, \Gamma(1 + \mathrm{i}\nu)\right), \label{eq:FnuSmallArgument} \\
    G_\nu(x) &= \sqrt{\frac{2 \coth(\pi \nu/2)}{\pi \nu}} \sin\left(\nu \ln(x/2) - \mathrm{arg} \, \Gamma(1 + \mathrm{i}\nu) \right),  \label{eq:GnuSmallArgument} 
\end{align}
as $x \to 0^+$ and large-argument asymptotics
\begin{align}
    F_\nu(x) &= \sqrt{\frac{2}{\pi x}} \cos\left(x - \pi/4\right), \\
    G_\nu(x) &= \sqrt{\frac{2}{\pi x}} \sin\left(x - \pi/4\right),
\end{align}
as $x \to + \infty$.
Furthermore, their Wronskian,
\begin{equation}
\partial_x F_\nu(x) G_\nu(x) - \partial_x G_\nu(x) F_\nu(x) = 2/\pi x,
\end{equation}
is identical to the Wronskian of the unmodified Bessel functions.
In the regime $k \eta_0 \ll \mathrm{e}^{N}$ we can use the small-argument asymptotics given by \cref{eq:FnuSmallArgument,eq:GnuSmallArgument} to write the power spectrum as
\begin{equation}
    \begin{aligned}
    &\frac{P_\psi(k,\eta_\mathrm{f})}{H k \eta_0} = \frac{\mathrm{e}^{-2N}}{64 \lvert \nu(k) \rvert} \frac{1}{\sqrt{1 + (k \eta_0 \sigma)^2}} \\
    &\times \lvert \tanh(\tfrac{1}{2}\pi \lvert \nu(k) \rvert) r \cos (\varphi_k) + \coth(\tfrac{1}{2} \pi \lvert \nu(k) \rvert) s \sin (\varphi_k) \rvert^2,
\end{aligned}
\end{equation}
where \cite{Dunster1990}
\begin{equation}
    \varphi_k = \lvert \nu(k) \rvert \ln(\tfrac{1}{2}\mathrm{e}^{-N} k \eta_0) - \mathrm{arg} \, \Gamma(1 + \mathrm{i}\lvert \nu(k) \rvert).
\end{equation}
For modes that are well superhorizon finally, $k \eta_0 \gg 1$, the order $\lvert \nu(k) \rvert \approx (k \eta_0)^2 / \sigma_\mathrm{i}$ becomes large such that we can further simplify
\begin{equation}
    \begin{aligned}
    P_\psi(k,\eta_\mathrm{f}) &= \frac{H \sigma_\mathrm{i} \mathrm{e}^{-2N}}{64 \lvert \nu(k) \rvert} \lvert  r \cos \left(\varphi_k\right) + s \sin\left(\varphi_k\right) \rvert^2. 
\end{aligned}
\end{equation}
To approximate the coefficients $r$ and $s$ we have to take into account that \emph{both} the argument $k \eta_0$ and the order $\lvert \nu(k) \rvert$ of the remaining Bessel functions becomes large. 
This kind of situation can be handled by the uniform asymptotic expansion that has been derived recently by Dunster \cite{Dunster2025}.
According to Dunster, one has 
\begin{equation}
    \lvert J_\mathrm{i\lvert \nu \rvert}(\lvert \nu \rvert x) \rvert \sim \frac{    \exp\left(\frac{1}{2} \lvert \nu \rvert \pi + \mathcal{O}(\lvert \nu \rvert^{-2}) \right)}{\sqrt{2\pi \lvert \nu \rvert} (x^2 + 1)^{1/4}}, \label{eq:AbsJ_UniformAsymp} \\
\end{equation}
and 
\begin{equation}
    \arg J_\mathrm{i\lvert \nu \rvert}(\lvert \nu \rvert x) \sim \lvert \nu \rvert \rho(x) - \frac{\pi}{4} + \mathcal{O}(x^{-3}),
    \label{eq:ArgJ_UniformAsymp} 
\end{equation}
for $\lvert \nu \rvert \gg 1$, uniformly in $x \in (0,\infty)$, where
\begin{equation}
    \rho(x) = \sqrt{x^2 + 1}- \ln \left(\frac{1 + \sqrt{x^2 + 1}}{x}\right),
\end{equation}
and $\sim$ denotes asymptotic equivalence.
We directly infer that 
\begin{align}
    F_\mathrm{\nu}(\nu x) &\sim \sqrt{\frac{2}{\pi \nu}} (x^2 + 1)^{-1/4} \cos \left(\nu \rho(x) - \frac{\pi}{4}  \right), \\
    G_\mathrm{\nu}(\nu x) &\sim \sqrt{\frac{2}{\pi \nu}} (x^2 + 1)^{-1/4} \sin \left(\nu \rho(x) - \frac{\pi}{4}  \right),
\end{align}
according to the definitions \eqref{eq:FnuDef} and \eqref{eq:GnuDef}.
Applied to the present case, this means that we can take
\begin{align}
    F_{(k \eta_0)^2/ \sigma_\mathrm{i}}(k \eta_0) &\sim \sqrt{\frac{2}{\pi\sigma}} (k \eta_0)^{-1} \cos (\mu_k), \\
    G_{(k \eta_0)^2/ \sigma_\mathrm{i}}(k \eta_0) &\sim \sqrt{\frac{2}{\pi\sigma}} (k \eta_0)^{-1} \sin (-\mu_k),
\end{align}
where 
\begin{equation}
    \mu_k = \frac{(k \eta_0)^2}{\sigma_\mathrm{i}} \ln(2 k \eta_0),
\end{equation}
to approximate the coefficients $r$ and $s$ at leading order in $1/(k \eta_0)$.
As a result, we have 
\begin{equation}
    r \sim \sqrt{\frac{2}{ \pi \sigma_\mathrm{i}}} (2\mathrm{i} k \eta_0) \sin(\mu_k), \quad  s \sim \sqrt{\frac{2}{\pi \sigma_\mathrm{i}}} (2\mathrm{i} k \eta_0) \cos(\mu_k),
\end{equation}
and therefore find the constant value
\begin{equation}
    \begin{aligned}
        P_\psi(k,\eta_\mathrm{f}) 
        &= \frac{H \sigma_\mathrm{i}}{4\pi} \mathrm{e}^{-2N},\\
    \end{aligned}
\end{equation}
where we averaged over the rapidly oscillating phases $\mu_k$ and $\varphi_k$.

\section{Density correlations: acoustic and ultraviolet limits}
\label{app:LimitingCases}

\subsection{Infrared (acoustic) limit }
The relation \eqref{eq:GnnSk} between the density-density correlations $\mathcal{G}_{nn}$ and the power spectrum $S(k)$ reduces to
\begin{equation}
    G_{nn}(t,L) = \frac{\hbar a_\mathrm{f}}{m n_0} \int_k J_0(kL) \, k \, S(k),
\end{equation}
which coincides with the acoustic result \cite{Tolosa2022}.

\subsection{Ultraviolet limit: Shot-noise}
As a consistency check, let us apply this limit to \cref{eq:GnnSk} for the density correlation function. 
As a consequence of abscent quasi-particles, the limit of the excitation spectrum assumes the value $S(k) \simeq 1/2$ and becomes static.
From the integrand of \cref{eq:GnnSk} we can read off that 
\begin{equation}
    \langle \delta_c(\bm{k}) \delta_c(\bm{k'}) \rangle = \frac{\hbar a_\mathrm{f}}{m n_0} \frac{k \, S(k)}{\sqrt{1 + \tfrac{1}{2}k^2 \xi_\mathrm{f}^2}} \delta(\bm{k} + \bm{k'}) \\
\end{equation}
which in the limit of large momenta $k \xi_\mathrm{f} \gg 1$ and $k' \xi_\mathrm{f} \gg 1$ can be approximated to
\begin{equation}
    \langle \delta_c(\bm{k}) \delta_c(\bm{k'}) \rangle \simeq \frac{1}{n_0} \delta(\bm{k} + \bm{k'}),
\end{equation}
or in terms of connected density-fluctuations
\begin{equation}
    \langle n(\bm{k}) n(\bm{k'}) \rangle_c = \frac{N}{V} \delta(\bm{k} + \bm{k'}).
\end{equation}
where we set $n_0 = N/V$, with $N$ as the particle number and $V$ as the volume of the atomic cloud. 
Thus, the density fluctuations obey a shot noise spectrum $ \Delta N \simeq \sqrt{N}$ in the ultraviolet regime,
as expected from considering that for high wavenumbers the kinetic energies of the atoms surpass the interaction energies by far, resulting in an equivalence to an ideal bose gas.
To realize why one expects a shot noise spectrum in the ultraviolet limit of the latter, let us consider a homogenenous, ideal bose gas contained in a volume $V$.
The effective bosonic field $\Psi$ is quantized via the mode expansion \cite{LandauLifshitz1980}
\begin{equation}
    \Phi(\bm{x}) = \sum_k \phi_{\bm{k}}(\bm{x}) \hat a_{\bm{k}}, \quad \phi_{\bm{k}}(\bm{x}) = \exp{\mathrm{i} \bm{k} \cdot \bm{x}}/\sqrt{V}
    \label{eq:NonRelModeExpansion}
\end{equation}
where one has the well-known commutation relations
\begin{equation}
    \begin{aligned}
    \relax    [\Phi(\bm{x}), \Phi^\dagger(\bm{y})] &= \delta(\bm{x} - \bm{y}), \\
    \relax    [\Phi(\bm{x}), \Phi(\bm{y})] &= [\Phi^\dagger(\bm{x}), \Phi^\dagger(\bm{y})] = 0.
    \end{aligned}
\end{equation}
The two-point density correlations in the atomic gas can be computed from the expectation values
\begin{equation}
    \begin{aligned}
        \langle n(\bm{x}) n(\bm{y}) \rangle_c &=  \langle n(\bm{x}) n(\bm{y}) \rangle - \langle n(\bm{x}) \rangle \langle n(\bm{y}) \rangle, \\
        \langle n(\bm{x}) n(\bm{y}) \rangle &= \langle \Phi^\dagger(\bm{x}) \Phi^\dagger(\bm{y}) \Phi(\bm{y}) \Phi(\bm{x}) \rangle, \\
        \langle n(\bm{x}) \rangle &= \langle \Phi^\dagger(\bm{x}) \Phi(\bm{x}) \rangle,
    \end{aligned}
\end{equation}
which are taken in the state that is annihilated by $\hat a_k$ from \cref{eq:NonRelModeExpansion}.
For the connected density two-point function, one finds \cite{LandauLifshitz1980}
\begin{equation}
    \langle n(\bm{x}) n(\bm{y}) \rangle_c = \sum_{\bm{k},\bm{k'}} \langle \hat a_{\bm{k}}^\dagger \hat a_{\bm{k'}} \hat a_{\bm{k'}}^\dagger a_{\bm{k}} \rangle \phi^*_{\bm{k}}(\bm{x}) \phi_{\bm{k'}}(\bm{x}) \phi^*_{\bm{k'}}(\bm{y}) \phi_{\bm{k}}(\bm{y}),
\end{equation} 
where one has
\begin{equation}
    \hat a_{\bm{k}}^\dagger \hat a_{\bm{k}} = \hat n_{\bm{k}}, \quad \hat a_{\bm{k}}\hat a_{\bm{k}}^\dagger = \hat n_{\bm{k}} + 1,
\end{equation}
in terms of the number-density operator $\hat n_{\bm{k}}$. 
Assuming that the system is in the non-condensed phase enables us to replace $ \, \sum_{\bm{k}} \to V \int_{\bm{k}}$ for a sufficiently large volume $V$ (the condensed case is for example discussed in \cite{Ziff1977,Naraschewski1999}).
We then find \cite{LandauLifshitz1980}
\begin{equation}
    \langle n(\bm{x}) n(\bm{y}) \rangle_c = \frac{N}{V} \delta(\bm{x}-\bm{y}) + \bigg \lvert \int_{\bm{k}} \mathrm{e}^{\mathrm{i} \bm{k} \cdot (\bm{x}-\bm{y})}\langle \hat n_{\bm{k}} \rangle \bigg \rvert^2.
\end{equation}
where we set $\langle n(\bm{x}) \rangle = N/V$.
Taking a Fourier-transform we find
\begin{equation}
    \langle n(\bm{k}) n(\bm{k'}) \rangle_c = \frac{N}{V} \delta(\bm{k} + \bm{k'}) + \int_{\bm{q}} \langle \hat n_{\bm{q}} \rangle \langle \hat n_{\bm{k+q}} \rangle \delta(\bm{k} + \bm{k'}),
\end{equation}
where the first term represents shot-noise and the second term autocorrelations.
If the Bose gas is in thermal equilibrium, $\langle \hat n_{\bm{q+k}} \rangle$ is the Bose-Einstein distribution which decays exponentially as $\lvert \bm{k} \rvert \to \infty$, such that only the shot-noise-term remains in the ultraviolet limit.  

In this section, we focussed on the case of an ideal gas, which is appropriate in the ultraviolet limit.
The non-ideal case can be found for example in \cite{Pitaevskii2016,Naraschewski1999} and references therein. 

\section{Thermal structure factor}
\label{app:ThermalStructureFactor}
An introduction to the spectral methods employed in this section can be found in chapter IV of \cite{LifshitzPitaevskii2013} where also the following calculation is carried out in a similar manner. 
The connection between the full quantum field of the condensate $\Phi$ and the full condensate density $n$ is given by the relation
\begin{equation}
\begin{split}
n (t,u,\varphi) &= \abs{\Phi(t,u,\varphi)}^2\\
&= n_0(u)+\sqrt{2n_0(u)}\phi_1(t,u,\varphi) \\
&+ \frac{1}{2} \phi_1^2(t,u,\varphi) + \frac{1}{2} \phi_2^2(t,u,\varphi),
\label{eq:FullDensity}
\end{split}
\end{equation}
where we considered the background field ${\phi_0\equiv\sqrt{n_0}}$, $\phi_1$ and $\phi_2$ to be real after some local $U(1)$ gauge transformation. 
When we take the expectation value of the full condensate density $\langle n(t,u,\varphi) \rangle$, the second term in Eq.~\eqref{eq:FullDensity} drops out, because it is linear in fluctuations.

Let us now determine the spectral density $\Delta_{ab}^\rho (\hbar\omega,\bm{k})$ in the present situation for the fields $\phi_1$ and $\phi_2$. In momentum space the quadratic action becomes
\begin{align}
\begin{split}
&\Gamma [\phi_1, \phi_2] \\
&=\int \frac{\text{d} \omega }{2\pi}\, \frac{\text{d}^2 k}{(2\pi)^2} \Bigg\{ -\frac{1}{2} 
\begin{pmatrix}
\tilde \phi_1 & \tilde \phi_2
\end{pmatrix}
\tilde G^{-1}(\omega,k)
\begin{pmatrix}
\tilde \phi_1 \\
\tilde \phi_2
\end{pmatrix}\Bigg\},
\end{split}
\end{align}
with the inverse propagator
\begin{equation}
\tilde{G}^{-1}_{ab}(\hbar \omega,\bm{k}) =\begin{pmatrix}
\frac{\hbar^2 k^2}{2m} + 2\lambda n_0 & -i\hbar\omega \\
i\hbar\omega  & \frac{\hbar^2k^2}{2m} 
\end{pmatrix}
\end{equation}
such that 
\begin{equation}
\tilde{G}_{ab}(\hbar \omega,\bm{k}) = \frac{1}{-\hbar^2\omega^2 + \hbar^2 \omega_\mathrm{B}^2(k)}
\begin{pmatrix}
\frac{\hbar^2k^2}{2m}& i\hbar\omega \\
-i\hbar\omega  & \frac{\hbar^2k^2}{2m}  + 2\lambda n_0  
\end{pmatrix},
\end{equation}
where we introduced the celebrated Bogoliubov dispersion relation
\begin{align}
    \hbar\omega_\mathrm{B}(k)=\pm\sqrt{\left(\frac{\hbar^2k^2}{2m}+2\lambda n_0\right)\frac{\hbar^2k^2}{2m}}.
\label{eq:BogoliubovDispersionRelation}
\end{align}
as the pole of the propagator,
\begin{align}
    \det \tilde{G}^{-1}(\hbar \omega_\mathrm{B},\bm{k})=0.
\end{align}
The spectral function follows from this as
\begin{equation}
\begin{split}
    &\Delta^\rho_{ab} (\hbar \omega, \bm{k}) = -i \tilde{G}_{ab} (\hbar \omega +i\epsilon, \bm{k}) + i \tilde{G}_{ab} (\hbar \omega -i\epsilon, \bm{k}).
\end{split}
\end{equation}
To evaluate this one needs the identity
\begin{equation}
    \frac{1}{x\pm i \epsilon} = \mp i\pi \delta(x) + \text{P.V.} \left( \frac{1}{x} \right)
\end{equation}
and $-\left(\hbar\omega\pm i \epsilon\right)^2 = -\omega^2 \mp i\epsilon \text{sign}(\omega)$ which results in the spectral function
\begin{equation}
\begin{aligned}
            \Delta^\rho_{ab} (\hbar \omega, \bm{k}) &= 2\pi \delta\left(\hbar^2\omega^2 - \hbar^2 \omega_\mathrm{B}^2\right) \text{sign}(\omega) \\& \quad \times\begin{pmatrix}
\frac{\hbar^2k^2}{2m}& i\hbar\omega \\
-i\hbar\omega  & \frac{\hbar^2k^2}{2m}  + 2\lambda n_0  
\end{pmatrix} .
    \end{aligned}
\end{equation}
One can now determine the static structure factor as
\begin{equation}
\begin{split}
    S_{ab}(\bm{k}) =& \hbar\int_{-\infty}^\infty \frac{\text{d}\omega}{2\pi} \left[\frac{1}{2} + n_B(\hbar\omega)\right] \Delta^\rho_{ab} (\hbar\omega, \bm{k})\\
    =&\frac{\hbar}{2}\int_{-\infty}^\infty \text{d}\omega \coth(\abs{\hbar\omega}/2) \delta\left(\hbar^2\omega^2 - \hbar^2 \omega_\mathrm{B}^2(k) \right) \\
    &\times \text{sign}(\hbar\omega) \begin{pmatrix}
\frac{\hbar^2k^2}{2m}& i\hbar\omega \\
-i\hbar\omega  & \frac{\hbar^2k^2}{2m}  + 2\lambda n_0  
\end{pmatrix} ,
\end{split}
\end{equation}
where we used
\begin{equation}
    \begin{aligned}
        \frac{1}{2} + n_B(\hbar\omega) &= \left[ \frac{1}{2} + n_B(\abs{\hbar\omega}) \right] \text{sign}(\hbar\omega) \\
        &= \frac{1}{2} \coth\left(\frac{\abs{\hbar\omega}}{2}\right) \, \text{sign}(\hbar\omega)
    \end{aligned}
\end{equation}
One further observes that the off-diagonal terms $i\hbar\omega$ do not contribute to the static structure factor, because they are odd with respect to $\hbar\omega\rightarrow -\hbar\omega$ while the rest of the integral is even. Otherwise one can perform the frequency integration using the Dirac deltas with zero-crossings are $\hbar\omega=\pm \epsilon_k$, which gives
\begin{equation}
    S_{ab}(\bm{k}) = \frac{1}{2} \coth\left(\frac{\abs{\hbar\omega}}{2}\right) 
    \begin{pmatrix}
    E_k/\epsilon_k & 0 \\
    0 & \epsilon_k/E_k  
    \end{pmatrix},
\end{equation}
with 
\begin{equation}
    \epsilon_k=\sqrt{\frac{\hbar^2\bm{k}^2}{2m}\left(\frac{\hbar^2\bm{k}^2}{2m} + 2\lambda n_0\right)}, \quad E_k = \frac{\hbar^2\bm{k}^2}{2m}.
\end{equation}
In terms of the healing length $\xi = \hbar/\sqrt{2m\lambda n_0}$ and the speed of sound $c=\sqrt{\lambda n_0/m}$ one can introduce the phase velocity 
$c_k = c \sqrt{1+\xi^2 \abs{\bm{k}}^2 /2}$, such that the Bogoliubov energy can be written as
\begin{equation}
    \epsilon_k= \hbar c_k \lvert \bm{k} \rvert  = \hbar c \abs{\bm{k}} \sqrt{1+\xi^2 \abs{\bm{k}} ^2 /2},
\end{equation}
and also
\begin{equation}
    \frac{E_k}{\epsilon_k} = \frac{\xi k}{\sqrt{2}} \frac{1}{\sqrt{1 + \tfrac{1}{2} k^2 \xi^2}}, \quad \frac{\epsilon_k}{E_k} = \frac{\sqrt{2}}{k\xi} \sqrt{1 + \tfrac{1}{2} k^2 \xi^2}.
    \label{eq:EnergieRatioIdentities}
\end{equation}
The components of the matrix are identical to expressions obtained from invoking a Bogoliubov transformation between the atom and the phonon basis as one can see from \cref{eq:VacuumDensityCorrelator,eq:VacuumPhaseCorrelator,eq:SqueezingFactors}. 

The leading terms in the acoustic approximations, as well as their corrections are now clear to see. 
In particular, we find for the Fourier transform of the density correlation function
\begin{equation}
\begin{aligned}
        S_{nn}(\bm{k}) &= 2 n_0 S_{11}(\bm{k}) \\ &= \sqrt{2} n_0 \left[\frac{1}{2} + n_B(E_{\bm{k}}) \right]  \frac{\xi \abs{\bm{k}}}{\sqrt{1+\frac{\xi^2 \abs{\bm{k}}^2}{ 2}}}.
\end{aligned}
\end{equation}

\bibliography{main}

@article{Sanchez2022,
  title = {Scalar quantum fields in cosmologies with $2+1$ spacetime dimensions},
  author = {S\'anchez-Kuntz, Natalia and Parra-L\'opez, \'Alvaro and Tolosa-Sime\'on, Mireia and Haas, Tobias and Floerchinger, Stefan},
  journal = {Phys. Rev. D},
  volume = {105},
  issue = {10},
  pages = {105020},
  numpages = {18},
  year = {2022},
  month = {05},
  publisher = {American Physical Society},
  doi = {10.1103/PhysRevD.105.105020},
  url = {https://link.aps.org/doi/10.1103/PhysRevD.105.105020}
}

@book{Lifshitz1980,
    author = {E. M. Lifshitz and Lev Pitaevskii},
    title = {{Statistical Physics Part 2}},
    year = {1980},
    publisher = {Pergamon Press},
    address = {Oxford},
    url = {https://www.elsevier.com/books/statistical-physics/lifshitz/978-0-08-050350-9}
}

@article{Madelung1927,
    Author = {E. Madelung},
    Title = {{Quantentheorie in hydrodynamischer Form}},
    Journal = {Z. Phys.},
    Year = {1927},
    Volume = {40},
    Number = {3},
    Pages = {322-326},
    url = {https://neo-classical-physics.info/uploads/3/4/3/6/34363841/madelung_-_hydrodynamical_interp..pdf}
}

@article{Tolosa2022,
  title = {Curved and expanding spacetime geometries in Bose-Einstein condensates},
  author = {Tolosa-Sime\'on, Mireia and Parra-L\'opez, \'Alvaro and S\'anchez-Kuntz, Natalia and Haas, Tobias and Viermann, Celia and Sparn, Marius and Liebster, Nikolas and Hans, Maurus and Kath, Elinor and Strobel, Helmut and Oberthaler, Markus K. and Floerchinger, Stefan},
  journal = {Phys. Rev. A},
  volume = {106},
  issue = {3},
  pages = {033313},
  numpages = {18},
  year = {2022},
  month = {05},
  publisher = {American Physical Society},
  doi = {10.1103/PhysRevA.106.033313},
  url = {https://link.aps.org/doi/10.1103/PhysRevA.106.033313}
}

@article{Viermann2022,
	author = {Viermann, Celia and Sparn, Marius and Liebster, Nikolas and Hans, Maurus and Kath, Elinor and Parra-L{\'o}pez, {\'A}lvaro and Tolosa-Sime{\'o}n, Mireia and S{\'a}nchez-Kuntz, Natalia and Haas, Tobias and Strobel, Helmut and Floerchinger, Stefan and Oberthaler, Markus K.},
	doi = {10.1038/s41586-022-05313-9},
    date = {2022-11-01},
	id = {Viermann2022},
	isbn = {1476-4687},
	journal = {Nature},
	number = {7935},
	pages = {260--264},
	title = {Quantum field simulator for dynamics in curved spacetime},
	url = {https://doi.org/10.1038/s41586-022-05313-9},
	volume = {611},
	year = {2022}}

@inbook{Brandenberger2000,
	address = {Dordrecht},
	author = {Brandenberger, Robert H.},
	editor = {Mansouri, Reza and Brandenberger, Robert},
	pages = {169--211},
	publisher = {Springer Netherlands},
	title = {Inflationary Cosmology: Progress and Problems},
	year = {2000}}

@article{MartinBrandenberger2001,
  title = {Trans-Planckian problem of inflationary cosmology},
  author = {Martin, J\'er\^ome and Brandenberger, Robert H.},
  journal = {Phys. Rev. D},
  volume = {63},
  issue = {12},
  pages = {123501},
  numpages = {16},
  year = {2001},
  month = {May},
  publisher = {American Physical Society},
  doi = {10.1103/PhysRevD.63.123501},
  url = {https://link.aps.org/doi/10.1103/PhysRevD.63.123501}
}

@article{Hassan2003,
  title = {Trans-Planckian effects in inflationary cosmology and the modified uncertainty principle},
  volume = {674},
  ISSN = {0550-3213},
  url = {http://dx.doi.org/10.1016/j.nuclphysb.2003.09.041},
  DOI = {10.1016/j.nuclphysb.2003.09.041},
  number = {1–2},
  journal = {Nuclear Physics B},
  publisher = {Elsevier BV},
  author = {Hassan,  S.F. and Sloth,  Martin S.},
  year = {2003},
  month = dec,
  pages = {434–458}
}

@article{Weinfurtner2009,
	author = {Silke Weinfurtner and Piyush Jain and Matt Visser and C W Gardiner},
	doi = {10.1088/0264-9381/26/6/065012},
	journal = {Classical and Quantum Gravity},
	month = {feb},
	number = {6},
	pages = {065012},
	title = {Cosmological particle production in emergent rainbow spacetimes},
	url = {https://dx.doi.org/10.1088/0264-9381/26/6/065012},
	volume = {26},
	year = {2009}}

@article{JacobsonMattingly2001,
  title = {Gravity with a dynamical preferred frame},
  author = {Jacobson, Ted and Mattingly, David},
  journal = {Phys. Rev. D},
  volume = {64},
  issue = {2},
  pages = {024028},
  numpages = {9},
  year = {2001},
  month = {Jun},
  publisher = {American Physical Society},
  doi = {10.1103/PhysRevD.64.024028},
  url = {https://link.aps.org/doi/10.1103/PhysRevD.64.024028}
}

@article{NiemeyerParentaniCampo2002,
  title = {Minimal modifications of the primordial power spectrum from an adiabatic short distance cutoff},
  author = {Niemeyer, Jens C. and Parentani, Renaud and Campo, David},
  journal = {Phys. Rev. D},
  volume = {66},
  issue = {8},
  pages = {083510},
  numpages = {5},
  year = {2002},
  month = {Oct},
  publisher = {American Physical Society},
  doi = {10.1103/PhysRevD.66.083510},
  url = {https://link.aps.org/doi/10.1103/PhysRevD.66.083510}
}

@inproceedings{MartinBrandenberger2000,
    author = "Martin, Jerome and Brandenberger, Robert H.",
    title = "{A Cosmological window on transPlanckian physics}",
    booktitle = "{9th Marcel Grossmann Meeting on Recent Developments in Theoretical and Experimental General Relativity, Gravitation and Relativistic Field Theories (MG 9)}",
    eprint = "astro-ph/0012031",
    archivePrefix = "arXiv",
    pages = "2001--2002",
    month = "7",
    year = "2000"
}

@article{Hung2013,
  title = {From Cosmology to Cold Atoms: Observation of Sakharov Oscillations in a Quenched Atomic Superfluid},
  volume = {341},
  ISSN = {1095-9203},
  url = {http://dx.doi.org/10.1126/science.1237557},
  DOI = {10.1126/science.1237557},
  number = {6151},
  journal = {Science},
  publisher = {American Association for the Advancement of Science (AAAS)},
  author = {Hung,  Chen-Lung and Gurarie,  Victor and Chin,  Cheng},
  year = {2013},
  month = sep,
  pages = {1213–1215}
}

@article{Chen2021,
	author = {Chen, Cheng-An and Khlebnikov, Sergei and Hung, Chen-Lung},
	doi = {10.1103/PhysRevLett.127.060404},
	issue = {6},
	journal = {Phys. Rev. Lett.},
	pages = {060404},
	title = {{Observation of quasiparticle pair production and quantum entanglement in atomic quantum gases quenched to an attractive interaction}},
	url = {https://link.aps.org/doi/10.1103/PhysRevLett.127.060404},
	volume = {127},
	year = {2021}}

@article{MukhanovChibisov1981,
    author = "Mukhanov, Viatcheslav F. and Chibisov, G. V.",
    title = "{Quantum Fluctuations and a Nonsingular Universe}",
    journal = "JETP Lett.",
    volume = "33",
    pages = "532--535",
    year = "1981"
}

@article{Starobinsky1979,
    author = "Starobinsky, Alexei A.",
    editor = "Khalatnikov, I. M. and Mineev, V. P.",
    title = "{Spectrum of relict gravitational radiation and the early state of the universe}",
    journal = "JETP Lett.",
    volume = "30",
    pages = "682--685",
    year = "1979"
}

@article{Planck2018,
    author = "Akrami, Y. and others",
    collaboration = "Planck",
    title = "{Planck 2018 results. X. Constraints on inflation}",
    eprint = "1807.06211",
    archivePrefix = "arXiv",
    primaryClass = "astro-ph.CO",
    doi = "10.1051/0004-6361/201833887",
    journal = "Astron. Astrophys.",
    volume = "641",
    pages = "A10",
    year = "2020"
}

@article{ChaFischer2017,
  title = {Probing the Scale Invariance of the Inflationary Power Spectrum in Expanding Quasi-Two-Dimensional Dipolar Condensates},
  author = {Ch\"a, Seok-Yeong and Fischer, Uwe R.},
  journal = {Phys. Rev. Lett.},
  volume = {118},
  issue = {13},
  pages = {130404},
  numpages = {6},
  year = {2017},
  month = {Mar},
  publisher = {American Physical Society},
  doi = {10.1103/PhysRevLett.118.130404},
  url = {https://link.aps.org/doi/10.1103/PhysRevLett.118.130404}
}

@article{RibeiroFischer2023,
  title = {Impact of trans-Planckian excitations on black-hole radiation in dipolar condensates},
  author = {Holanda Ribeiro, Caio C. and Fischer, Uwe R.},
  journal = {Phys. Rev. D},
  volume = {107},
  issue = {12},
  pages = {L121502},
  numpages = {6},
  year = {2023},
  month = {Jun},
  publisher = {American Physical Society},
  doi = {10.1103/PhysRevD.107.L121502},
  url = {https://link.aps.org/doi/10.1103/PhysRevD.107.L121502}
}

@article{MukhanovFeldmanBrandenberger1992,
	author = {V.F. Mukhanov and H.A. Feldman and R.H. Brandenberger},
	doi = {https://doi.org/10.1016/0370-1573(92)90044-Z},
	issn = {0370-1573},
	journal = {Physics Reports},
	number = {5},
	pages = {203-333},
	title = {Theory of cosmological perturbations},
	url = {https://www.sciencedirect.com/science/article/pii/037015739290044Z},
	volume = {215},
	year = {1992}}

@article{BrandenbergerJorasMartin2002,
  title = {Trans-Planckian physics and the spectrum of fluctuations in a bouncing universe},
  author = {Brandenberger, Robert H. and Jor\'as, Sergio E. and Martin, J\'er\^ome},
  journal = {Phys. Rev. D},
  volume = {66},
  issue = {8},
  pages = {083514},
  numpages = {9},
  year = {2002},
  month = {Oct},
  publisher = {American Physical Society},
  doi = {10.1103/PhysRevD.66.083514},
  url = {https://link.aps.org/doi/10.1103/PhysRevD.66.083514}
}

@article{Starobinsky2001,
  title = {Robustness of the inflationary perturbation spectrum to trans-Planckian physics},
  volume = {73},
  ISSN = {1090-6487},
  url = {http://dx.doi.org/10.1134/1.1381588},
  DOI = {10.1134/1.1381588},
  number = {8},
  journal = {Journal of Experimental and Theoretical Physics Letters},
  publisher = {Pleiades Publishing Ltd},
  author = {Starobinsky,  A. A.},
  year = {2001},
  month = apr,
  pages = {371–374}
}

@article{CorleyJacobson1996,
  title = {Hawking spectrum and high frequency dispersion},
  author = {Corley, Steven and Jacobson, Ted},
  journal = {Phys. Rev. D},
  volume = {54},
  issue = {2},
  pages = {1568--1586},
  numpages = {0},
  year = {1996},
  month = {Jul},
  publisher = {American Physical Society},
  doi = {10.1103/PhysRevD.54.1568},
  url = {https://link.aps.org/doi/10.1103/PhysRevD.54.1568}
}

@book{Volovik2009,
	title = {The universe in a helium droplet},
	isbn = {978-0-19-956484-2},
	url = {https://doi.org/10.1093/acprof:oso/9780199564842.001.0001},
	publisher = {Oxford University Press},
	author = {Volovik, Grigory E.},
	month = feb,
	year = {2009},
	doi = {10.1093/acprof:oso/9780199564842.001.0001},
}

@article{Liberati2013,
	author = {Liberati, S},
	doi = {10.1088/0264-9381/30/13/133001},
	journal = {Classical and Quantum Gravity},
	month = {jun},
	number = {13},
	pages = {133001},
	publisher = {IOP Publishing},
	title = {Tests of Lorentz invariance: a 2013 update},
	url = {https://doi.org/10.1088/0264-9381/30/13/133001},
	volume = {30},
	year = {2013}}

@article{MacherParentani2009a,
  title = {Black-hole radiation in Bose-Einstein condensates},
  author = {Macher, Jean and Parentani, Renaud},
  journal = {Phys. Rev. A},
  volume = {80},
  issue = {4},
  pages = {043601},
  numpages = {26},
  year = {2009},
  month = {Oct},
  publisher = {American Physical Society},
  doi = {10.1103/PhysRevA.80.043601},
  url = {https://link.aps.org/doi/10.1103/PhysRevA.80.043601}
}

@article{MacherParentani2009b,
  title = {Black/white hole radiation from dispersive theories},
  author = {Macher, Jean and Parentani, Renaud},
  journal = {Phys. Rev. D},
  volume = {79},
  issue = {12},
  pages = {124008},
  numpages = {23},
  year = {2009},
  month = {Jun},
  publisher = {American Physical Society},
  doi = {10.1103/PhysRevD.79.124008},
  url = {https://link.aps.org/doi/10.1103/PhysRevD.79.124008}
}

@article{Volovik2009HLG,
  title = {z = 3 Lifshitz-Hořava model and fermi-point scenario of emergent gravity},
  volume = {89},
  ISSN = {1090-6487},
  url = {http://dx.doi.org/10.1134/S0021364009110010},
  DOI = {10.1134/s0021364009110010},
  number = {11},
  journal = {JETP Letters},
  publisher = {Pleiades Publishing Ltd},
  author = {Volovik,  G. E.},
  year = {2009},
  month = aug,
  pages = {525–528}
}

@article{Steinhauer2016,
	author = {Jeff Steinhauer},
	doi = {10.1038/nphys3863},
	journal = {Nat. Phys.},
	number = {10},
	pages = {959-965},
	title = {{Observation of quantum Hawking radiation and its entanglement in an analogue black hole}},
	url = {https://doi.org/10.1038/nphys3863},
	volume = {12},
	year = {2016}}

@article{Torres2017,
  title = {Rotational superradiant scattering in a vortex flow},
  volume = {13},
  ISSN = {1745-2481},
  url = {http://dx.doi.org/10.1038/nphys4151},
  DOI = {10.1038/nphys4151},
  number = {9},
  journal = {Nature Physics},
  publisher = {Springer Science and Business Media LLC},
  author = {Torres,  Theo and Patrick,  Sam and Coutant,  Antonin and Richartz,  Maurício and Tedford,  Edmund W. and Weinfurtner,  Silke},
  year = {2017},
  month = jun,
  pages = {833–836}
}

@article{Jaskula2012,
	author = {J.-C. Jaskula and G. B. Partridge and M. Bonneau and R. Lopes and J. Ruaudel and D. Boiron and C. I. Westbrook},
	doi = {10.1103/PhysRevLett.109.220401},
	issue = {22},
	journal = {Phys. Rev. Lett.},
	pages = {220401},
	title = {{Acoustic analog to the dynamical Casimir effect in a Bose-Einstein condensate}},
	url = {https://link.aps.org/doi/10.1103/PhysRevLett.109.220401},
	volume = {109},
	year = {2012}}

@article{AmelinoCamelia2013,
  title = {Quantum-Spacetime Phenomenology},
  volume = {16},
  ISSN = {1433-8351},
  url = {http://dx.doi.org/10.12942/lrr-2013-5},
  DOI = {10.12942/lrr-2013-5},
  number = {1},
  journal = {Living Reviews in Relativity},
  publisher = {Springer Science and Business Media LLC},
  author = {Amelino-Camelia,  Giovanni},
  year = {2013},
  month = jun 
}

@article{Moore1970,
  title = {Quantum Theory of the Electromagnetic Field in a Variable-Length One-Dimensional Cavity},
  volume = {11},
  ISSN = {1089-7658},
  url = {http://dx.doi.org/10.1063/1.1665432},
  DOI = {10.1063/1.1665432},
  number = {9},
  journal = {Journal of Mathematical Physics},
  publisher = {AIP Publishing},
  author = {Moore,  Gerald T.},
  year = {1970},
  month = sep,
  pages = {2679–2691}
}

@article{FullingDavies1976,
  volume = {348},
  ISSN = {0080-4630},
  url = {http://dx.doi.org/10.1098/rspa.1976.0045},
  DOI = {10.1098/rspa.1976.0045},
  number = {1654},
  journal = {Proceedings of the Royal Society of London. A. Mathematical and Physical Sciences},
  publisher = {The Royal Society},
  year = {1976},
  month = mar,
  pages = {393–414}
}

@article{Braidotti2022,
  title = {Measurement of Penrose Superradiance in a Photon Superfluid},
  author = {Braidotti, Maria Chiara and Prizia, Radivoje and Maitland, Calum and Marino, Francesco and Prain, Angus and Starshynov, Ilya and Westerberg, Niclas and Wright, Ewan M. and Faccio, Daniele},
  journal = {Phys. Rev. Lett.},
  volume = {128},
  issue = {1},
  pages = {013901},
  numpages = {6},
  year = {2022},
  month = {Jan},
  publisher = {American Physical Society},
  doi = {10.1103/PhysRevLett.128.013901},
  url = {https://link.aps.org/doi/10.1103/PhysRevLett.128.013901}
}

@article{Belgiorno2010,
  title = {Hawking Radiation from Ultrashort Laser Pulse Filaments},
  author = {Belgiorno, F. and Cacciatori, S. L. and Clerici, M. and Gorini, V. and Ortenzi, G. and Rizzi, L. and Rubino, E. and Sala, V. G. and Faccio, D.},
  journal = {Phys. Rev. Lett.},
  volume = {105},
  issue = {20},
  pages = {203901},
  numpages = {4},
  year = {2010},
  month = {Nov},
  publisher = {American Physical Society},
  doi = {10.1103/PhysRevLett.105.203901},
  url = {https://link.aps.org/doi/10.1103/PhysRevLett.105.203901}
}

@article{Steinhauer2022,
  title = {Analogue cosmological particle creation in an ultracold quantum fluid of light},
  volume = {13},
  ISSN = {2041-1723},
  url = {http://dx.doi.org/10.1038/s41467-022-30603-1},
  DOI = {10.1038/s41467-022-30603-1},
  number = {1},
  journal = {Nature Communications},
  publisher = {Springer Science and Business Media LLC},
  author = {Steinhauer,  Jeff and Abuzarli,  Murad and Aladjidi,  Tangui and Bienaimé,  Tom and Piekarski,  Clara and Liu,  Wei and Giacobino,  Elisabeth and Bramati,  Alberto and Glorieux,  Quentin},
  year = {2022},
  month = may 
}

@article{Wittemer2019,
	author = {Matthias Wittemer and Frederick Hakelberg and Philip Kiefer and Jan-Philipp Schr\"oder and Christian Fey and Ralf Sch\"utzhold and Ulrich Warring and Tobias Schaetz},
	doi = {10.1103/PhysRevLett.123.180502},
	issue = {18},
	journal = {Phys. Rev. Lett.},
	pages = {180502},
	title = {{Phonon pair creation by inflating quantum fluctuations in an ion trap}},
	url = {https://link.aps.org/doi/10.1103/PhysRevLett.123.180502},
	volume = {123},
	year = {2019}}

@article{Jacquet2020,
  title = {Polariton fluids for analogue gravity physics},
  volume = {378},
  ISSN = {1471-2962},
  url = {http://dx.doi.org/10.1098/rsta.2019.0225},
  DOI = {10.1098/rsta.2019.0225},
  number = {2177},
  journal = {Philosophical Transactions of the Royal Society A: Mathematical,  Physical and Engineering Sciences},
  publisher = {The Royal Society},
  author = {Jacquet,  M. J. and Boulier,  T. and Claude,  F. and Maître,  A. and Cancellieri,  E. and Adrados,  C. and Amo,  A. and Pigeon,  S. and Glorieux,  Q. and Bramati,  A. and Giacobino,  E.},
  year = {2020},
  month = jul,
  pages = {20190225}
}

@article{Sotiriou2011,
	author = {Sotiriou, Thomas P},
	doi = {10.1088/1742-6596/283/1/012034},
	journal = {Journal of Physics: Conference Series},
	month = {feb},
	number = {1},
	pages = {012034},
	title = {Ho{\v r}ava-Lifshitz gravity: a status report},
	url = {https://doi.org/10.1088/1742-6596/283/1/012034},
	volume = {283},
	year = {2011}}

@article{BarceloGarayJannes2009,
  title = {Sensitivity of Hawking radiation to superluminal dispersion relations},
  author = {Barcel\'o, C. and Garay, L. J. and Jannes, G.},
  journal = {Phys. Rev. D},
  volume = {79},
  issue = {2},
  pages = {024016},
  numpages = {13},
  year = {2009},
  month = {Jan},
  publisher = {American Physical Society},
  doi = {10.1103/PhysRevD.79.024016},
  url = {https://link.aps.org/doi/10.1103/PhysRevD.79.024016}
}

@article{DelPorro2023,
  title = {Hawking radiation in Lorentz violating gravity: a tale of two horizons},
  volume = {2023},
  ISSN = {1029-8479},
  url = {http://dx.doi.org/10.1007/JHEP12(2023)094},
  DOI = {10.1007/jhep12(2023)094},
  number = {12},
  journal = {Journal of High Energy Physics},
  publisher = {Springer Science and Business Media LLC},
  author = {Del Porro,  F. and Herrero-Valea,  M. and Liberati,  S. and Schneider,  M.},
  year = {2023},
  month = dec 
}

@article{Girelli2007b,
   title={Planck-scale modified dispersion relations and Finsler geometry},
   volume={75},
   ISSN={1550-2368},
   url={http://dx.doi.org/10.1103/PhysRevD.75.064015},
   DOI={10.1103/physrevd.75.064015},
   number={6},
   journal={Physical Review D},
   publisher={American Physical Society (APS)},
   author={Girelli, F. and Liberati, S. and Sindoni, L.},
   year={2007},
   month=mar }

@inbook{JacobsonLiberatiMattingly2005b,
   title={Quantum Gravity Phenomenology and Lorentz Violation},
   ISBN={3540228039},
   url={http://dx.doi.org/10.1007/3-540-26798-0_8},
   DOI={10.1007/3-540-26798-0_8},
   booktitle={Particle Physics and the Universe},
   publisher={Springer-Verlag},
   author={Jacobson, Ted and Liberati, Stefano and Mattingly, David},
   pages={83–98}}

@article{CoutantWeinfurtner2017,
  title={Low-frequency analogue Hawking radiation: The Bogoliubov-de Gennes model},
  author={Antonin Coutant and Silke E. Ch. Weinfurtner},
  journal={Physical Review D},
  year={2017},
  volume={97},
  pages={025006},
  url={https://api.semanticscholar.org/CorpusID:55216581}
}

@article{Visser2009,
  title = {Lorentz symmetry breaking as a quantum field theory regulator},
  author = {Visser, Matt},
  journal = {Phys. Rev. D},
  volume = {80},
  issue = {2},
  pages = {025011},
  numpages = {6},
  year = {2009},
  month = {Jul},
  publisher = {American Physical Society},
  doi = {10.1103/PhysRevD.80.025011},
  url = {https://link.aps.org/doi/10.1103/PhysRevD.80.025011}
}

@article{SotiriouVisserWeinfurtner2009,
  title = {Phenomenologically Viable Lorentz-Violating Quantum Gravity},
  author = {Sotiriou, Thomas P. and Visser, Matt and Weinfurtner, Silke},
  journal = {Phys. Rev. Lett.},
  volume = {102},
  issue = {25},
  pages = {251601},
  numpages = {4},
  year = {2009},
  month = {Jun},
  publisher = {American Physical Society},
  doi = {10.1103/PhysRevLett.102.251601},
  url = {https://link.aps.org/doi/10.1103/PhysRevLett.102.251601}
}

@article{Weinfurtner2007,
title = {Trans-Planckian physics and signature change events in Bose gas hydrodynamics},
author = {Weinfurtner, Silke and White, Angela and Visser, Matt},
journal = {Phys. Rev. D},
volume = {76},
issue = {12},
pages = {124008},
numpages = {19},
year = {2007},
month = {Dec},
publisher = {American Physical Society},
doi = {10.1103/PhysRevD.76.124008},
url = {https://link.aps.org/doi/10.1103/PhysRevD.76.124008}
}

@article{Schmidt2024,
  title = {Cosmological particle production in a quantum field simulator as a quantum mechanical scattering problem},
  author = {Schmidt, Christian F. and Parra-L\'opez, \'Alvaro and Tolosa-Sime\'on, Mireia and Sparn, Marius and Kath, Elinor and Liebster, Nikolas and Duchene, Jelte and Strobel, Helmut and Oberthaler, Markus K. and Floerchinger, Stefan},
  journal = {Phys. Rev. D},
  volume = {110},
  issue = {12},
  pages = {123523},
  numpages = {34},
  year = {2024},
  month = {Dec},
  publisher = {American Physical Society},
  doi = {10.1103/PhysRevD.110.123523},
  url = {https://link.aps.org/doi/10.1103/PhysRevD.110.123523}
}

@article{Jacobson1991,
  title = {Black-hole evaporation and ultrashort distances},
  author = {Jacobson, Theodore},
  journal = {Phys. Rev. D},
  volume = {44},
  issue = {6},
  pages = {1731--1739},
  numpages = {0},
  year = {1991},
  month = {Sep},
  publisher = {American Physical Society},
  doi = {10.1103/PhysRevD.44.1731},
  url = {https://link.aps.org/doi/10.1103/PhysRevD.44.1731}
}

@article{Unruh1995,
  title = {Sonic analogue of black holes and the effects of high frequencies on black hole evaporation},
  author = {Unruh, W. G.},
  journal = {Phys. Rev. D},
  volume = {51},
  issue = {6},
  pages = {2827--2838},
  numpages = {0},
  year = {1995},
  month = {Mar},
  publisher = {American Physical Society},
  doi = {10.1103/PhysRevD.51.2827},
  url = {
  https://link.aps.org/doi/10.1103/PhysRevD.51.2827}
}

@article{Lobo2017,
  title = {Rainbows without unicorns: metric structures in theories with modified dispersion relations},
  volume = {77},
  ISSN = {1434-6052},
  url = {http://dx.doi.org/10.1140/epjc/s10052-017-5017-0},
  DOI = {10.1140/epjc/s10052-017-5017-0},
  number = {7},
  journal = {The European Physical Journal C},
  publisher = {Springer Science and Business Media LLC},
  author = {Lobo,  Iarley P. and Loret,  Niccoló and Nettel,  Francisco},
  year = {2017},
  month = jul 
}

@article{Jacobson1999,
   title={Trans-Planckian Redshifts and the Substance of the Space-Time River},
   volume={136},
   ISSN={0375-9687},
   url={http://dx.doi.org/10.1143/PTPS.136.1},
   DOI={10.1143/ptps.136.1},
   journal={Progress of Theoretical Physics Supplement},
   publisher={Oxford University Press (OUP)},
   author={Jacobson, Ted},
   year={1999},
   pages={1–17} }

@article{NiemeyerParentani2001,
  title = {Trans-Planckian dispersion and scale invariance of inflationary perturbations},
  author = {Niemeyer, Jens C. and Parentani, Renaud},
  journal = {Phys. Rev. D},
  volume = {64},
  issue = {10},
  pages = {101301},
  numpages = {4},
  year = {2001},
  month = {Oct},
  publisher = {American Physical Society},
  doi = {10.1103/PhysRevD.64.101301},
  url = {https://link.aps.org/doi/10.1103/PhysRevD.64.101301}
}

@book{MukhanovWinitzki2007,
	author = {Mukhanov, Viatcheslav and Winitzki, Sergei},
	doi = {10.1017/CBO9780511809149},
	place = {Cambridge},
	publisher = {Cambridge University Press},
	title = {Introduction to Quantum Effects in Gravity},
	year = {2007}}

@article{UnruhSchuetzhold2005,
  title = {Universality of the Hawking effect},
  author = {Unruh, William G. and Sch\"utzhold, Ralf},
  journal = {Phys. Rev. D},
  volume = {71},
  issue = {2},
  pages = {024028},
  numpages = {11},
  year = {2005},
  month = {Jan},
  publisher = {American Physical Society},
  doi = {10.1103/PhysRevD.71.024028},
  url = {https://link.aps.org/doi/10.1103/PhysRevD.71.024028}
}

@article{BrandenbergerMartin2002,
   title={ON SIGNATURES OF SHORT DISTANCE PHYSICS IN THE COSMIC MICROWAVE BACKGROUND},
   volume={17},
   ISSN={1793-656X},
   url={http://dx.doi.org/10.1142/S0217751X02010765},
   DOI={10.1142/s0217751x02010765},
   number={25},
   journal={International Journal of Modern Physics A},
   publisher={World Scientific Pub Co Pte Lt},
   author={Brandenberger, Robert H. and Martin, Jérome},
   year={2002},
   month=oct, pages={3663–3680} }

@article{MacherParentani2008,
  title = {Signatures of trans-Planckian dispersion in inflationary spectra},
  author = {Macher, Jean and Parentani, Renaud},
  journal = {Phys. Rev. D},
  volume = {78},
  issue = {4},
  pages = {043522},
  numpages = {15},
  year = {2008},
  month = {Aug},
  publisher = {American Physical Society},
  doi = {10.1103/PhysRevD.78.043522},
  url = {https://link.aps.org/doi/10.1103/PhysRevD.78.043522}
}

@article{CoutantParentani2014,
  title = {Hawking radiation with dispersion: The broadened horizon paradigm},
  author = {Coutant, Antonin and Parentani, Renaud},
  journal = {Phys. Rev. D},
  volume = {90},
  issue = {12},
  pages = {121501},
  numpages = {5},
  year = {2014},
  month = {Dec},
  publisher = {American Physical Society},
  doi = {10.1103/PhysRevD.90.121501},
  url = {https://link.aps.org/doi/10.1103/PhysRevD.90.121501}
}

@article{Cropp2014,
  title = {Ray tracing Einstein-\AE{}ther black holes: Universal versus Killing horizons},
  author = {Cropp, Bethan and Liberati, Stefano and Mohd, Arif and Visser, Matt},
  journal = {Phys. Rev. D},
  volume = {89},
  issue = {6},
  pages = {064061},
  numpages = {16},
  year = {2014},
  month = {Mar},
  publisher = {American Physical Society},
  doi = {10.1103/PhysRevD.89.064061},
  url = {https://link.aps.org/doi/10.1103/PhysRevD.89.064061}
}

@article{DelPorro2025,
   title={Tunneling method for Hawking quanta in analogue gravity},
   volume={25},
   ISSN={1878-1535},
   url={http://dx.doi.org/10.5802/crphys.239},
   DOI={10.5802/crphys.239},
   number={S2},
   journal={Comptes Rendus. Physique},
   publisher={Cellule MathDoc/Centre Mersenne},
   author={Del Porro, Francesco and Liberati, Stefano and Schneider, Marc},
   year={2025},
   month=feb, pages={1–27} }

@article{Corley1998,
  title = {Computing the spectrum of black hole radiation in the presence of high frequency dispersion: An analytical approach},
  author = {Corley, Steven},
  journal = {Phys. Rev. D},
  volume = {57},
  issue = {10},
  pages = {6280--6291},
  numpages = {0},
  year = {1998},
  month = {May},
  publisher = {American Physical Society},
  doi = {10.1103/PhysRevD.57.6280},
  url = {https://link.aps.org/doi/10.1103/PhysRevD.57.6280}
}

@article{Liberati2006,
  title = {Analogue quantum gravity phenomenology from a two-component Bose–Einstein condensate},
  volume = {23},
  ISSN = {1361-6382},
  url = {http://dx.doi.org/10.1088/0264-9381/23/9/023},
  DOI = {10.1088/0264-9381/23/9/023},
  number = {9},
  journal = {Classical and Quantum Gravity},
  publisher = {IOP Publishing},
  author = {Liberati,  Stefano and Visser,  Matt and Weinfurtner,  Silke},
  year = {2006},
  month = apr,
  pages = {3129–3154}
}

@article{Niemeyer2001,
  title = {Inflation with a Planck-scale frequency cutoff},
  author = {Niemeyer, Jens C.},
  journal = {Phys. Rev. D},
  volume = {63},
  issue = {12},
  pages = {123502},
  numpages = {7},
  year = {2001},
  month = {May},
  publisher = {American Physical Society},
  doi = {10.1103/PhysRevD.63.123502},
  url = {https://link.aps.org/doi/10.1103/PhysRevD.63.123502}
}

@article{KMM1995,
  title = {Hilbert space representation of the minimal length uncertainty relation},
  author = {Kempf, Achim and Mangano, Gianpiero and Mann, Robert B.},
  journal = {Phys. Rev. D},
  volume = {52},
  issue = {2},
  pages = {1108--1118},
  numpages = {0},
  year = {1995},
  month = {Jul},
  publisher = {American Physical Society},
  doi = {10.1103/PhysRevD.52.1108},
  url = {https://link.aps.org/doi/10.1103/PhysRevD.52.1108}
}

@article{Kempf2001,
  title = {Mode generating mechanism in inflation with a cutoff},
  author = {Kempf, Achim},
  journal = {Phys. Rev. D},
  volume = {63},
  issue = {8},
  pages = {083514},
  numpages = {5},
  year = {2001},
  month = {Mar},
  publisher = {American Physical Society},
  doi = {10.1103/PhysRevD.63.083514},
  url = {https://link.aps.org/doi/10.1103/PhysRevD.63.083514}
}

@article{KempfNiemeyer2001,
  title = {Perturbation spectrum in inflation with a cutoff},
  author = {Kempf, Achim and Niemeyer, Jens C.},
  journal = {Phys. Rev. D},
  volume = {64},
  issue = {10},
  pages = {103501},
  numpages = {6},
  year = {2001},
  month = {Oct},
  publisher = {American Physical Society},
  doi = {10.1103/PhysRevD.64.103501},
  url = {https://link.aps.org/doi/10.1103/PhysRevD.64.103501}
}

@article{Brout1999,
  title = {Minimal length uncertainty principle and the trans-Planckian problem of black hole physics},
  author = {Brout, R. and Gabriel, Cl. and Lubo, M. and Spindel, Ph.},
  journal = {Phys. Rev. D},
  volume = {59},
  issue = {4},
  pages = {044005},
  numpages = {6},
  year = {1999},
  month = {Jan},
  publisher = {American Physical Society},
  doi = {10.1103/PhysRevD.59.044005},
  url = {https://link.aps.org/doi/10.1103/PhysRevD.59.044005}
}

@article{Garay1995,
   title={QUANTUM GRAVITY AND MINIMUM LENGTH},
   volume={10},
   ISSN={1793-656X},
   url={http://dx.doi.org/10.1142/S0217751X95000085},
   DOI={10.1142/s0217751x95000085},
   number={02},
   journal={International Journal of Modern Physics A},
   publisher={World Scientific Pub Co Pte Lt},
   author={GARAY, LUIS J.},
   year={1995},
   month=jan, pages={145–165} }

@article{Parentani2002,
  title = {WHAT DID WE LEARN FROM STUDYING ACOUSTIC BLACK HOLES?},
  volume = {17},
  ISSN = {1793-656X},
  url = {http://dx.doi.org/10.1142/S0217751X02011679},
  DOI = {10.1142/s0217751x02011679},
  number = {20},
  journal = {International Journal of Modern Physics A},
  publisher = {World Scientific Pub Co Pte Lt},
  author = {Parentani,  Renaud},
  year = {2002},
  month = aug,
  pages = {2721–2725}
}

@article{Brout1995,
  title = {Hawking radiation without trans-Planckian frequencies},
  author = {Brout, R. and Massar, S. and Parentani, R. and Spindel, Ph.},
  journal = {Phys. Rev. D},
  volume = {52},
  issue = {8},
  pages = {4559--4568},
  numpages = {0},
  year = {1995},
  month = {Oct},
  publisher = {American Physical Society},
  doi = {10.1103/PhysRevD.52.4559},
  url = {https://link.aps.org/doi/10.1103/PhysRevD.52.4559}
}

@article{MassarParentani1997,
  title={Particle creation and non-adiabatic transitions in quantum cosmology},
  author={Serge Massar and Renaud Parentani},
  journal={Nuclear Physics},
  year={1997},
  volume={513},
  pages={375-401},
  url={https://api.semanticscholar.org/CorpusID:119428197}
}

@article{FinazziParentani2012,
  title = {Hawking radiation in dispersive theories, the two regimes},
  author = {Finazzi, Stefano and Parentani, Renaud},
  journal = {Phys. Rev. D},
  volume = {85},
  issue = {12},
  pages = {124027},
  numpages = {11},
  year = {2012},
  month = {Jun},
  publisher = {American Physical Society},
  doi = {10.1103/PhysRevD.85.124027},
  url = {https://link.aps.org/doi/10.1103/PhysRevD.85.124027}
}

@article{MacherParentani2009,
  title = {Black/white hole radiation from dispersive theories},
  author = {Macher, Jean and Parentani, Renaud},
  journal = {Phys. Rev. D},
  volume = {79},
  issue = {12},
  pages = {124008},
  numpages = {23},
  year = {2009},
  month = {Jun},
  publisher = {American Physical Society},
  doi = {10.1103/PhysRevD.79.124008},
  url = {https://link.aps.org/doi/10.1103/PhysRevD.79.124008}
}

@article{CoutantParentaniFinazzi2012,
  title = {Black hole radiation with short distance dispersion, an analytical $S$-matrix approach},
  author = {Coutant, Antonin and Parentani, Renaud and Finazzi, Stefano},
  journal = {Phys. Rev. D},
  volume = {85},
  issue = {2},
  pages = {024021},
  numpages = {22},
  year = {2012},
  month = {Jan},
  publisher = {American Physical Society},
  doi = {10.1103/PhysRevD.85.024021},
  url = {https://link.aps.org/doi/10.1103/PhysRevD.85.024021}
}

@article{FinazziParentani2012b,
	author = {Finazzi, S and Parentani, R},
	doi = {10.1088/1742-6596/314/1/012030},
	journal = {Journal of Physics: Conference Series},
	month = {sep},
	number = {1},
	pages = {012030},
	title = {On the robustness of acoustic black hole spectra},
	url = {https://doi.org/10.1088/1742-6596/314/1/012030},
	volume = {314},
	year = {2011}}

@article{MartinBrandenberger2002,
  title = {Corley-Jacobson dispersion relation and trans-Planckian inflation},
  author = {Martin, J\'er\^ome and Brandenberger, Robert H.},
  journal = {Phys. Rev. D},
  volume = {65},
  issue = {10},
  pages = {103514},
  numpages = {5},
  year = {2002},
  month = {May},
  publisher = {American Physical Society},
  doi = {10.1103/PhysRevD.65.103514},
  url = {https://link.aps.org/doi/10.1103/PhysRevD.65.103514}
}

@article{Weinfurtner2006,
	author = {Weinfurtner, Silke and Liberati, Stefano and Visser, Matt},
	journal = {Journal of Physics A: Mathematical and General},
	number = {21},
	pages = {6807},
	publisher = {IOP Publishing},
	title = {Analogue model for quantum gravity phenomenology},
	volume = {39},
	year = {2006}}

@article{MartinBrandenberger2013,
  title = {Trans-Planckian issues for inflationary cosmology},
  volume = {30},
  ISSN = {1361-6382},
  url = {http://dx.doi.org/10.1088/0264-9381/30/11/113001},
  DOI = {10.1088/0264-9381/30/11/113001},
  number = {11},
  journal = {Classical and Quantum Gravity},
  publisher = {IOP Publishing},
  author = {Brandenberger,  Robert H and Martin,  Jér\^ome},
  year = {2013},
  month = apr,
  pages = {113001}
}

@article{FinelliBrandenberger2002,
  title = {Generation of a scale-invariant spectrum of adiabatic fluctuations in cosmological models with a contracting phase},
  author = {Finelli, Fabio and Brandenberger, Robert},
  journal = {Phys. Rev. D},
  volume = {65},
  issue = {10},
  pages = {103522},
  numpages = {8},
  year = {2002},
  month = {May},
  publisher = {American Physical Society},
  doi = {10.1103/PhysRevD.65.103522},
  url = {https://link.aps.org/doi/10.1103/PhysRevD.65.103522}
}

@article{Grass2025,
  title = {Colloquium: Synthetic quantum matter in nonstandard geometries},
  author = {Grass, Tobias and Bercioux, Dario and Bhattacharya, Utso and Lewenstein, Maciej and Nguyen, Hai Son and Weitenberg, Christof},
  journal = {Rev. Mod. Phys.},
  volume = {97},
  issue = {1},
  pages = {011001},
  numpages = {31},
  year = {2025},
  month = {Mar},
  publisher = {American Physical Society},
  doi = {10.1103/RevModPhys.97.011001},
  url = {https://link.aps.org/doi/10.1103/RevModPhys.97.011001}
}

@misc{Brandenberger2012,
      author={Robert H. Brandenberger},
      year={2012},
      eprint={1206.4196},
      archivePrefix={arXiv},
      primaryClass={astro-ph.CO},
      url={https://arxiv.org/abs/1206.4196}, 
}

@article{MartinBrandenberger2001b,
  title = {THE ROBUSTNESS OF INFLATION TO CHANGES IN SUPER-PLANCK-SCALE PHYSICS},
  volume = {16},
  ISSN = {1793-6632},
  url = {http://dx.doi.org/10.1142/S0217732301004170},
  DOI = {10.1142/s0217732301004170},
  number = {15},
  journal = {Modern Physics Letters A},
  publisher = {World Scientific Pub Co Pte Lt},
  author = {Brandenberger,  Robert H. and Martin,  Jérome},
  year = {2001},
  month = may,
  pages = {999–1006}
}

@article{Horava2009,
  title = {Quantum gravity at a Lifshitz point},
  author = {Ho\ifmmode \check{r}\else \v{r}\fi{}ava, Petr},
  journal = {Phys. Rev. D},
  volume = {79},
  issue = {8},
  pages = {084008},
  numpages = {15},
  year = {2009},
  month = {Apr},
  publisher = {American Physical Society},
  doi = {10.1103/PhysRevD.79.084008},
  url = {https://link.aps.org/doi/10.1103/PhysRevD.79.084008}
}

@article{Jain2007,
  title = {Analog model of a Friedmann-Robertson-Walker universe in Bose-Einstein condensates: Application of the classical field method},
  author = {Jain, Piyush and Weinfurtner, Silke and Visser, Matt and Gardiner, C. W.},
  journal = {Phys. Rev. A},
  volume = {76},
  issue = {3},
  pages = {033616},
  numpages = {24},
  year = {2007},
  month = {Sep},
  publisher = {American Physical Society},
  doi = {10.1103/PhysRevA.76.033616},
  url = {https://link.aps.org/doi/10.1103/PhysRevA.76.033616}
}

@article{Robertson2017,
  title = {Assessing degrees of entanglement of phonon states in atomic Bose gases through the measurement of commuting observables},
  author = {Robertson, Scott and Michel, Florent and Parentani, Renaud},
  journal = {Phys. Rev. D},
  volume = {96},
  issue = {4},
  pages = {045012},
  numpages = {16},
  year = {2017},
  month = {Aug},
  publisher = {American Physical Society},
  doi = {10.1103/PhysRevD.96.045012},
  url = {https://link.aps.org/doi/10.1103/PhysRevD.96.045012}
}

@article{Eckel2018,
	author = {S. Eckel and A. Kumar and T. Jacobson and I. B. Spielman and G. K. Campbell},
	doi = {10.1103/PhysRevX.8.021021},
	issue = {2},
	journal = {Phys. Rev. X},
	pages = {021021},
	title = {{A Rapidly Expanding Bose-Einstein Condensate: An Expanding Universe in the Lab}},
	url = {https://link.aps.org/doi/10.1103/PhysRevX.8.021021},
	volume = {8},
	year = {2018}}

@article{Gondret2024,
     author = {Victor Gondret and Rui Dias and Clothilde Lamirault and L\'ea Camier and Amaury Micheli and Charlie Leprince and Quentin Marolleau and Scott Robertson and Denis Boiron and Christoph I. Westbrook},
     title = {Parametric pair production of collective excitations in a {Bose{\textendash}Einstein} condensate},
     journal = {Comptes Rendus. Physique},
     year = {2024},
     publisher = {Acad\'emie des sciences, Paris},
     doi = {10.5802/crphys.266},
     language = {en},
     note = {Online first},
}

@article{Svancara2024,
  title = {Rotating curved spacetime signatures from a giant quantum vortex},
  volume = {628},
  ISSN = {1476-4687},
  url = {http://dx.doi.org/10.1038/s41586-024-07176-8},
  DOI = {10.1038/s41586-024-07176-8},
  number = {8006},
  journal = {Nature},
  publisher = {Springer Science and Business Media LLC},
  author = {Švančara,  Patrik and Smaniotto,  Pietro and Solidoro,  Leonardo and MacDonald,  James F. and Patrick,  Sam and Gregory,  Ruth and Barenghi,  Carlo F. and Weinfurtner,  Silke},
  year = {2024},
  month = mar,
  pages = {66–70}
}

@article{Weinfurtner2011,
	author = {Silke Weinfurtner and Edmund W. Tedford and Matthew C. J. Penrice and William G. Unruh and Gregory A. Lawrence},
	doi = {10.1103/PhysRevLett.106.021302},
	issue = {2},
	journal = {Phys. Rev. Lett.},
	pages = {021302},
	title = {{Measurement of stimulated Hawking emission in an analogue system}},
	url = {https://link.aps.org/doi/10.1103/PhysRevLett.106.021302},
	volume = {106},
	year = {2011}}

@article{MagueijoSmolin2004,
  title = {Gravity’s rainbow},
  volume = {21},
  ISSN = {1361-6382},
  url = {http://dx.doi.org/10.1088/0264-9381/21/7/001},
  DOI = {10.1088/0264-9381/21/7/001},
  number = {7},
  journal = {Classical and Quantum Gravity},
  publisher = {IOP Publishing},
  author = {Magueijo,  João and Smolin,  Lee},
  year = {2004},
  month = mar,
  pages = {1725–1736}
}

@article{Ling2007,
  title = {Rainbow universe},
  volume = {2007},
  ISSN = {1475-7516},
  url = {http://dx.doi.org/10.1088/1475-7516/2007/08/017},
  DOI = {10.1088/1475-7516/2007/08/017},
  number = {08},
  journal = {Journal of Cosmology and Astroparticle Physics},
  publisher = {IOP Publishing},
  author = {Ling,  Yi},
  year = {2007},
  month = aug,
  pages = {017–017}
}

@article{Duval2023,
  title = {Quantum kinetics of quenched two-dimensional Bose superfluids},
  author = {Duval, Cl\'ement and Cherroret, Nicolas},
  journal = {Phys. Rev. A},
  volume = {107},
  issue = {4},
  pages = {043305},
  numpages = {13},
  year = {2023},
  month = {Apr},
  publisher = {American Physical Society},
  doi = {10.1103/PhysRevA.107.043305},
  url = {https://link.aps.org/doi/10.1103/PhysRevA.107.043305}
}

@article{Sparn2024,
  title = {Experimental Particle Production in Time-Dependent Spacetimes: A One-Dimensional Scattering Problem},
  author = {Sparn, Marius and Kath, Elinor and Liebster, Nikolas and Duchene, Jelte and Schmidt, Christian F. and Tolosa-Sime\'on, Mireia and Parra-L\'opez, \'Alvaro and Floerchinger, Stefan and Strobel, Helmut and Oberthaler, Markus K.},
  journal = {Phys. Rev. Lett.},
  volume = {133},
  issue = {26},
  pages = {260201},
  numpages = {6},
  year = {2024},
  month = {Dec},
  publisher = {American Physical Society},
  doi = {10.1103/PhysRevLett.133.260201},
  url = {https://link.aps.org/doi/10.1103/PhysRevLett.133.260201}
}

@article{Naraschewski1999,
  title = {Spatial coherence and density correlations of trapped Bose gases},
  author = {Naraschewski, M. and Glauber, R. J.},
  journal = {Phys. Rev. A},
  volume = {59},
  issue = {6},
  pages = {4595--4607},
  numpages = {0},
  year = {1999},
  month = {Jun},
  publisher = {American Physical Society},
  doi = {10.1103/PhysRevA.59.4595},
  url = {https://link.aps.org/doi/10.1103/PhysRevA.59.4595}
}

@article{Floerchinger2008,
	author = {S. Floerchinger and C. Wetterich},
	doi = {10.1103/PhysRevA.77.053603},
	issue = {5},
	journal = {Phys. Rev. A},
	pages = {053603},
	title = {{Functional renormalization for Bose-Einstein condensation}},
	url = {https://link.aps.org/doi/10.1103/PhysRevA.77.053603},
	volume = {77},
	year = {2008}
}

@book{Pitaevskii2016,
author = {Pitaevskii, Lev and Stringari, Sandro},
month = jan,
publisher = {Oxford University PressOxford},
title = {Bose-Einstein Condensation and Superfluidity},
year = {2016}
}

@article{Esteve2006,
  title = {Observations of Density Fluctuations in an Elongated Bose Gas: Ideal Gas and Quasicondensate Regimes},
  author = {Esteve, J. and Trebbia, J.-B. and Schumm, T. and Aspect, A. and Westbrook, C. I. and Bouchoule, I.},
  journal = {Phys. Rev. Lett.},
  volume = {96},
  issue = {13},
  pages = {130403},
  numpages = {4},
  year = {2006},
  month = {Apr},
  publisher = {American Physical Society},
  doi = {10.1103/PhysRevLett.96.130403},
  url = {https://link.aps.org/doi/10.1103/PhysRevLett.96.130403}
}

@incollection{LandauLifshitz1980,
	address = {Oxford},
	author = {L. D. Landau and E.M. Lifshitz},
	booktitle = {Statistical Physics (Third Edition)},
	edition = {Third Edition},
	editor = {L.D.Landau and E.M. Lifshitz},
	pages = {333-400},
	publisher = {Butterworth-Heinemann},
	title = {CHAPTER XII - FLUCTUATIONS},
	year = {1980}}

@book{LifshitzPitaevskii2013,
  title = {Statistical Physics: {{Theory}} of the Condensed State},
  author = {Lifshitz, E.M. and Pitaevskii, L.P.},
  year = 2013,
  series = {Course of Theoretical Physics},
  number = {Bd. 9},
  publisher = {Butterworth-Heinemann},
  isbn = {978-0-08-050350-9}
}

@article{Ziff1977,
	author = {Robert M Ziff and George E Uhlenbeck and Mark Kac},
	journal = {Physics Reports},
	number = {4},
	pages = {169-248},
	title = {The ideal Bose-Einstein gas, revisited},
	volume = {32},
	year = {1977}}

@software{Crameri2021,
	author = {Crameri, Fabio},
	doi = {10.5281/zenodo.5501399},
	month = sep,
	publisher = {Zenodo},
	title = {Scientific colour maps},
	url = {https://doi.org/10.5281/zenodo.5501399},
	version = {7.0.1},
	year = 2021}

@book{birrell_davies_1982,
	author = {Birrell, N. D. and Davies, P. C. W.},
	collection = {Cambridge Monographs on Mathematical Physics},
	doi = {10.1017/CBO9780511622632},
	place = {Cambridge},
	publisher = {Cambridge University Press},
	series = {Cambridge Monographs on Mathematical Physics},
	title = {Quantum Fields in Curved Space},
	year = {1982}}

@article{Ford2021,
	author = {Ford, L H},
	journal = {Reports on Progress in Physics},
	month = {oct},
	number = {11},
	pages = {116901},
	title = {Cosmological particle production: a review},
	volume = {84},
	year = {2021}}

@article{Dunster1990,
	author = {Dunster, T. M.},
	doi = {10.1137/0521055},
	eprint = {https://doi.org/10.1137/0521055},
	journal = {SIAM Journal on Mathematical Analysis},
	number = {4},
	pages = {995-1018},
	title = {Bessel Functions of Purely Imaginary Order, with an Application to Second-Order Linear Differential Equations Having a Large Parameter},
	url = {https://doi.org/10.1137/0521055},
	volume = {21},
	year = {1990}}

@misc{Dunster2025,
      author={T. M. Dunster},
      year={2025},
      eprint={2412.12595},
      archivePrefix={arXiv},
      primaryClass={math.CA},
      url={https://arxiv.org/abs/2412.12595}, 
}

@article{Wands1999,
  title = {Duality invariance of cosmological perturbation spectra},
  author = {Wands, David},
  journal = {Phys. Rev. D},
  volume = {60},
  issue = {2},
  pages = {023507},
  numpages = {4},
  year = {1999},
  month = {Jun},
  publisher = {American Physical Society},
  doi = {10.1103/PhysRevD.60.023507},
  url = {https://link.aps.org/doi/10.1103/PhysRevD.60.023507}
}

@article{Hawking1974,
	author = {Hawking, S.  W. },
	date = {1974/03/01},
	doi = {10.1038/248030a0},
	id = {HAWKING1974},
	isbn = {1476-4687},
	journal = {Nature},
	number = {5443},
	pages = {30--31},
	title = {Black hole explosions?},
	url = {https://doi.org/10.1038/248030a0},
	volume = {248},
	year = {1974}}

@article{Hawking1975,
    author = {Hawking, S. W.},
    title = {{Particle creation by black holes}},
    journal = {Comm. Math. Phys.},
    volume = {43},
    number = {3},
    pages = {199-220},
    doi = {10.1007/BF02345020},
    url = {https://doi.org/10.1007/BF02345020},
    year = {1975}
}

@article{Parker1969,
  title = {Quantized Fields and Particle Creation in Expanding Universes. I},
  author = {Parker, Leonard},
  journal = {Phys. Rev.},
  volume = {183},
  issue = {5},
  pages = {1057--1068},
  numpages = {0},
  year = {1969},
  month = {Jul},
  publisher = {American Physical Society},
  doi = {10.1103/PhysRev.183.1057},
  url = {https://link.aps.org/doi/10.1103/PhysRev.183.1057}
}

@ARTICLE{Zeldovich1971,
       author = {{Zel'Dovich}, Ya. B.},
        title = "{Generation of Waves by a Rotating Body}",
      journal = {Soviet Journal of Experimental and Theoretical Physics Letters},
         year = 1971,
        month = aug,
       volume = {14},
        pages = {180},
       adsurl = {https://ui.adsabs.harvard.edu/abs/1971JETPL..14..180Z}
}

@article{Unruh1974,
  title = {Second quantization in the Kerr metric},
  author = {Unruh, W. G.},
  journal = {Phys. Rev. D},
  volume = {10},
  issue = {10},
  pages = {3194--3205},
  numpages = {0},
  year = {1974},
  month = {Nov},
  publisher = {American Physical Society},
  doi = {10.1103/PhysRevD.10.3194},
  url = {https://link.aps.org/doi/10.1103/PhysRevD.10.3194}
}

@article{Unruh1981,
	author = {Unruh, W. G.},
	doi = {10.1103/PhysRevLett.46.1351},
	issue = {21},
	journal = {Phys. Rev. Lett.},
	month = {May},
	numpages = {0},
	pages = {1351--1353},
	publisher = {American Physical Society},
	title = {Experimental Black-Hole Evaporation?},
	url = {https://link.aps.org/doi/10.1103/PhysRevLett.46.1351},
	volume = {46},
	year = {1981}}

@misc{BarceloLiberatiVisser2024,
  author        = {Barcelo, Carlos and Liberati, Stefano and Visser, Matt},
  year          = {2024},
  eprint        = {gr-qc/0505065v4},
  archivePrefix = {arXiv},
  primaryClass  = {gr-qc},
  url={https://arxiv.org/abs/gr-qc/0505065v4}
}

@misc{Schuetzhold2025,
      author={Ralf Schützhold},
      year={2025},
      eprint={2501.03785},
      archivePrefix={arXiv},
      primaryClass={quant-ph},
      url={https://arxiv.org/abs/2501.03785}, 
}

@article{ChandranFischer2025,
  title = {Expansion-contraction duality breaking in a Planck-scale sensitive cosmological quantum simulator},
  volume = {85},
  ISSN = {1434-6052},
  url = {http://dx.doi.org/10.1140/epjc/s10052-025-15187-6},
  DOI = {10.1140/epjc/s10052-025-15187-6},
  number = {12},
  journal = {The European Physical Journal C},
  publisher = {Springer Science and Business Media LLC},
  author = {Chandran,  S. Mahesh and Fischer,  Uwe R.},
  year = {2025},
  month = dec 
}
\bibliographystyle{apsrev4-2}

\end{document}